\documentclass[%
 reprint,
 amsmath,amssymb,
 aps,
]{revtex4-2}

\usepackage{graphicx}
\usepackage{dcolumn}
\usepackage{bm}
\usepackage{subcaption}
\usepackage{makecell}
\usepackage{textcomp}
\usepackage{multirow}
\usepackage[outdir=./]{epstopdf}

\begin{document}

\preprint{APS/123-QED}

\title{Energetics, electronic states, and magnetism of iron phthalocyanine on pristine and defected graphene layers}

\author{Aleksei Koshevarnikov}
 \email{aleksei.koshevarnikov@fuw.edu.pl}
\author{Jacek A. Majewski}
 \email{jacek.majewski@fuw.edu.pl}

\affiliation{%
Institute of Theoretical Physics, Faculty of Physics,
University of Warsaw, Pasteura 5, 02-093 Warsaw, Poland
}%

\begin{abstract}
Transition metal phthalocyanines (TMPc's) are under intense scrutiny in the field of spintronics, as they may be promising storage devices. The simplicity and cheapness of such molecules increase their commercial potential. There is an active study of how the magnetic moment of the metal centre of such molecules can be changed. 
Here, we particularly consider the iron phthalocyanine molecule (FePc) on a graphene layer as a substrate. We study how graphene defects  (the Stone-Wales defect, B-doping, N-doping, S-doping, and combined B (N, S)-doped Stone-Wales defects) change the FePc electronic structure. 
We present \textit{ab initio} study of the systems, which is done using several approaches: based on periodic plane wave density functional theory (DFT), a linear combination of atomic orbitals (LCAO) DFT with a cluster representation of graphene, and multiconfigurational methods with the pyrene molecule presented as a miniaturised graphene cluster. 
The treatment of the FePc/Graphene hybrid system using multiconfigurational methods was done for the first time. It was found that the hybrid systems with B- and N- dopings have quasi-degenerate ground states and it is necessary to go beyond the approximation of one Slater determinant.
\end{abstract}

\maketitle
\section{Introduction}

The development of spintronics is inextricably linked with the commercial component. The commercial sphere is always interested in miniaturising devices and reducing the cost of production. Recently, for such purposes, magnetic tetrapyrrole molecules (such as porphyrin and phthalocyanine) with a metalic atom in the centre have been studied. Such molecules have been known for a long time,\cite{linstead1934212} are easy to manufacture, and were originally used as a colour pigment. Nowadays, they have an implementation in spintronics and optoelectronics,\cite{lian2010printed,barraud2016phthalocyanine,warner2013potential} organic photodetector,\cite{kaneko2003fast} and as a field-effect transistor.\cite{bao1996organic} Tetrapyrrole molecules with a $3d$-transition metal in the centre have a stable flat structure that allows to layer molecules on the substrate as an additional thin layer or tightly pack them vertically.\cite{zemla2020graphene} A substrate for planting such organometallic molecules is important, since upon contact the magnetic properties of molecules can change for better or worse. The nature of the adhesion and the change of molecules morphology also depend on the properties of the substrate.

Graphene is a layer of one carbon atom thick, consisting of condensed six-membered rings, constituting a honeycomb lattice. Carbon atoms in graphene are linked by $sp^2$-bonds in a hexagonal two-dimensional (2D) lattice. The first description of the production of graphene was published in 2004.\cite{novoselov2004electric} A separate layer of graphene was obtained by mechanical peeling a graphite rod on a silicon dioxide surface. 
Recent studies have shown the possibility of changing the magnetisation of a single metal-organic molecule. The control of the magnetic moment of the manganese phthalocyanine (MnPc) molecule lying on the BiAg\textsubscript{2} nanostructure using an external electric field was demonstrated.\cite{kugel2018reversible} It was theoretically predicted that control of the spin state of a structurally similar iron porphyrin molecule by stretching and compressing a graphene substrate would be also possible.\cite{bhandary2011graphene} 
In both studies, the central metal atom has two adaptive states with different magnetic moments. Computational predictions \cite{ferriani2017designing} also indicate that ZrPc or HfPc deposited on the graphene/Ni(111) substrate have two different structural conformations, for which the molecules attain different magnetic states depending on the position of the centre metallic ion, either above the Pc or between the Pc and the substrate.
Such bistability lets us represent these molecules as elementary keepers of a bit of information. 
Creating a controlled array of such molecules will bring us closer to creating a storage device based on single-molecular excitations, which in turn will lead to a significant increase in the density of information.


Recently, the TMPc/2D-material hybrid structures have also been actively investigated as catalysts for redox reactions. The emerging field of hydrogen energetics needs cheaper catalysts for fuel cells to reduce the cost of commercial production. Currently, the most popular catalysts are based on platinum group metals,\cite{steele2011materials} and the cost of producing such cells is quite high. Recent laboratory studies already show that the use of the FePc/2D materials structures shows a higher specific activity of redox reactions. In such catalysts, the iron atom is used as a centre for trapping oxygen. Then, this atomic oxygen attaches to itself two protons (the result of the splitting of a hydrogen molecule) and forms water. Among 2D materials, the best results were obtained using Ti\textsubscript{3}C\textsubscript{2}X\textsubscript{2} MXene layer,\cite{li2018marriage} but there are also many studies with graphene as a substrate for FePc molecules.\cite{park2020iron, calmeiro2020supramolecular}

Also, FePc/Graphene structures can be tuned in various ways. The use of substituents on the periphery of the phthalocyanine isoindole rings helps to change the electron density around the central iron atom.\cite{yu2016novel} Si-FePc-Graphene sandwich structures \cite{wang2020graphene} can also serve as catalysts, with the outer graphene layer protecting the device from external poisons. Axial Fe-O coordination improves oxygen adsorption and thus increases redox productivity.\cite{chen2020iron} The use of defects in the graphene layer increases the catalytic efficiency of the elements. For example, FePc/Graphene systems with nitrogen impurities on the surface demonstrated \cite{zhang2012iron} better specific activity than platinum catalysts, while such systems were characterised by a higher current density.


The interaction between graphene and transition metal phthalocyanine (TMPc) molecules is well understood. The results show that phthalocyanine molecules are attracted to the surface by van der Waals forces, and the electronic configuration of metal atoms in the centre does not change significantly. The presence of the TMPc molecule does not open the graphene band gap.\cite{feng2019phthalocyanine}  
FePc and CoPc molecules on the top of MoS\textsubscript{2} and graphene 2D layers were studied theoretically \cite{haldar2018comparative} and it was found that while the adsorption energy of a molecule to a surface in the MoS\textsubscript{2} cases is about 2.5 eV higher than in the graphene cases, both layers do not significantly change magnetic anisotropy parameters and  metallic $d$-orbitals distributions. 
When using graphene as a layer between the FePc molecule and a metal surface, graphene weakens their ferromagnetic interaction.\cite{candini2014ferromagnetic} The perpendicular stuck of FePc on graphene is also possible.\cite{zemla2020graphene} It was found that such a combination is stable and graphene barely influences on FePc magnetic properties. 

In experiments, FePc molecules found on pure graphene tend to form self-connecting structures based on the attraction of molecules to each other by van der Waals forces.\cite{de2018non} For artificial isolation of a single molecule from the film, injection of defects on graphene can be used. Theoretical calculations of the FePc molecule adsorped to the defected graphene were carried out for single vacancy, double vacancy,\cite{wang2019electronic} and (B, N, S)-dopings,\cite{sarmah2019computational} whereas the N-doping was also experimentally studied.\cite{de2018non} The above studies show a little bit higher adsorption energy of FePc to defected graphene compared to pristine graphene. Scanning tunnelling microscopy images clearly show the FePc molecule adsorped on top of the N-doping. Moreover, it turns out that the magnetic moment of the system depends on the type of defect. 
In particular, B-dopants induce the increase of the magnetic moment, N-doping leads to a decrease of the magnetic moment, whereas introducing of S-impurities causes the quenching of the magnetic moment. 

In the literature, a similar interaction of TMPc molecules with a Stone-Wales defect \cite{podlivaev2015dynamics} in graphene was considered. This defect, which consists of two pairs of five and seven carbon polygons, occurs due to the rotation of two adjacent carbon atoms relative to their centre by 90 degrees. Thus, when a defect is formed, there is no change in the chemical composition of the material. This fact may lead to the idea of maintaining the magnetic moment of the TMPc - defected graphene system. The adsorption energy of TMPc to graphene with the Stone-Wales defect was calculated \cite{mendoza2021distortion} to be 6\% higher for Zn as TM in TMPc and 10\% higher for Cu. Studies of the iron porphyrin/graphene/Ni(111) revealed \cite{bhandary2013defect} that when the Stone-Wales defect is formed the graphene layer is not flat anymore, just exhibiting a wavelike shape.

Most of the studies of systems consisting of a two-dimensional surface and a metal-organic molecule were carried out using the Kohn-Sham realisation of the density functional theory (DFT) employing plane-waves as the basis set. This method reproduces well the geometry of a two-dimensional surface due to the fact that periodic conditions are used but does not allow one to study the $d$, $f$ - orbitals of metal atoms, which are of direct interest for such structures.

Nowadays, the computational capabilities allow us to carry out studies of 2D-surface - metal-organic molecule complexes using multireference methods. For example, the iron porphyrin molecule and a graphene ribbon were treated separately with  very high-level accuracy.\cite{levine2020casscf} Porphyrin without a central metal on the graphene oxide was studied using multireference methods with 8 orbitals as active ones.\cite{siklitskaya2021lerf}
To employ multireference methods, the problem has to be reformulated in terms of finite, non-periodic systems, commonly used in quantum molecular chemistry.
This problem can be easily solved in the case of graphene. A limited piece of the graphene layer, hereinafter referred to as a cluster, can be limited by functionalising the extreme carbon atoms with hydrogen. With a sufficiently large graphene layer in the area, it is possible to create a structure very similar in physical and chemical properties to pure graphene.

Here, we use both DFT and multireference approaches to investigate the FePc/Graphene hybrid system.
In addition to the interaction of FePc with the pure sheet of graphene, the interaction of FePc with defects in graphene is also of interest. One of the main defect selection factors was the compliance of geometric parameters between the system with boundary conditions and the cluster. For example, it was found that graphene clusters, in the centre of which one (single valence) or two (double valence) atoms are missing, do not repeat the flat structure of analogous periodic systems; stable states obtained after optimization have strong curvatures. Therefore, further comparison of periodic and cluster systems with these defects is not possible.

Stone-Wales defects and doping of atoms are of particular interest for studying. The interaction of the Stone-Wales graphene defect with the FePc molecule has not been studied thoroughly enough, limiting itself to describing similar structures.\cite{mendoza2021distortion, bhandary2013defect}
Doping of atoms in the case of a graphene cluster requires special consideration, because, in contrast to a doped periodic structure, in which there may be no magnetic moment, the cluster must have an initial multiplicity in the case of adding an atom with an odd number of electrons. When a molecule with an intrinsic spin moment is added to a doped cluster, several options for choosing the spin moment arise. To study this situation, boron, nitrogen, and sulfur atoms were considered. Systems with the same doped atoms were studied previously \cite{sarmah2019computational} using density functional theory and these results could be compared with obtained results from cluster representation and multireference analysis.

The combined effects formed by replacing one of the atoms in the Stone-Wales defect were also studied. A theoretical study of such a multilevel defect using density functional theory predicts a slight broadening of the band gap, as well as an improvement in the accumulation of surface charge.\cite{zhou2020effect} Also boron, nitrogen and sulfur were considered as dopants.

The multireference analysis was done by replacing the graphene cluster with a smaller pyrene molecule. This change barely influences on geometrical and energetical properties of the hybrid system but allows us to perform the multireference analysis on a much higher level of theory. The FePc/Pyrene hybrid systems with the defects were also studied and discussed.  

\section{Theoretical Methods}

The periodic DFT calculations were performed employing the Quantum Espresso 6.5 package\cite{giannozzi2009quantum} on the level of the generalized gradient approximation (GGA) \cite{langreth1983beyond} using the Perdew-Burke-Ernzerhof (PBE) exchange-correlation functional.\cite{perdew1996generalized} The Van der Waals interaction between a molecule and a layer was included using the Grimme DFT-D3 methodology.\cite{grimme2010consistent} To treat the strong on-site Coulomb interaction of TM d-electrons, we used DFT+U approach within the Hubbard model.\cite{anisimov1997first} The U parameter value for the iron atom was taken from the results of linear response calculations for TMPc molecules.\cite{brumboiu2019ligand} The kinetic energy cutoff for wavefunctions was taken to be 45 Ry, and a corresponding parameter for charge density and potential was 450 Ry. Preliminary estimations, optimization, and calculation of energy parameters were performed at the $\Gamma$ point. Calculations of densities of states were performed using the 2 × 2 × 1 Monkhorst–Pack k-point mesh. Optimisation was carried out until a force value of less than 0.001 Ry/a.u on every atom and in each cartesian direction was achieved, and simultaneously stress tensor components reached values smaller than 0.5 kbar.

The calculations of cluster systems were performed using the \textit{ORCA} package,\cite{neese2012orca, neese2018software} where the GGA-PBE functional and the D3 correction were used as in computations with periodic boundary conditions.
 The triple-$\zeta$ polarized def2-TZVP basis set \cite{weigend2005balanced} was implemented and a semi-empirical counterpoise-type correction \cite{kruse2012geometrical} was used to decrease the basis set superposition error (BSSE). 

The following equation was used to estimate the adsorption energy $E_a$: 
\begin{equation}
 E_a = E_{FePc+Gr} - E_{FePc} - E_{Gr},
 \label{eq:Adsorption}
\end{equation}
where $E_{FePc+Gr}$ denotes the total energy of the FePc/graphene hybrid system, whereas $E_{FePc}$, and $E_{Gr}$ are energies of the free molecule, and the surface, respectively.

Cohesive energies for defected graphene clusters and pyrene were calculated using the formula
\begin{equation}
E_{coh} =\frac{E_{doped\ gr} -\ \sum ^{N}_{i} E_{i}}{N},
\label{eq:Cohesive}
\end{equation}
where $E_{doped\ gr}$ is the total energy of the doped system, $E_{i}$ is the total energy of the individual elements i (i = C, B, N, S, H), and N is the total number of atoms in the cluster. Cohesive energy shows the energy difference between the energy of atoms in the molecular state and in the gas state. This parameter allows for comparing the stability of defected structures. Similar calculations have also been done before.\cite{ricca2018b}

The multireference calculations of cluster systems were performed using the \textit{ORCA} package.\cite{neese2012orca, neese2018software} Optimisation of systems geometry was performed using DFT methods. Multireference calculations were done using different basis sets for different types of atoms: polarized valence double-zeta basis set def2-SVP for hydrogen and carbon atoms, diffuse polarized triple-zeta basis set def2-TZVPD for nitrogen, boron and sulfur atoms and diffuse doubly polarised triple-zeta basis set def2-TZVPPD for the iron atom. Also, the RIJCOSX \cite{neese2009efficient} algorithm that treats the Coulomb term via RI (repulsion integrals) and the exchange term via seminumerical integration was implemented. Corrected energetic states of the hybrid systems were found using strongly contracted NEVPT2 \cite{angeli2001introduction} perturbation theory.

One of the most important aspects at the beginning of CASSCF calculations is the choice of the orbitals for the active space. 
The method of constructing initial orbitals for further analysis that works quite well is the fragment derived guess which is implemented in \textit{ORCA}.
This method assumes the fragmentation of the molecular complex into ligand, and metal centre parts. Molecular orbitals are obtained for each fragment, where ligand and metal orbitals are found using DFT and CASSCF methods, respectively. Then the resulting orbitals are merged for further calculations. This method allows for the use of orbitals based on pure metal once in the analysis.

The fragment derived guess works well for systems where a multiplicity of a ligand is odd. In this case, a first CASSCF calculation can be performed using only metal $d$-orbitals as active CAS(6,5), where 6 is the number of electrons and 5 is the number of orbitals. Then the active space can be expanded using the ligand orbitals to CAS(10,9). The difficulty is that the CASSCF method requires full occupation of core orbitals. It means that FePc/Pyrene systems with an odd number of electrons in defects (B- and N-dopings) can not be treated using the CASSCF method with only metal $d$-orbitals in the active space. The solution to this problem was the gradual inclusion of orbitals in space. After the fragment derived guess analysis, such systems were studied using CASSCF(5,5), where one electron was removed from the $d$-orbital active space. Then the active space was expanded systematically to CAS(7,6) $\rightarrow$ CAS(9,7) $\rightarrow$ CAS(9,8) $\rightarrow$ CAS(11,9). The results of final calculations usually show the occupation of iron $d$-orbitals as in the usual FePc molecule, and the auxiliary assumption of the removal of one electron from the iron $d$-shell disappeared. 
The results of the final calculations show that the auxiliary assumption about the removal of one electron from the iron $d$-shell has disappeared and the electron density in the $d$-shell is quantitatively similar to the electron density for the free standing FePc molecule.

\section{Results and discussion}

\subsection{FePc}

\begin{figure}
\centering
\includegraphics[width=\columnwidth]{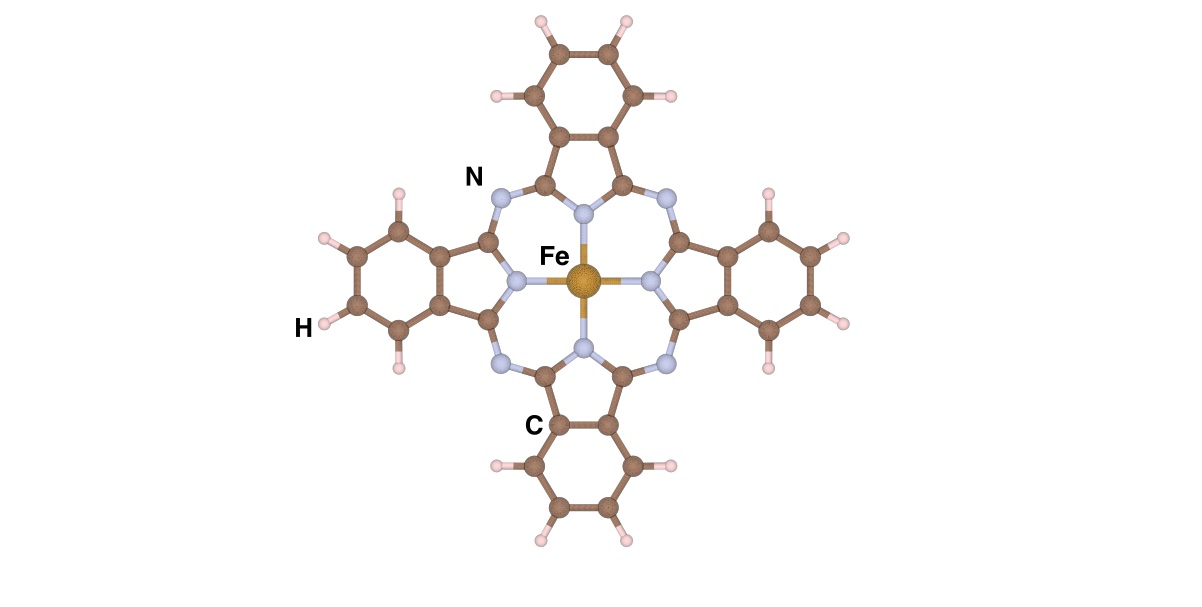}
\caption{The FePc molecule}
\label{fig:FePc}
\end{figure}

An isolated iron phthalocyanine molecule (Fig. \ref{fig:FePc}) has a tetragonal D\textsubscript{4h} symmetry. The central iron atom is in the square planar ligand field which is created by nitrogen atoms. The strong ligand field makes the iron $d_{x^{2} -y^{2}}$ orbital unfavourable, and, therefore, the ground state of the molecule is triplet. The exact ground state is a topic to study due to the energetic proximity of the two states. While it seems to be commonly accepted that the ground state of FePc is $E_{g}$ with the iron $3d$-shell configuration $d_{xy}^{2} d_{xz}^{2} d_{yz}^{1} d_{z^{2}}^{1} d_{x^{2} -y^{2}}^{0}$, there exist computations predicting $A_{2g}$ state (with the $d_{xy}^{2} d_{xz}^{1} d_{yz}^{1} d_{z^{2}}^{2} d_{x^{2} -y^{2}}^{0}$ iron $3d$-shell configuration) as the ground.\cite{ichibha2017new} 

The excited quintet and singlet states can be modelled using one-Slater-determinant DFT methods. Using a linear combination of atomic orbitals (LCAO) basis sets for the free FePc molecule, it was found that quintet and singlet states are higher by 1.24 and 1.43 eV, respectively, in comparison to the ground state energy. The Fe-N distances in the molecule are similar for all states and are about 1.935 \AA. Plane-wave methods with the molecule in the cubic cell with the 20 \AA \hspace{0pt} face show that the Fe-N bond lengths are 1.95 \AA\hspace{0pt} in the ground state, 2.01 \AA\hspace{0pt} in the quintet and 1.92 \AA\hspace{0pt} in the singlet. This elongation of the bonds was previously found for molecules with the Fe-N\textsubscript{4} centre.\cite{bhandary2011graphene} The Fe-N bonds lengthening weakens the ligand field created by nitrogen atoms of tetragonal symmetry. Thus, the electrons of the d-shell of the iron atom occupy the higher energy orbitals.
The quintet excited state is 0.44 eV higher than the triplet state and the singlet state is 2.68 eV higher. The precise multideterminant NeVPT2 method (Table \ref{table:FePcTrEnergies}) gives lower values of the transition energies than DFT. A more detailed discussion of the results will be made below in the part about multireference calculation analysis.

\begin{table}
        \centering
        \caption{Energy differences between the FePc ground triplet state and excited quintet and singlet states were calculated by different quantum chemistry methods.}
        \label{table:FePcTrEnergies}        
\begin{tabular}{ccc}
\hline
  &  Triplet-Quintet, eV & Triplet-Singlet, eV \\
\hline 
 DFT, LCAO & 1.24 & 1.43 \\
 DFT, Plane Wave & 0.44 & 2.68 \\
 NeVPT2(6,5) & 0.41 & 1.28 \\
 NeVPT2(10,9) & 0.4 & 0.85 \\
 \hline
\end{tabular}

        \end{table}

\subsection{Graphene}

\subsubsection{Graphene Cell Choice for the Calculations with Periodic Boundary Conditions}

The choice of the graphene surface size was dictated by a compromise between computational time and the accuracy of calculations. In the case of the periodic systems, we wanted to get rid of the eventually possible spurious interaction between FePc molecules, both in lateral plane and the vertical direction, in order to study the interaction of the strictly single molecule with the graphene layer. Such spurious interactions between the FePc molecule and its images in the neighbouring cell can be induced by periodic boundary conditions imposed on the system if the dimensions of the employed supercell are not sufficiently large. 
The estimations of the effect mentioned above (Table \ref{table:FePcGrDiffCells}) were performed using different graphene supersells, from the smallest possible supercell where FePc molecules can lie on the surface without overlapping each other up to considerably larger supercells. For purpose of such estimations, only the molecule was relaxed, while the graphene layer was frozen. That is why the estimations slightly differ from the results presented below.
Gradually increasing the lattice size, it was found that when using a cell consisting of 9x9 unit cells of graphene, the intermolecular interaction is practically absent and does not affect the geometry and energy parameters of the complex. This supercell was chosen for further studies.

\begin{table}
        \centering
        \caption{Adsorption energies and distances between FePc molecules (defined as the distance between the two closest hydrogen atoms of different molecules) for FePc/Gr systems as a function of supercell size.}
\begin{tabular}{ccc}
\hline
{Supercell Size} & {$E_a$, eV} & {Distance between FePc's, \AA} \\
\hline 
 6x6 & -6.73 & 1.60 \\
 7x7 & -4.72 & 2.27 \\
 8x8 & -3.73 & 4.62 \\
 9x9 & -3.01 & 7.07 \\
 10x10 & -3.03 & 9.53 \\
 15x15 & -3.33 & 21.68 \\
 \hline
\end{tabular}
        \label{table:FePcGrDiffCells}
        \end{table}

\subsubsection{Graphene Cluster Choice}

\begin{figure}
\centering
\begin{subfigure}{0.49\linewidth}
  \centering
  \includegraphics[width=\linewidth]{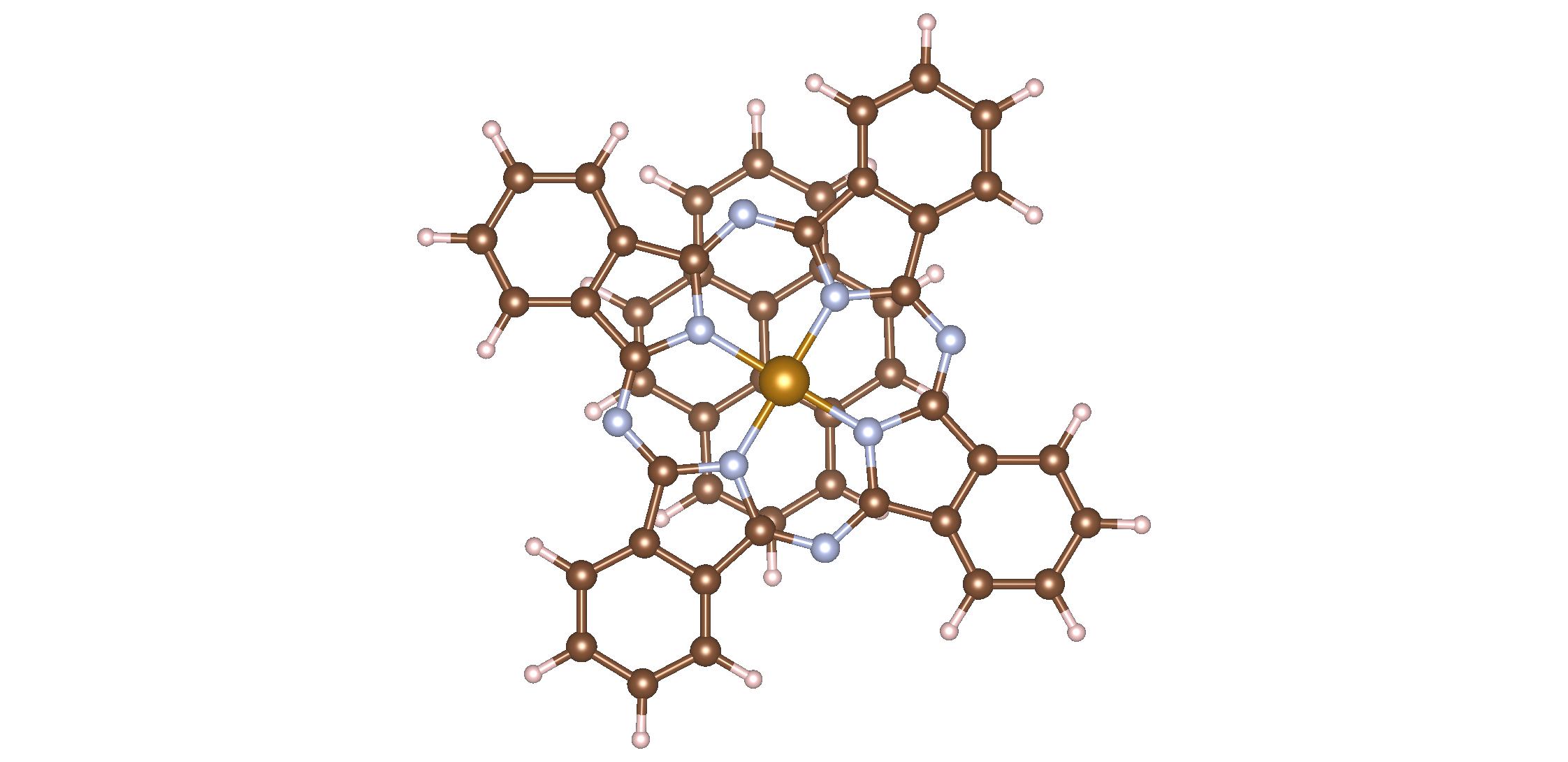}
  \caption{}
  \label{fig:FePcPyrene}
\end{subfigure}
\hfill
\begin{subfigure}{0.49\linewidth}
  \centering
  \includegraphics[width=\linewidth]{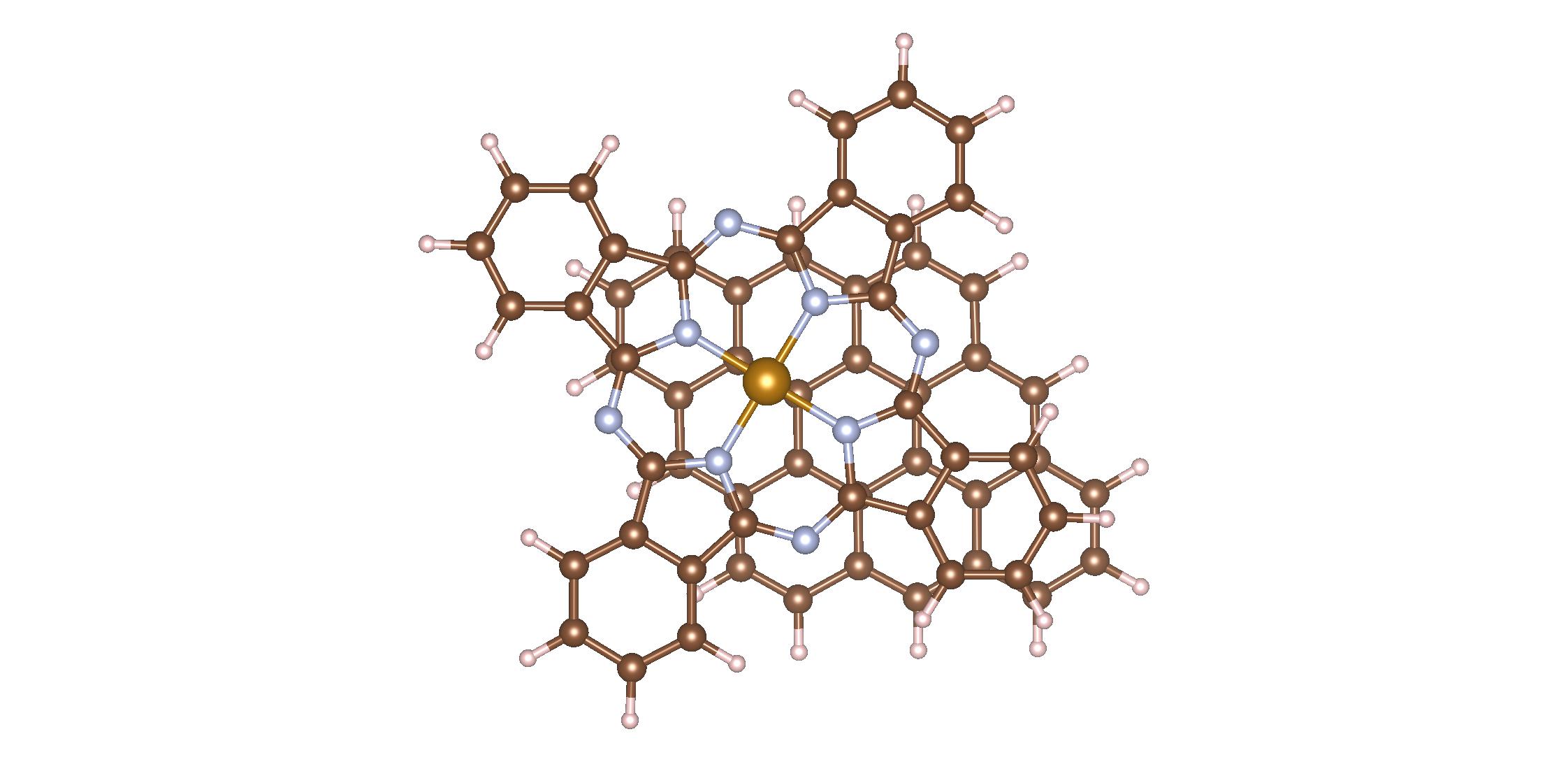}
  \caption{}
  \label{fig:FePcGr3x3}
\end{subfigure}
\begin{subfigure}{0.49\linewidth}
  \centering
  \includegraphics[width=\linewidth]{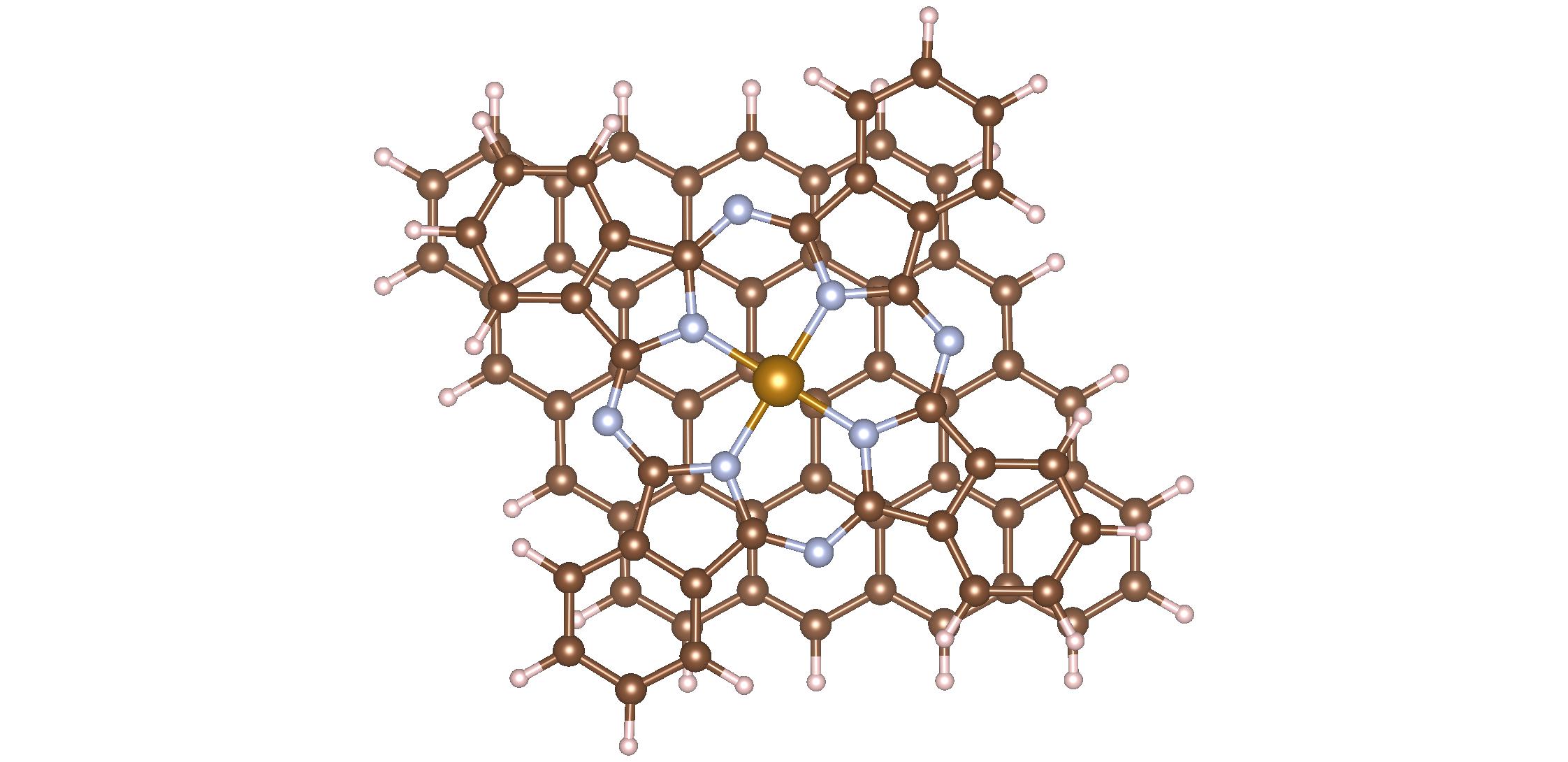}
  \caption{}
  \label{fig:FePcGr4x4}
\end{subfigure}
\begin{subfigure}{0.49\linewidth}
  \centering
  \includegraphics[width=\linewidth]{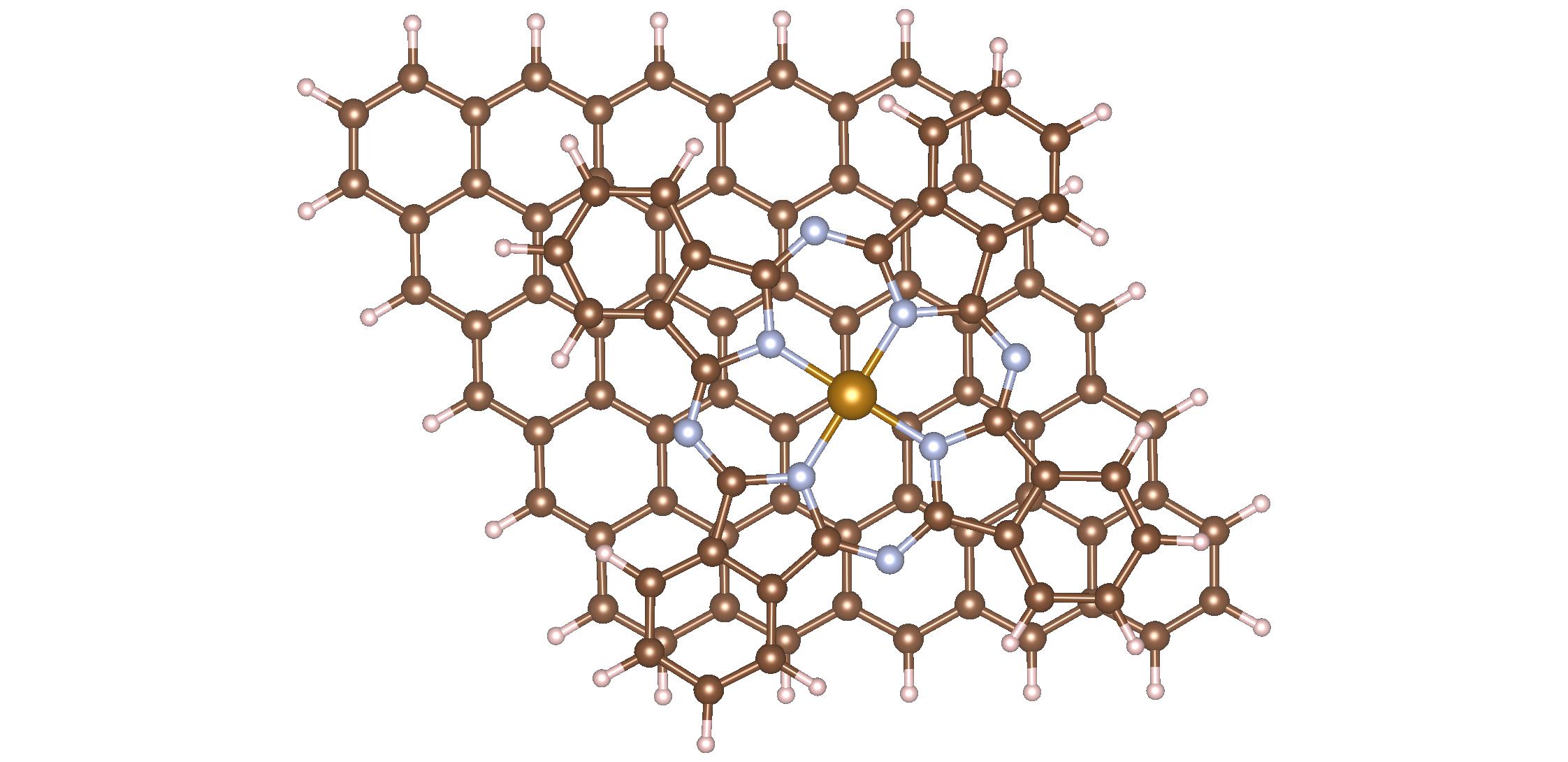}
  \caption{}
  \label{fig:FePcGr5x5}
\end{subfigure}

\caption{ Structures of FePc/Graphene hybrid systems with various sizes of graphene flakes:
different graphene areas: (a) FePc/Graphene2x2(Pyrene), (b) FePc/Graphene3x3, (c) FePc/Graphene4x4, (d) FePc/Graphene5x5.}
\label{fig:FePcGrFlakes}
\end{figure}

In the case of the cluster calculations, the main test parameters were, like in the case of supercell calculations, geometry and physisorption energy. The structure with a cluster of 25 (5x5) rings (Fig. \ref{fig:FePcGr5x5}) shows similar physisorption energy (Table \ref{table:FePcGrDiffClusters}) and also completely covers the surface under the molecule. An enlargement of the graphene surface significantly increases the computational burden. The FePc/Gr 5x5 model is good for comparing the possibility of representing the FePc/Gr system as a cluster but too big to perform multireference NeVPT2 calculations. Even in the small def2-SVP basis set more than 2 TB of RAM is needed, and available at the University of Warsaw computer resources do not allow us to perform such calculations with sufficient accuracy.
Therefore, we have also performed the DFT calculations with the smaller sizes of graphene flakes consisting of 16 (4x4), 9 (3x3), and 4 (2x2) carbon rings. These systems are depicted in
Fig. \ref{fig:FePcGr4x4}, \ref{fig:FePcGr3x3}, and \ref{fig:FePcPyrene}, respectively. The smallest flake (the 2x2 graphene cluster) is just the pyrene molecule. The values of adsorption energy and geometrical parameters of hybrid systems are shown in the table \ref{table:FePcGrDiffClusters}. It is clearly seen that adsorption energy decreases with the cluster size, which means that the adhesion of the FePc molecule to the graphene flake gets stronger with the flake's size. We ascribe this effect to the vdW interaction between carbon rings ("petals") in the molecule and the flakes. 
FePc-graphene and Fe-graphene distances differ for different complexes but the differences do not exceed standard deviations of FePc and clusters from the $xy$-plane. Both energetics and geometric parameters show that the van der Waals interaction determines the adsorption in all cases. These results strongly suggest that the system of FePc molecule on the smaller graphene flake can be taken for extremely costly computations with a multiconfigurational approach, particularly, in the case when one focuses on the iron atom and its closest surrounding.

\begin{table}
        \centering
        \caption{Adsorption energies and geometrical parameters of FePc/Graphene hybrid structures with various sizes of graphene flakes. The averaged distances between the graphene layer and FePc molecule are given in the third column, whereas the averaged distance between Fe atom and C atoms belonging to the graphene flake is depicted in the fourth column. Absolute error values have been calculated as a standard deviation from a mean z-coordinate of the layer and the molecule.}
\begin{tabular}{cccc}
\hline
  & $E_a$, eV &FePc-Gr dist, \AA &Fe-Gr dist, \AA\\
\hline 
 FePc/Gr5x5 & -2.03 & 3.44\textpm0.14 & 3.41\textpm0.09 \\
 FePc/Gr4x4 & -1.44 & 3.36\textpm0.26 & 3.43\textpm0.14 \\
 FePc/Gr3x3 & -1.37 & 3.39\textpm0.38 & 3.46\textpm0.21 \\
 FePc/Pyrene & -1.00 & 3.28\textpm0.19 & 3.36\textpm0.07 \\
\hline
\end{tabular}
        \label{table:FePcGrDiffClusters}
        \end{table}
 
\subsubsection{Pyrene}

\begin{figure}
\centering
\begin{subfigure}{0.49\linewidth}
  \centering
  \includegraphics[width=\linewidth]{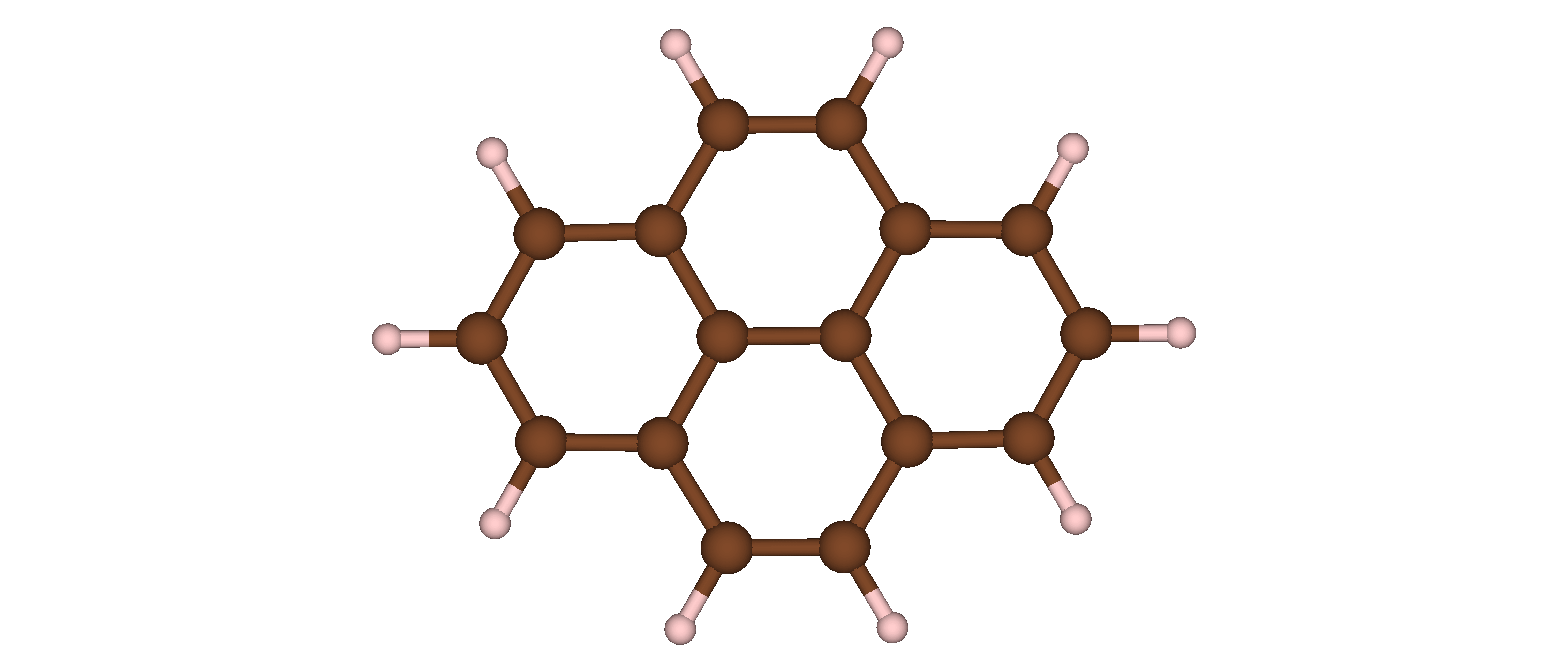}
  \caption{}
  \label{fig:Pyrene}
\end{subfigure}
\begin{subfigure}{0.49\linewidth}
  \centering
  \includegraphics[width=\linewidth]{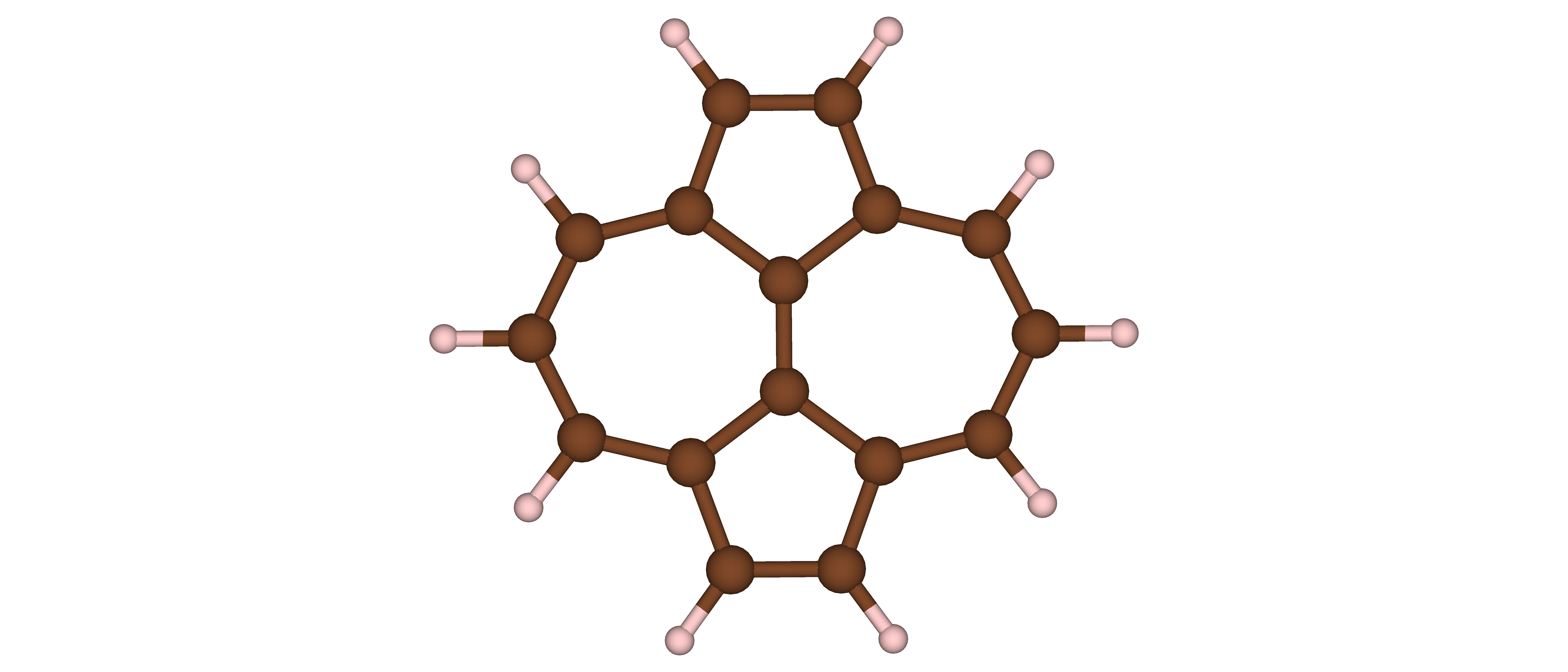}
  \caption{}
  \label{fig:PyreneSW}
\end{subfigure}
\caption{ Structures of (a) pyrene and (b) pyrene with a Stone-Wales defect}
\label{fig:PyreneStructures}
\end{figure}

Pyrene (C\textsubscript{16}H\textsubscript{10}, Fig. \ref{fig:Pyrene}) consists of four tight aromatic rings. It is the smallest carbon structure that can contain studied defects. Some of these defects have already been studied. The DFT study of the Stone-Wales defect in pyrene (Fig. \ref{fig:PyreneSW}) shows\cite{campisi2020defects} that the 8.26 eV energy barrier should be overcome to form a bond while the energy difference between pyrene and Stone-Wales pyrene is 2.35 eV. It has also been shown that the additional hydrogen atom on one of the 
central pyrene carbon atoms lowers the energy barrier to 6.6 eV. The substitution of two central carbon atoms into boron and nitrogen atoms has been done experimentally\cite{shi2021nbn, https://doi.org/10.1002/anie.200703535} and this system exhibits excellent stability. 

\subsubsection{Defects in Graphene}

\begin{figure}
\centering
\begin{subfigure}{0.49\linewidth}
  \centering
  \includegraphics[width=\linewidth]{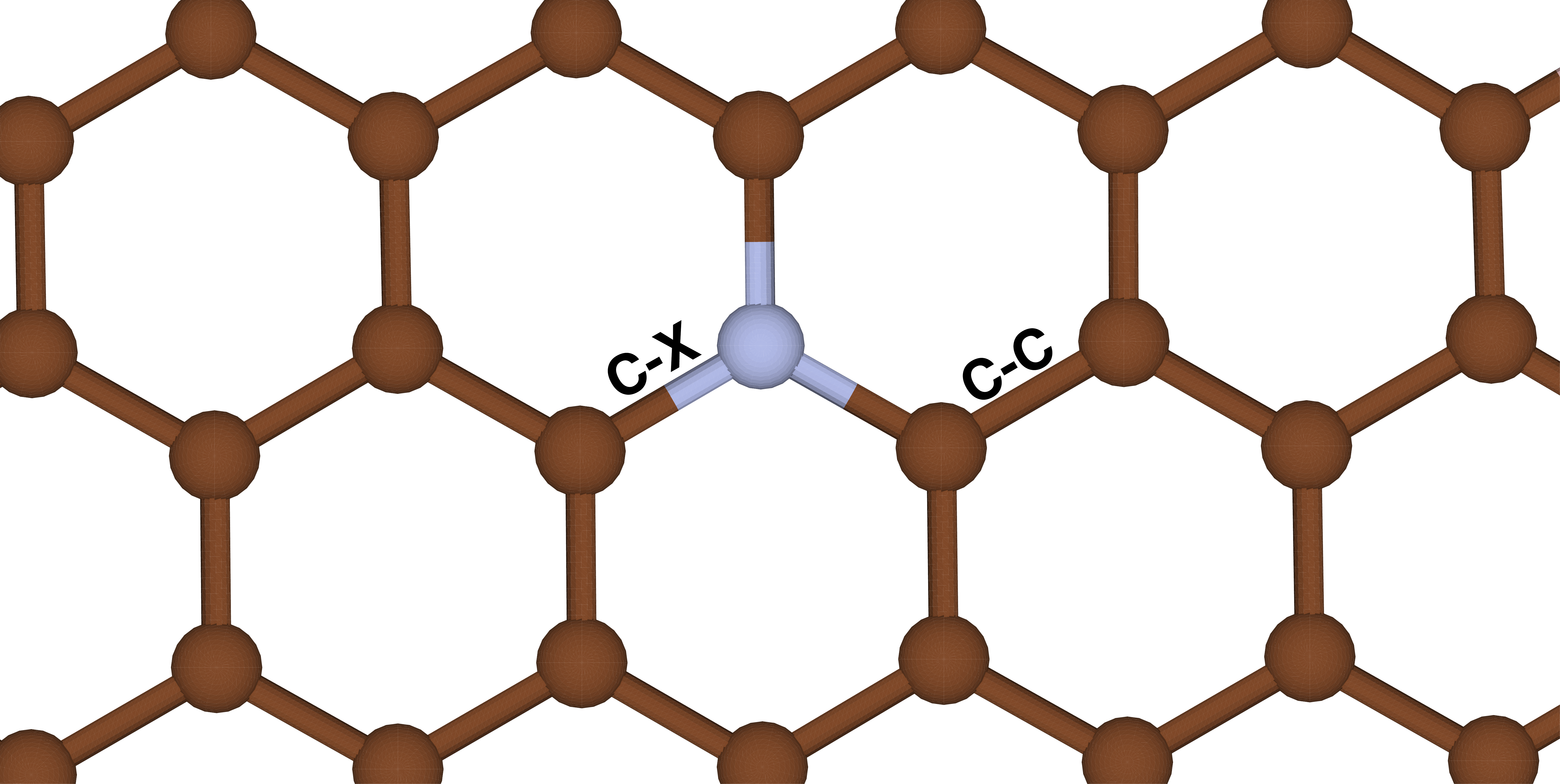}
  \caption{}
  \label{fig:GrX}
\end{subfigure}
\hfill
\begin{subfigure}{0.49\linewidth}
  \centering
  \includegraphics[width=\linewidth]{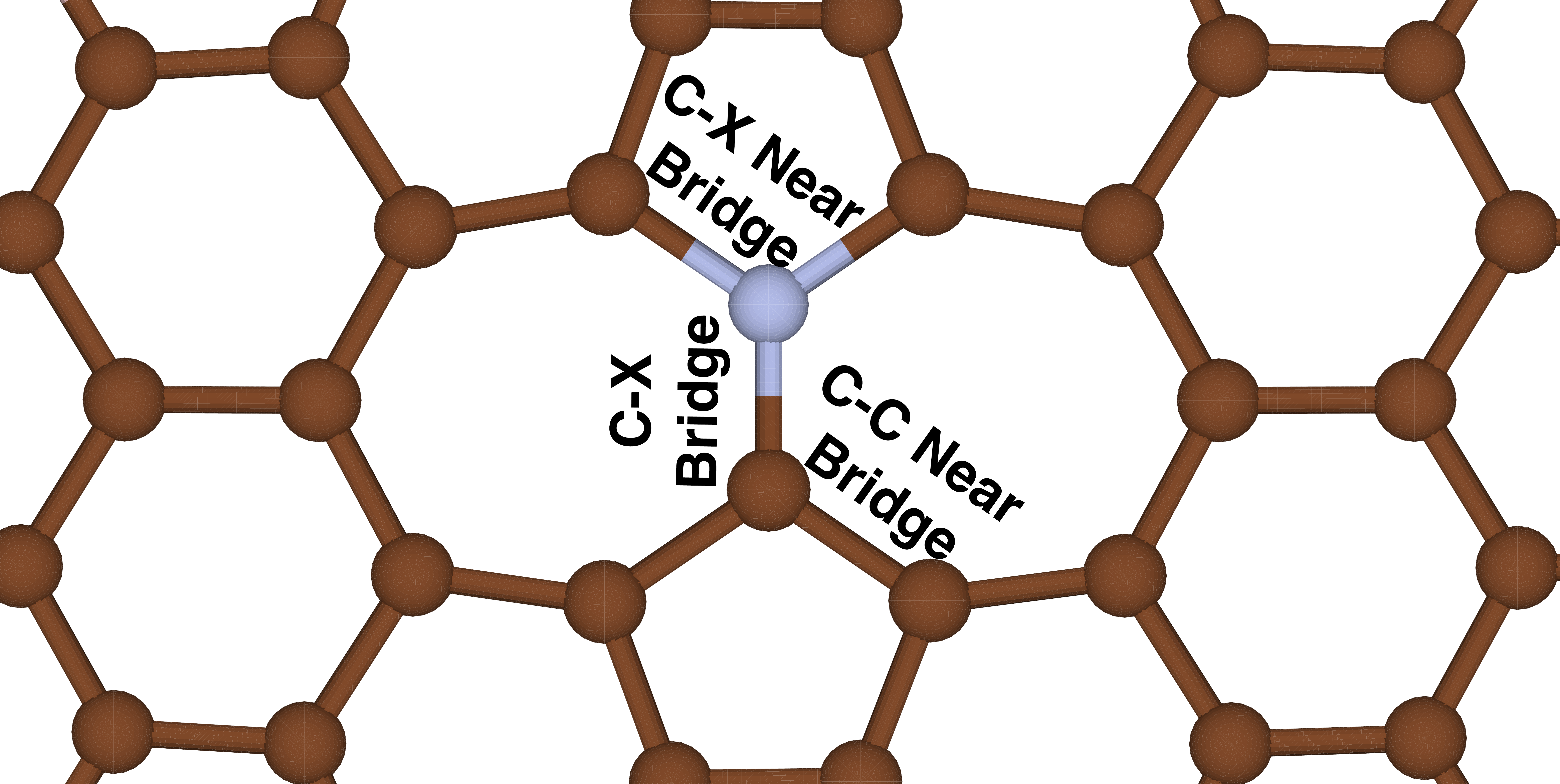}
  \caption{}
  \label{fig:Gr57X}
\end{subfigure}
\caption{Structures of studied graphene defects; (a) doped graphene, (b) graphene with the Stone-Wales graphene and doping.}
\label{fig:GrDefects}
\end{figure}

Let us now analyse various defects in graphene. First, we consider Stone-Wales 5-7 defect (hereafter indicated as graphene-SW), and substitutional impurities B, N, and S on the carbon site. Further on, we consider more complex defects where B, N, or S substitute the carbon atom in the Stone-Wales defect area. These defects are indicated as B-SW, N-SW, and S-SW. The simple substitutional defects, as well as X-SW defects, are depicted in Fig. \ref{fig:GrX} and \ref{fig:Gr57X}, respectively. 

As the host of these defects we consider three systems: (i) graphene supercell with periodic boundary conditions, (ii) the graphene cluster with edge carbon atoms saturated by hydrogen, and (iii) pyrene molecule. For pristine graphene, in the DFT calculations for all three systems listed above, the carbon-carbon bond length is 1.42 \AA. 
The optimised geometry of all 7 types of defects in three host types is summarised in Tables \ref{table:GeomDopDef} and \ref{table:GeomSWDopDef}.
It should be noted that all defected structures have a flat geometry. Calculations using cluster and periodic conditions, as well as the pyrene molecule generally show similar results. In the case of a doping atom, the bonds of the added atom with neighbouring carbon atoms are lengthened, whereas the bonds closest to the C-C defect are shortened. The Stone-Wales defect forms a bridge between pentagons and this bond is shorter than the standard bond in graphene. With the addition of a boron, nitrogen, or sulphur atom, this bridge bond increases, but remains shorter than the bonds of the doped atom with the rest of the carbon atoms. These effects are most visible in the case of sulphur doping, whereas for boron and nitrogen atoms these changes are rather tiny. 

\begin{table}
        \centering
        \caption{Geometric parameters of substitutional impurities of boron, nitrogen, and sulfur atom. "C-X" bond length is an average length between the impurity atom and adjacent carbon atoms; the "C-C" bond length is the average length of the closest \ to the defect carbon bonds.}
        \label{table:GeomDopDef}        
\begin{tabular}{l ccc ccc}
\hline
  & \multicolumn{3}{c}{ 
C-X bond length, \AA } & \multicolumn{3}{c}{
C-C bond length, \AA} \\
\hline 
  & Periodic  & Cluster & Pyrene
 & Periodic & Cluster & Pyrene \\

 Gr-B & 1.49 & 1.49 & 1.53 & 1.41 & 1.41 & 1.41 \\

 Gr-N & 1.41 & 1.4 & 1.43 & 1.42 & 1.42 & 1.42 \\

 Gr-S & 1.6 & 1.6 & 1.76 & 1.4 & 1.4 & 1.39 \\
 \hline
\end{tabular}
        \end{table}

\begin{table}
        \centering
        \caption{Geometric parameters of the Stone-Wales defect, as well as combined Stone-Wales defects with doped boron, nitrogen, and a sulfur atom. "C-X Bridge" bond length is a length between the pentagons; "C-X Near Bridge" bond length is an average length of a doped atom with neighbouring carbon atoms; "C-C Near Bridge" bond length is an average length between a carbon atom located on the bridge with other carbon atoms.}
        \label{table:GeomSWDopDef}
\begin{tabular}{l ccc}
\hline
  & Periodic & Cluster & Pyrene \\
\hline 
 \multicolumn{4}{c}{C-X Bridge bond length, \AA} \\
\hline 
 Gr-SW & 1.33 & 1.35 & 1.38 \\
 Gr-B-SW & 1.4 & 1.43 & 1.48 \\
 Gr-N-SW & 1.33 & 1.36 & 1.4 \\
 Gr-S-SW & 1.56 & 1.6 & 1.71 \\
\hline 
 \multicolumn{4}{c}{C-X Near Bridge bond length, \AA} \\
\hline 
 Gr-B-SW & 1.52 & 1.53 & 1.55 \\
 Gr-N-SW & 1.42 & 1.42 & 1.42 \\
 Gr-S-SW & 1.63 & 1.65 & 1.70 \\
\hline 
 \multicolumn{4}{c}{C-C Near Bridge bond length, \AA} \\
\hline 
 Gr-SW & 1.46 & 1.47 & 1.47 \\
 Gr-B-SW & 1.43 & 1.44 & 1.44 \\
 Gr-N-SW & 1.44 & 1.45 & 1.45 \\
 Gr-S-SW & 1.41 & 1.42 & 1.40 \\
 \hline
\end{tabular}

        \end{table}


To assess the stability of the studied defects, cohesion energies (Eq. \ref{eq:Cohesive}) were calculated for cluster and pyrene models; the results of how much the energies of these models are higher than the energy of the cluster (pyrene) without defects are shown in Table \ref{table:Cohesive}. It can be seen from the results that the formation of defects leads to a decrease in stability; nevertheless, the Stone-Wales defect and the implantation of a boron atom lead to minor decreases. Doping with a sulfur atom somewhat lowers the stability of the surface. Nevertheless, it has been shown that a formation with a doped sulfur atom is thermodynamically more favourable than a vacancy.\cite{lu2017sulfur}

\begin{table}
        \centering
        \caption{Cohesive energies relative to pristine graphene for the cluster and pyrene with studied defects.}
        \label{table:Cohesive}        
\begin{tabular}{ccc}
\hline
  & $E_{coh Cluster}$, eV & $E_{coh Pyrene}$, eV \\
\hline 
 Gr-SW & 0.035 & 0.08 \\
 Gr-B & 0.029 & 0.13 \\
 Gr-SW-B & 0.059 & 0.17 \\
 Gr-N & 0.037 & 0.20 \\
 Gr-SW-N & 0.069 & 0.16 \\
 Gr-S & 0.111 & 0.40 \\
 Gr-SW-S & 0.132 & 0.39 \\
 \hline
\end{tabular}
        \end{table}

\subsection{FePc/Graphene}

\subsubsection{FePc on Graphene and Graphene-SW}

The hybrid system FePc/Gr with the FePc molecule placed over the graphene layer has been studied theoretically.\cite{wang2019electronic} One of the important issues is the molecule position relative to the graphene layer.
Typically, one considers the location of the iron atom above the centre of the carbon ring (Hex), above the carbon atom (Top), and between the carbon atoms at the hexagon's edge (Bridge). It turned out that the Top and Bridge positions are more favourable and have extremely similar physisorption energies. Optimization performed in our QE computations does not allow us to really accurately point out the energetically most favourable structure, but the geometries of the Top and Bridge structures are nearly identical.

\begin{figure}
\centering
\begin{subfigure}{0.49\linewidth}
  \centering
  \includegraphics[width=\linewidth]{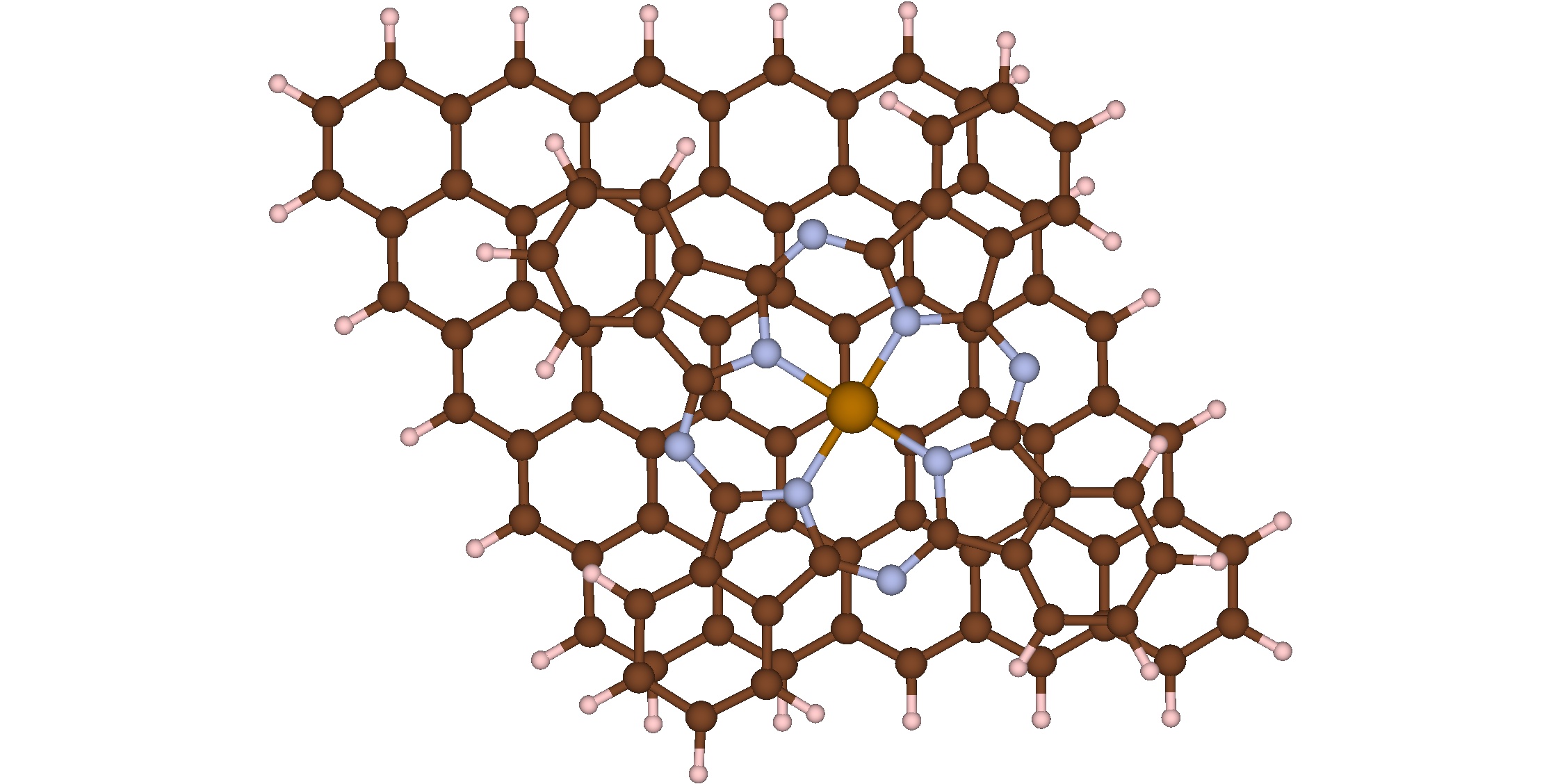}
  \caption{}
  \label{fig:FePcGrOpt}
\end{subfigure}
\hfill
\begin{subfigure}{0.49\linewidth}
  \centering
  \includegraphics[width=\linewidth]{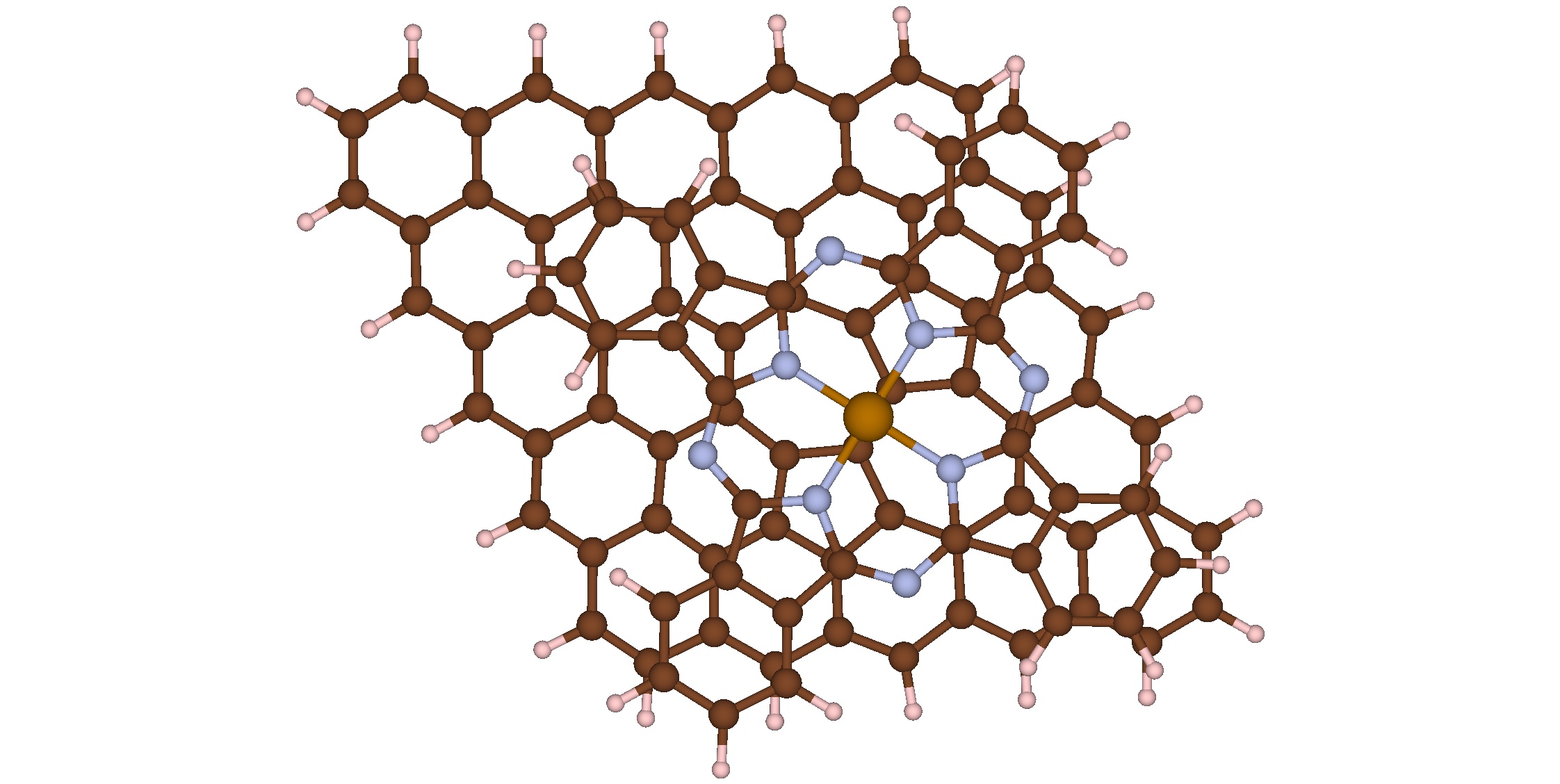}
  \caption{}
  \label{fig:FePcGr57Opt}
\end{subfigure}
\begin{subfigure}{0.49\linewidth}
  \centering
  \includegraphics[width=\linewidth]{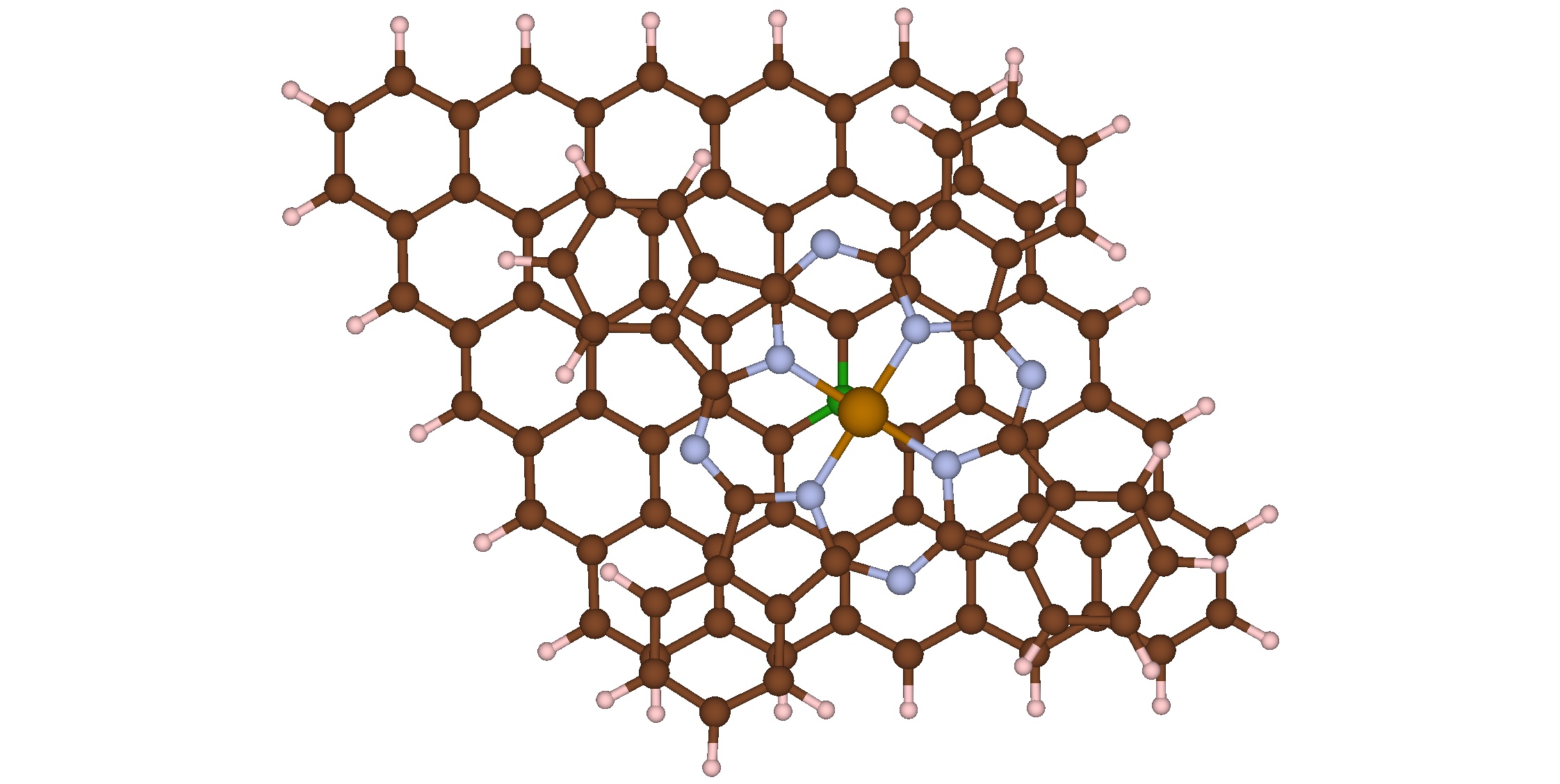}
  \caption{}
  \label{fig:FePcGrBOpt}
\end{subfigure}
\begin{subfigure}{0.49\linewidth}
  \centering
  \includegraphics[width=\linewidth]{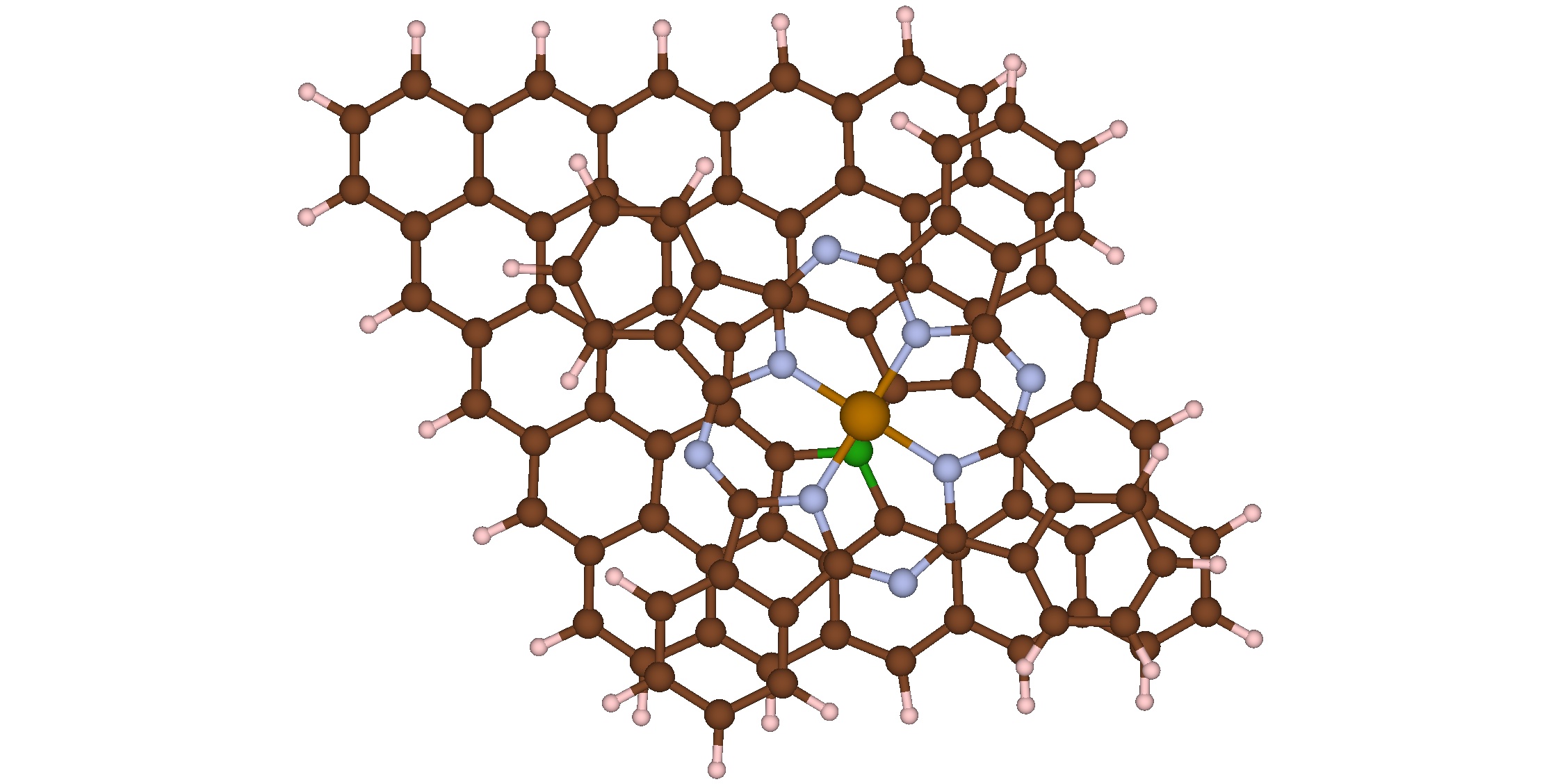}
  \caption{}
  \label{fig:FePcGr57BOpt}
\end{subfigure}
\begin{subfigure}{0.49\linewidth}
  \centering
  \includegraphics[width=\linewidth]{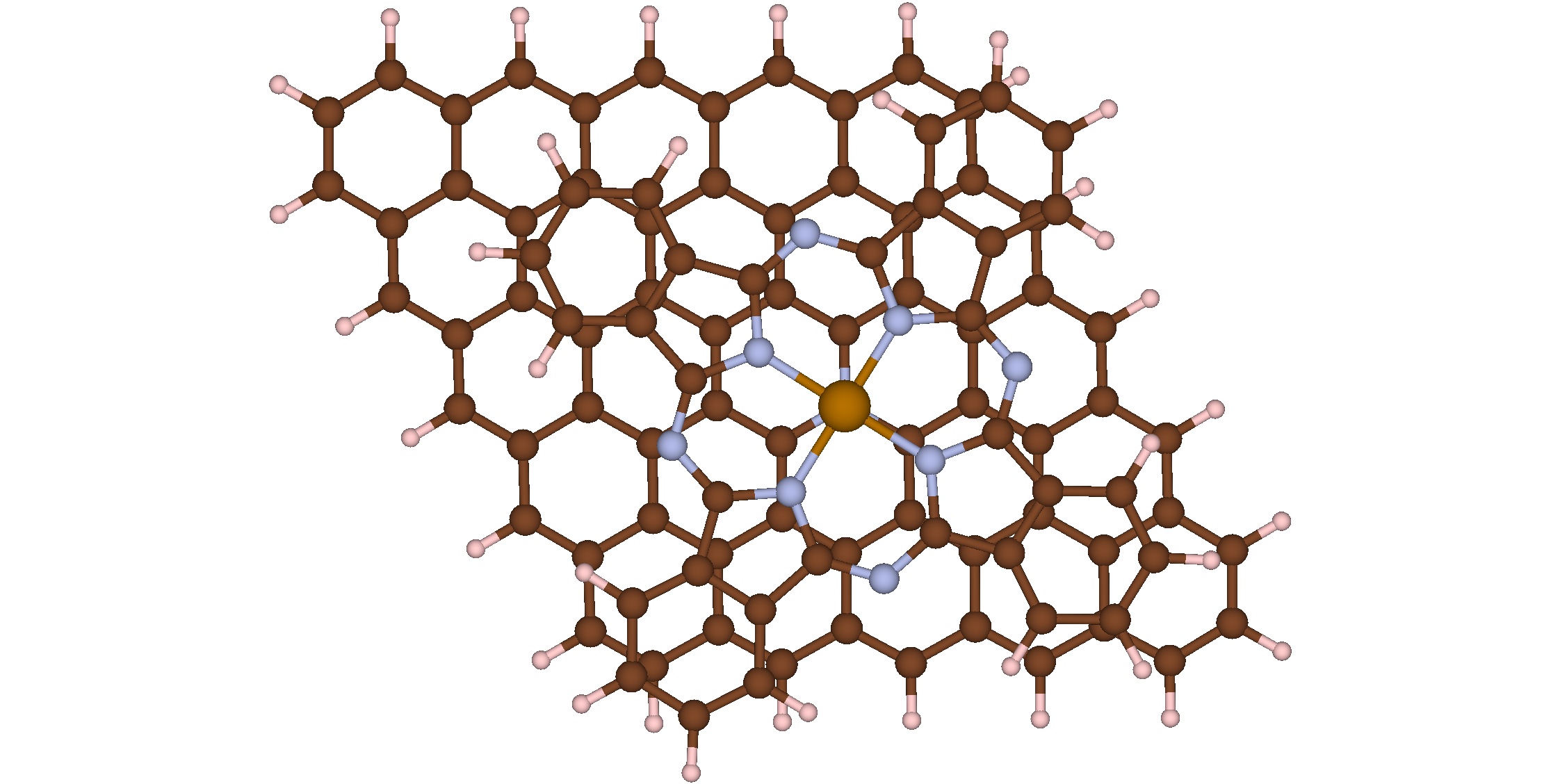}
  \caption{}
  \label{fig:FePcGrNOpt}
\end{subfigure}
\begin{subfigure}{0.49\linewidth}
  \centering
  \includegraphics[width=\linewidth]{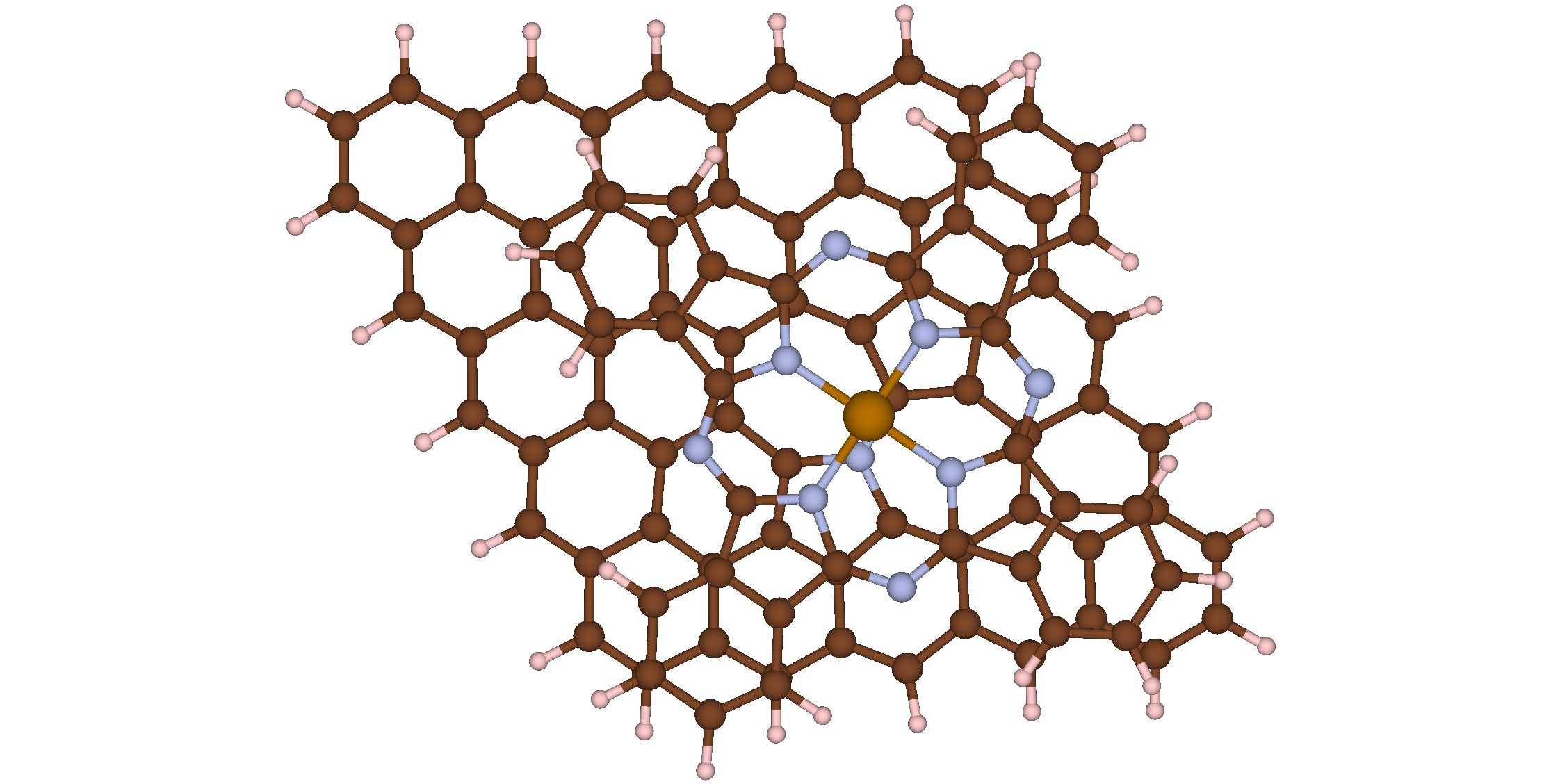}
  \caption{}
  \label{fig:FePcGr57NOpt}
\end{subfigure}
\begin{subfigure}{0.49\linewidth}
  \centering
  \includegraphics[width=\linewidth]{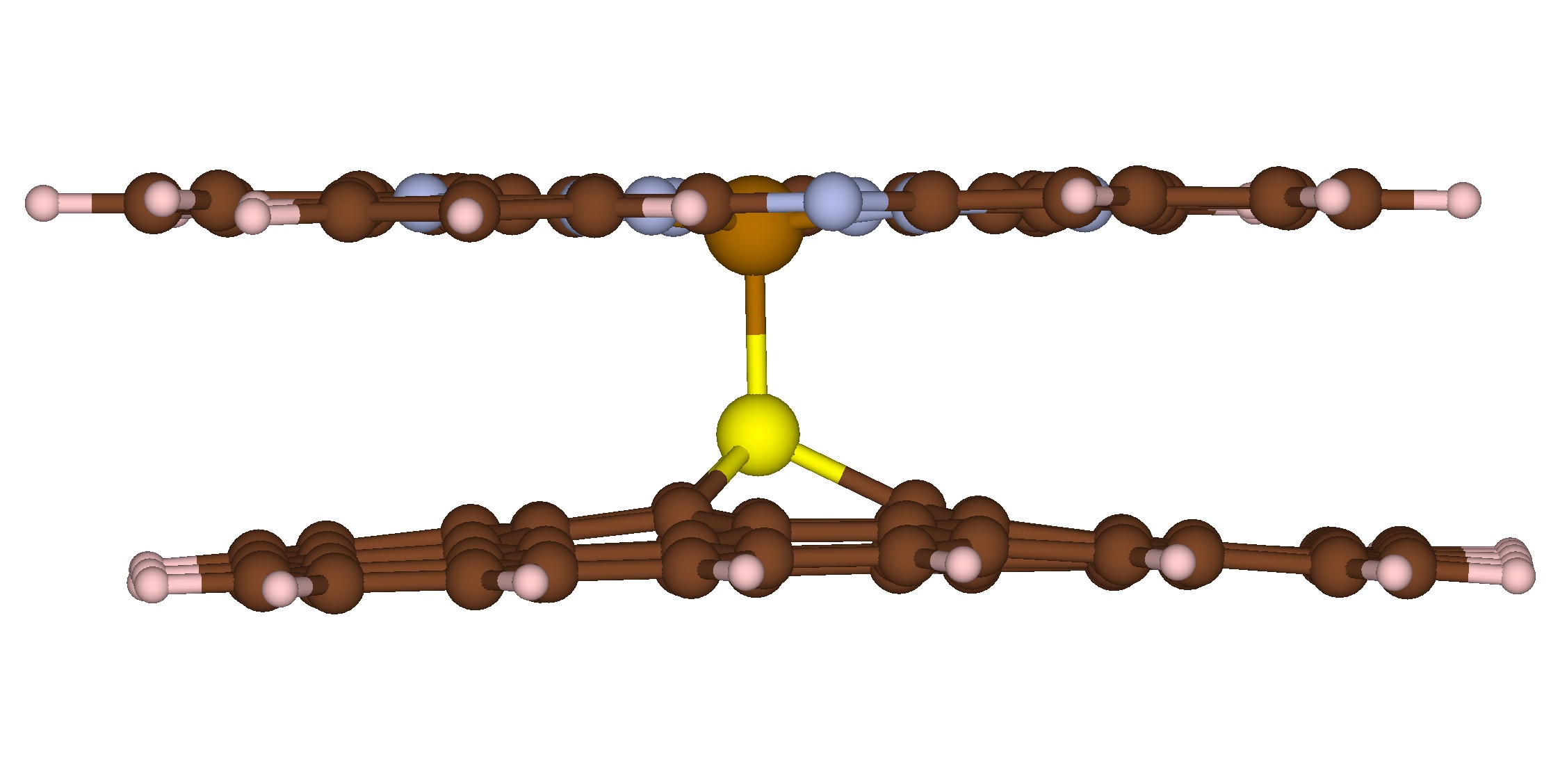}
  \caption{}
  \label{fig:FePcGrSOpt}
\end{subfigure}
\begin{subfigure}{0.49\linewidth}
  \centering
  \includegraphics[width=\linewidth]{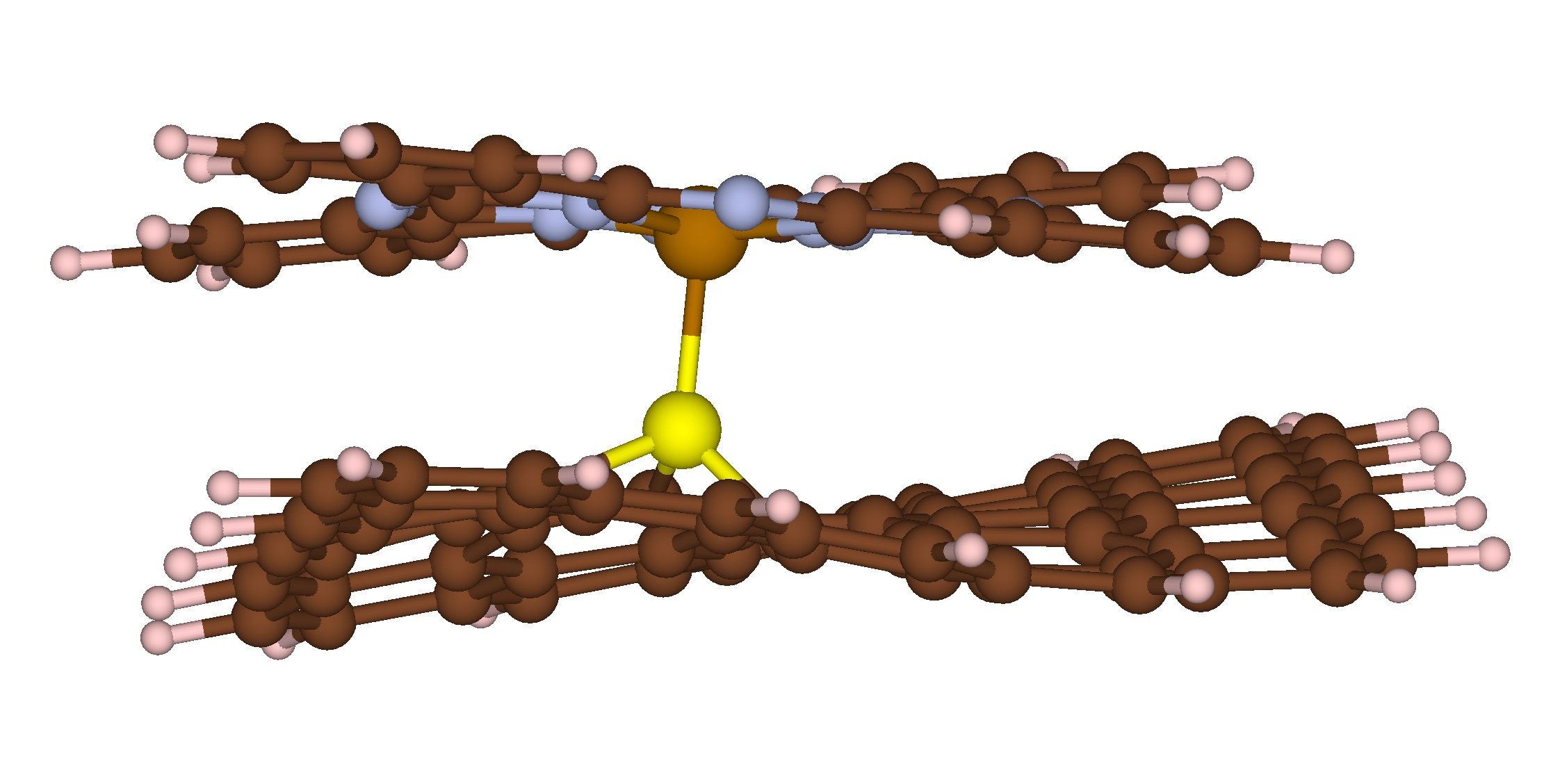}
  \caption{}
  \label{fig:FePcGr57SOpt}
\end{subfigure}
\caption{Views of FePc/Graphene cluster complexes with various types of graphene layer: a) pristine graphene; and graphene with b) the Stone-Wales defect, c) B-doping, d) B-doping + the Stone-Wales defect, e) N-doping, f) N-doping + the Stone-Wales defect, g) S-doping, h) S-doping + the Stone-Wales defect. }
\label{fig:FePcGrGeometries}
\end{figure}

\begin{figure}
\centering
\begin{subfigure}{0.49\linewidth}
  \centering
  \includegraphics[width=\linewidth]{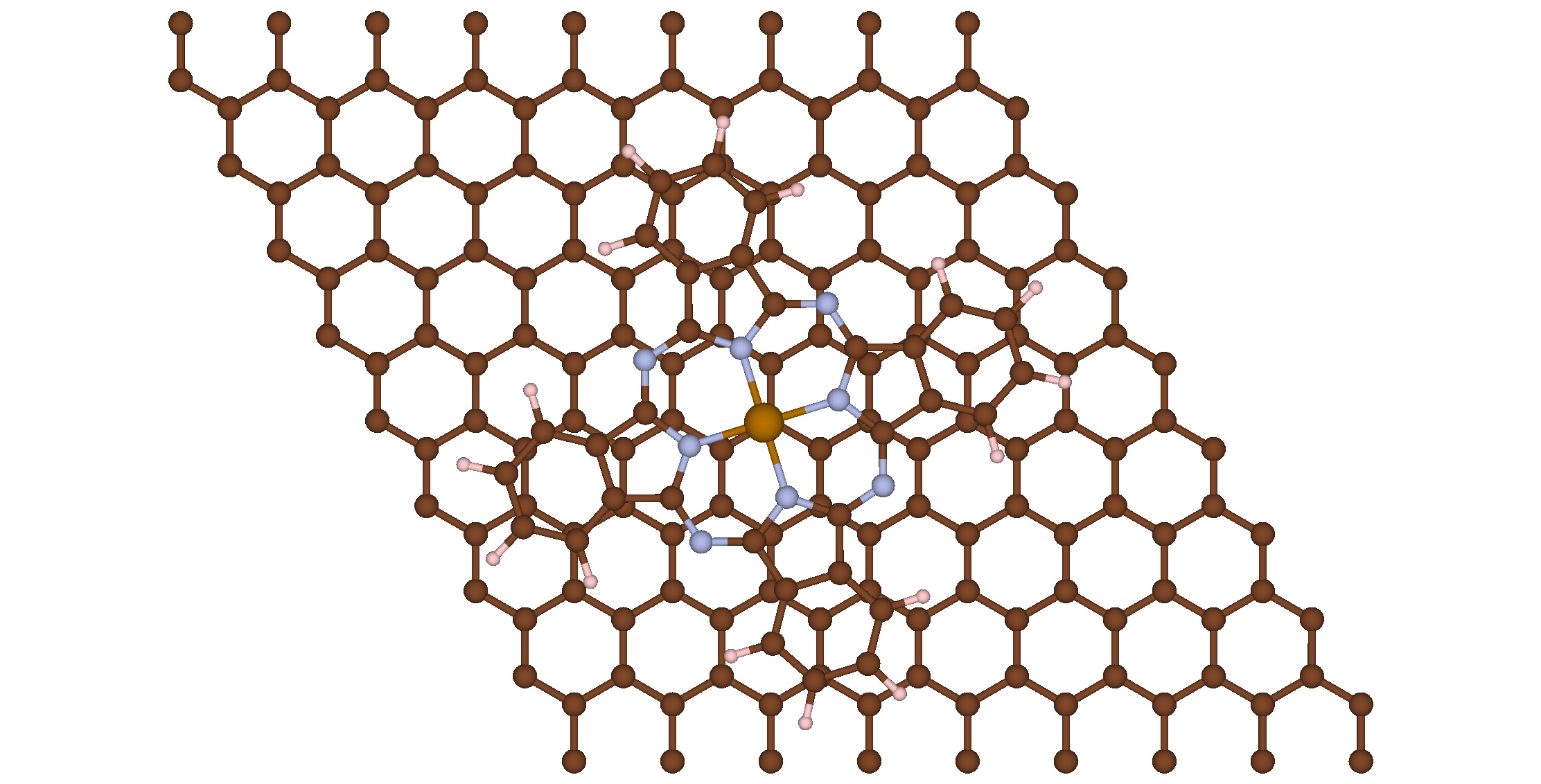}
  \caption{}
  \label{fig:FePcGrPer}
\end{subfigure}
\hfill
\begin{subfigure}{0.49\linewidth}
  \centering
  \includegraphics[width=\linewidth]{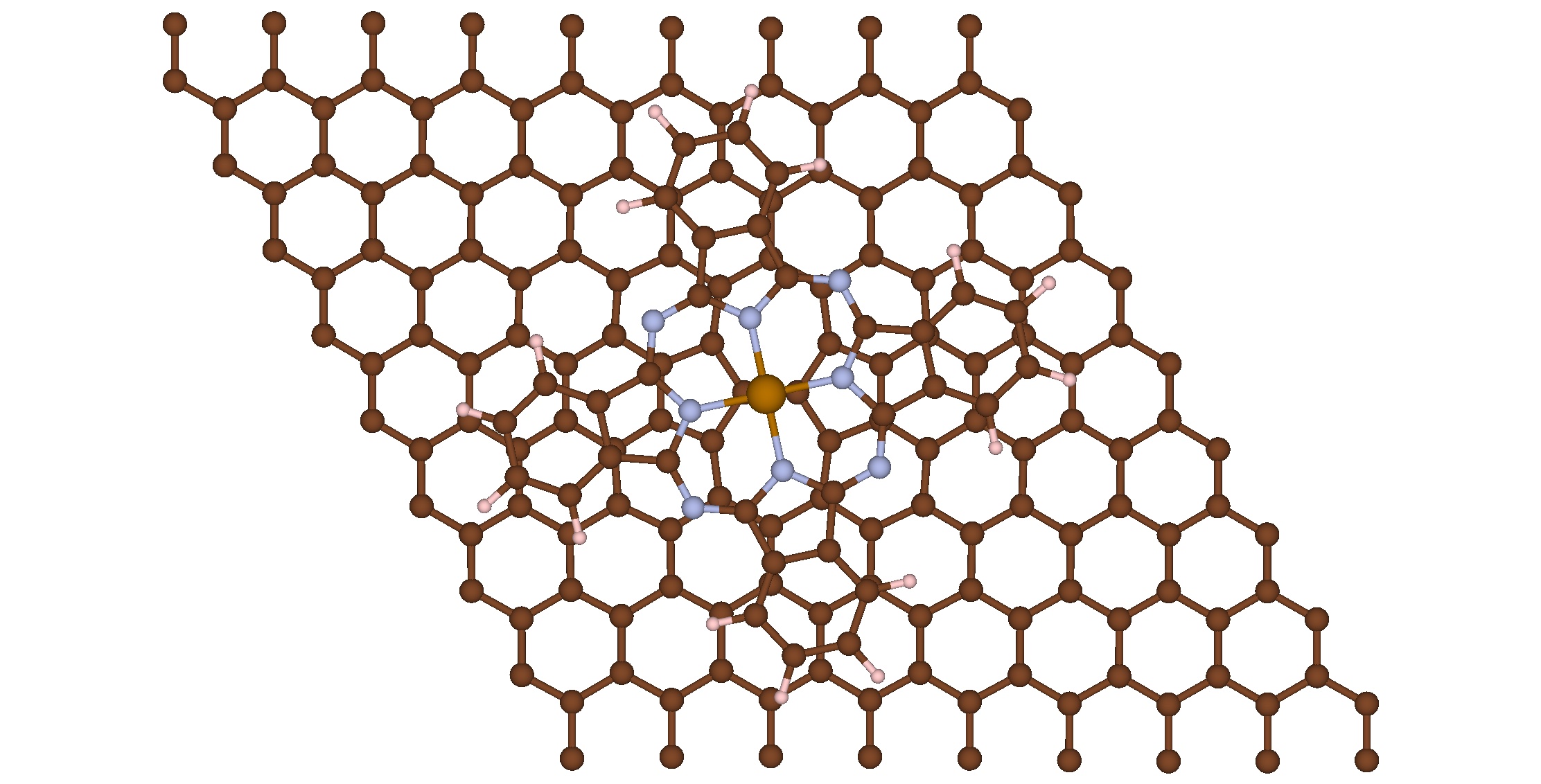}
  \caption{}
  \label{fig:FePcGr57Per}
\end{subfigure}
\begin{subfigure}{0.49\linewidth}
  \centering
  \includegraphics[width=\linewidth]{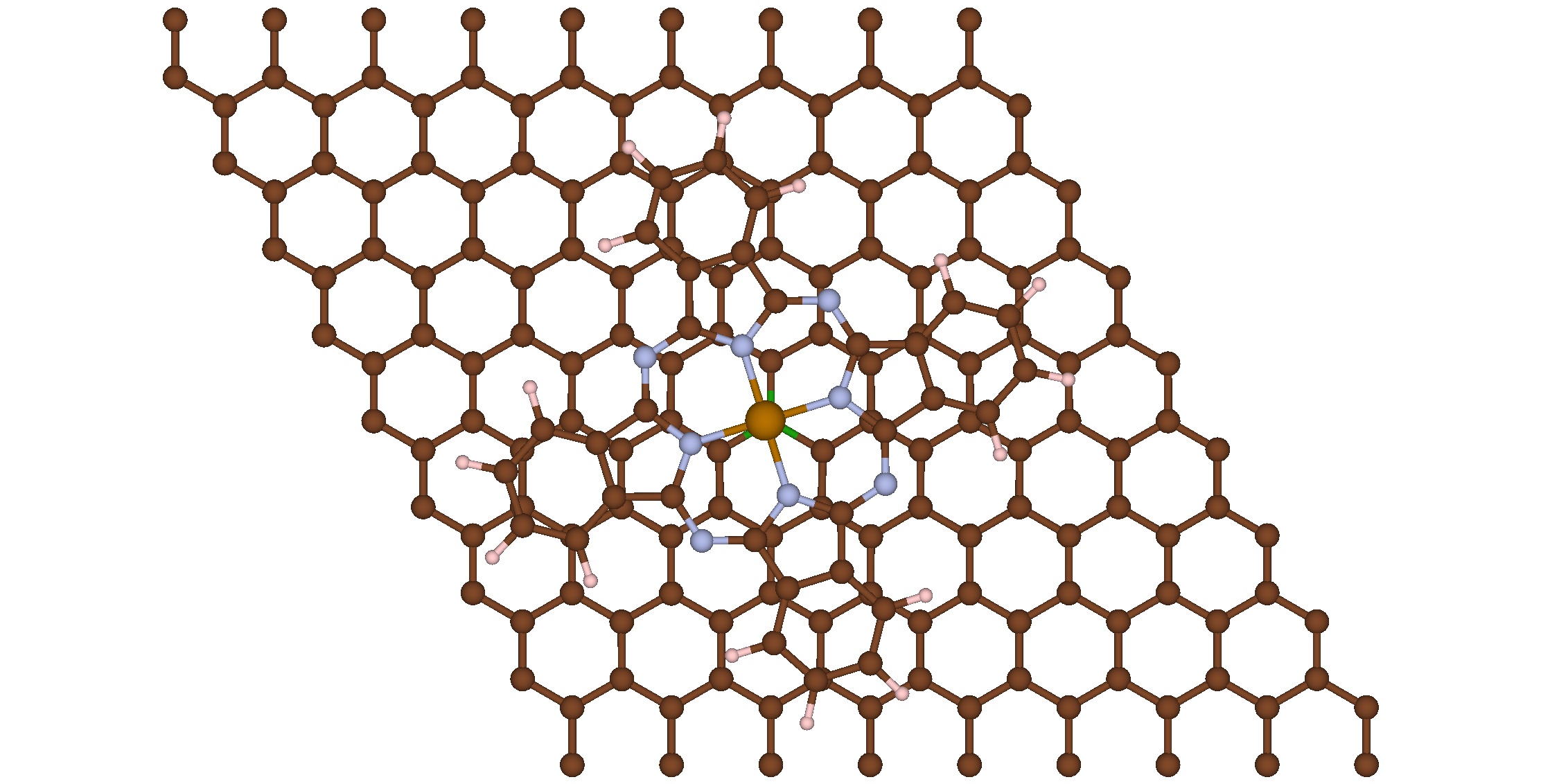}
  \caption{}
  \label{fig:FePcGrBPer}
\end{subfigure}
\begin{subfigure}{0.49\linewidth}
  \centering
  \includegraphics[width=\linewidth]{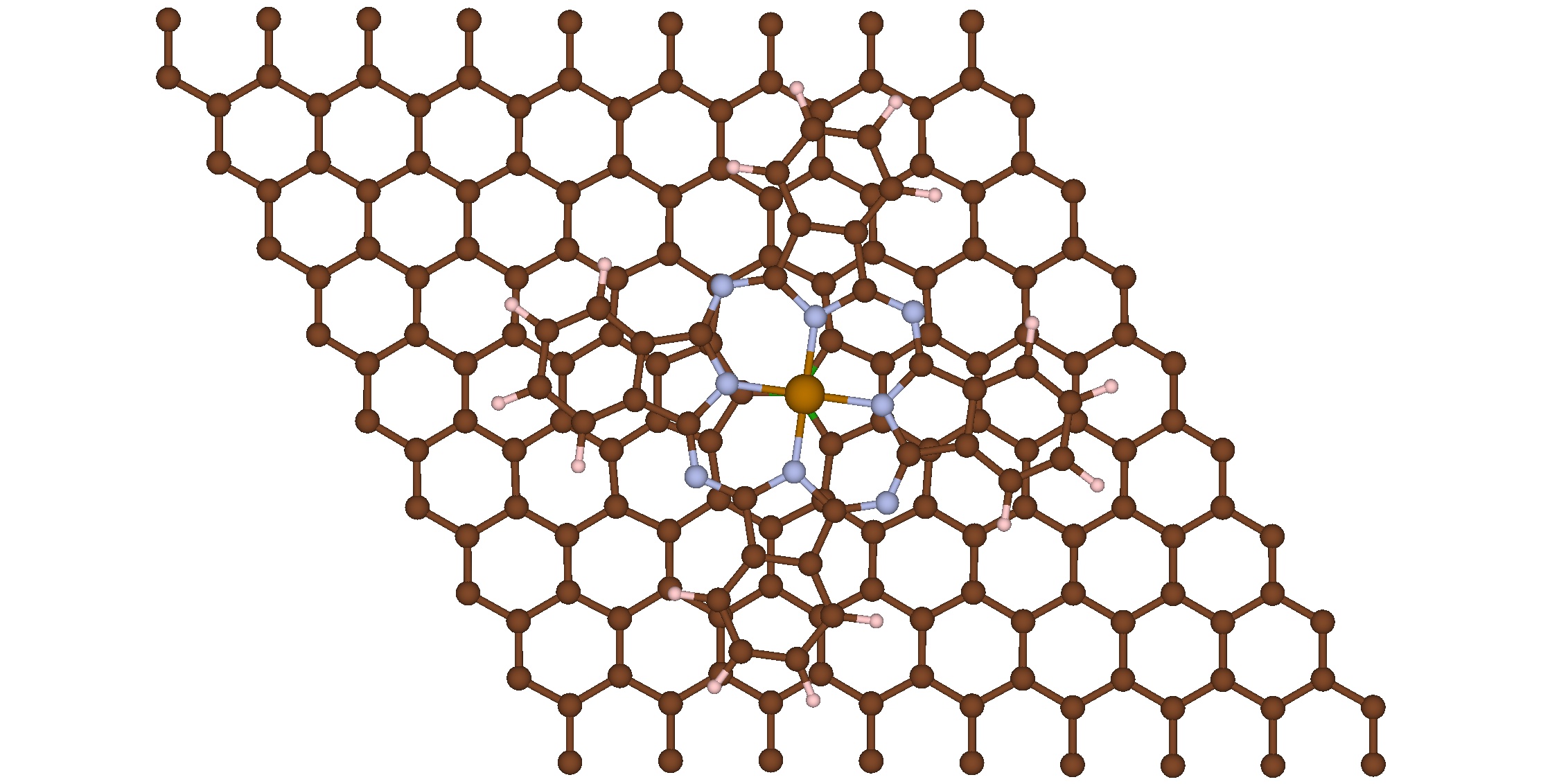}
  \caption{}
  \label{fig:FePcGr57BPer}
\end{subfigure}
\begin{subfigure}{0.49\linewidth}
  \centering
  \includegraphics[width=\linewidth]{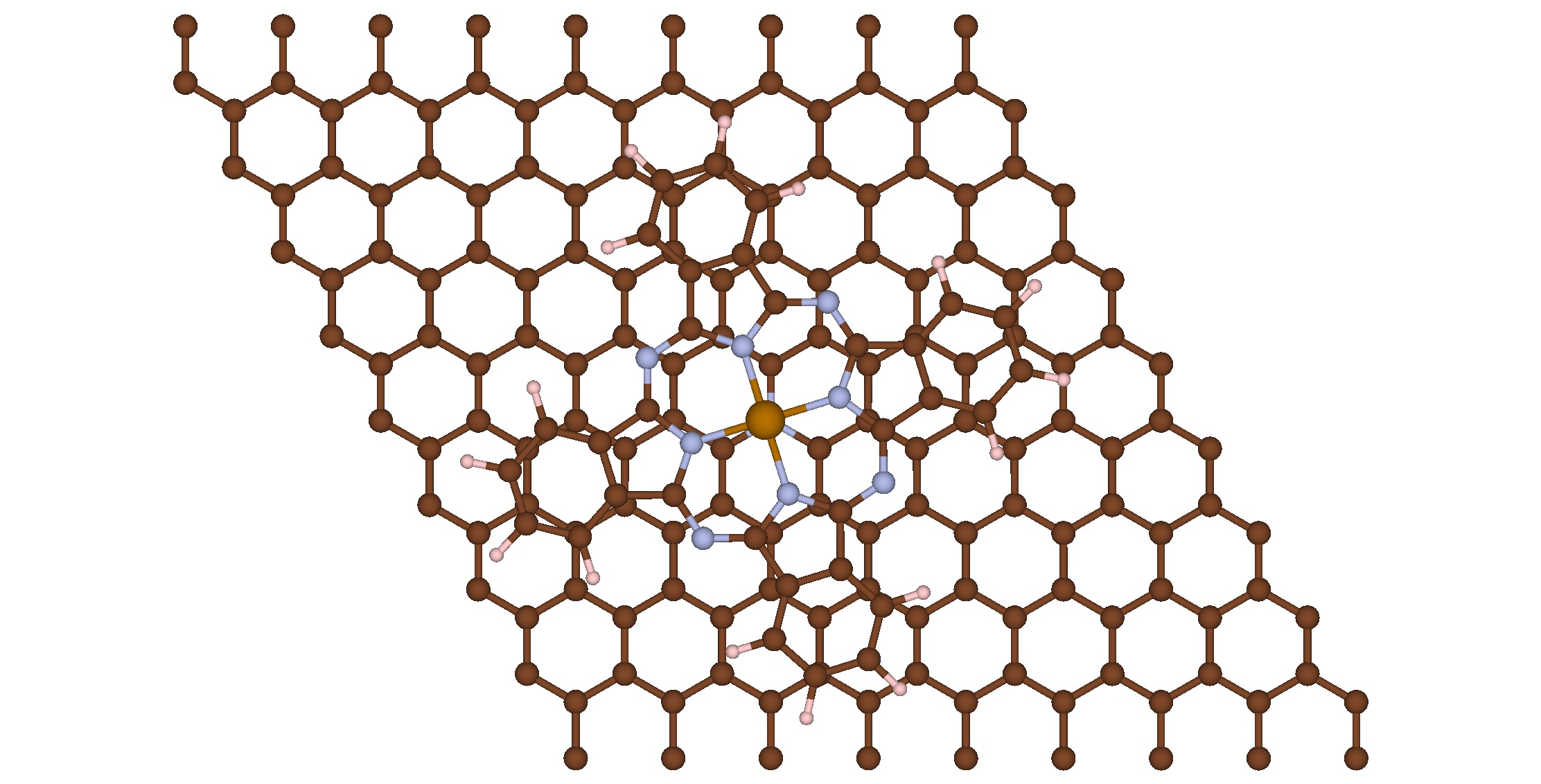}
  \caption{}
  \label{fig:FePcGrNPer}
\end{subfigure}
\begin{subfigure}{0.49\linewidth}
  \centering
  \includegraphics[width=\linewidth]{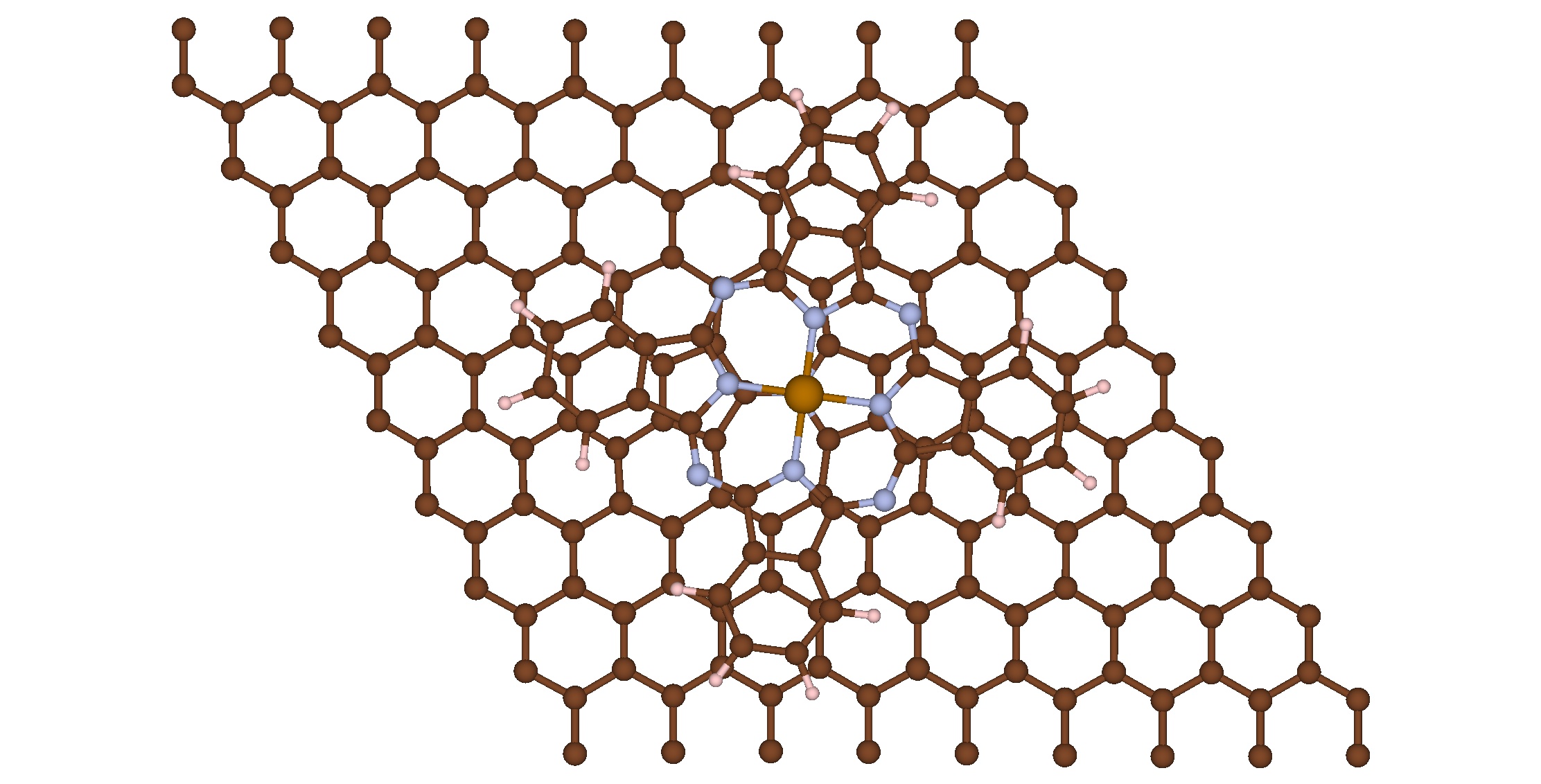}
  \caption{}
  \label{fig:FePcGr57NPer}
\end{subfigure}
\begin{subfigure}{0.49\linewidth}
  \centering
  \includegraphics[width=\linewidth]{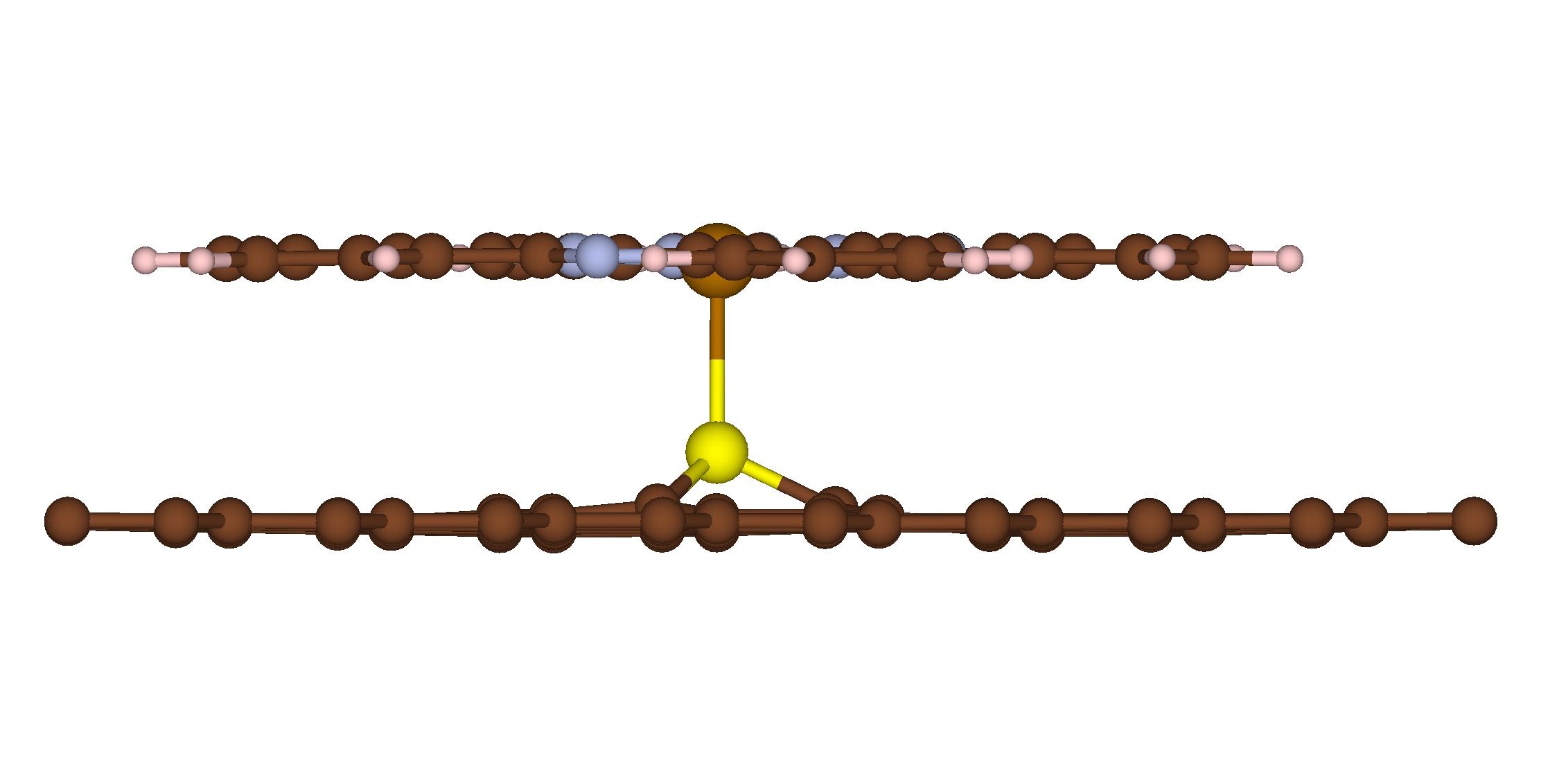}
  \caption{}
  \label{fig:FePcGrSPer}
\end{subfigure}
\begin{subfigure}{0.49\linewidth}
  \centering
  \includegraphics[width=\linewidth]{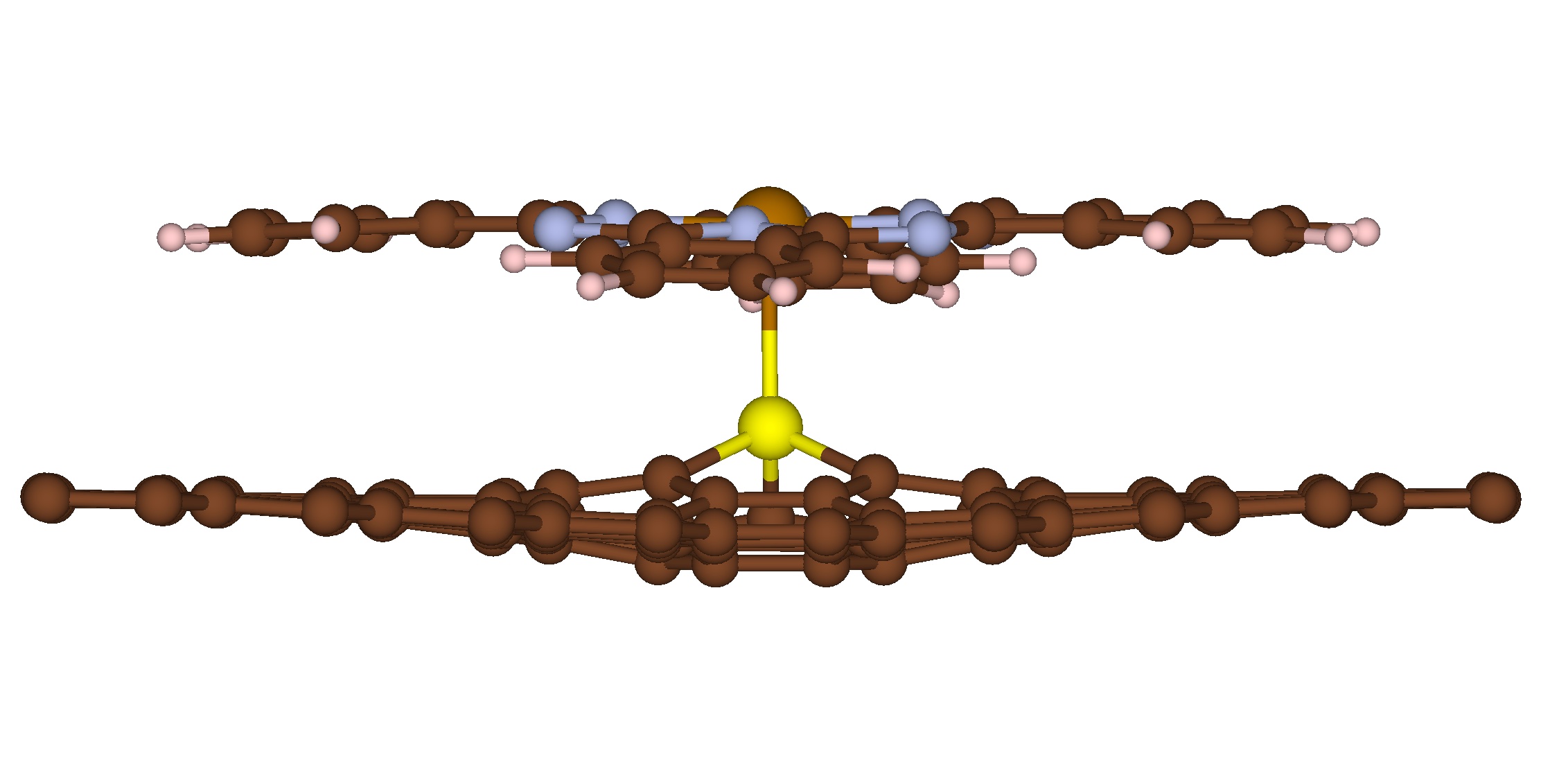}
  \caption{}
  \label{fig:FePcGr57SPer}
\end{subfigure}
\caption{Periodic cell views of FePc/Graphene complexes with various types of graphene layer: a) pristine graphene; and graphene with b) the Stone-Wales defect, c) B-doping, d) B-doping + the Stone-Wales defect, e) N-doping, f) N-doping + the Stone-Wales defect, g) S-doping, h) S-doping + the Stone-Wales defect. }
\label{fig:FePcGrPerGeometries}
\end{figure}

The geometrically optimised structure of the FePc/Graphene-SW system (see Fig. \ref{fig:FePcGr57Opt}, and \ref{fig:FePcGr57Per}) is similar to the geometry for a structure without the defect (Fig. \ref{fig:FePcGrOpt}, \ref{fig:FePcGrPer}) i.e. FePc/Graphene. The FePc molecule is flat and parallel to the graphene layer.  The distances between the molecule and the surface are shown in Table \ref{table:GeomFePcGrDist}. 
The cluster and periodic models give similar results. A decrease in the cluster size till the pyrene molecule leads to a reduction in the distance between the molecule and the cluster. The Stone-Wales defect does not significantly change the distance between the molecule and the layer. The FePc iron atom is slightly higher than the FePc middle plane in the FePc/Pyrene case. This is explained by the fact that the pyrene area is smaller than the area of FePc, the van der Waals forces do not act on the entire FePc molecule, and because of this, FePc bends slightly.

Energetic characteristics (Table \ref{table:FePcGrEnergies}) between the structures with and without the Stone-Wales defect also do not differ much. 
This allows us to predict that in experimental studies this defect will not play an important role in the localization of the molecule on the graphene's surface. Of course, this analysis does not take into account the influence of a possible material that is used to support the graphene layer.

\begin{table}
        \centering
          \caption{Distances between the graphene layer and the FePc molecule and the Fe atom in FePc/Gr hybrid systems. The absolute error value was calculated as a standard deviation from a mean z-coordinate of the layer and the molecule.}            
\begin{tabular}{l|ccc}
  & Periodic & Cluster & Pyrene \\
\hline 
  \multicolumn{4}{c}{FePc - Graphene distance, \AA} \\
\hline 
 FePcGr & 3.45±0.01 & 3.44±0.13 & 3.28±0.19 \\
 FePcGr-SW & 3.49±0.01 & 3.44±0.1 & 3.3±0.1 \\
 FePcGr-B & 3.49±0.01 & 3.43±0.15 & 3.23±0.21 \\
 FePcGr-SW-B & 3.49±0.03 & 3.38±0.42 & 3.26±0.32 \\
 FePcGr-N & 3.49±0.02 & 3.41±0.2 & 3.04±0.28 \\
 FePcGr-SW-N & 3.50±0.02 & 3.39±0.44 & 3.23±0.23 \\
 FePcGr-S & 3.49±0.01 & 3.50±0.25 & 3.35±0.42 \\
 FePcGr57-SW-S & 3.44±0.51 & 3.33±0.88 & 3.36±0.46 \\
\hline 
  \multicolumn{4}{c}{Fe - Graphene distance, \AA} \\
\hline 
 FePcGr & 3.50±0.01 & 3.41±0.09 & 3.37±0.07 \\
 FePcGr-SW & 3.48±0.01 & 3.38±0.07 & 3.40±0.02 \\
 FePcGr-B & 3.46±0.01 & 3.29±0.9 & 3.14±0.13 \\
 FePcGr-SW-B & 3.45±0.02 & 3.19±0.26 & 3.28±0.25 \\
 FePcGr-N & 3.5±0.01 & 3.38±0.14 & 3.21±0.08 \\
 FePcGr-SW-N & 3.5±0.01 & 3.32±0.23 & 3.27±0.11 \\
 FePcGr-S & 3.44±0.01 & 3.29±0.2 & 3.11±0.31 \\
 FePcGr57-SW-S & 3.66±0.28 & 3.09±0.50 & 3.08±0.34 \\
 \hline
\end{tabular}
  
        \label{table:GeomFePcGrDist}
        \end{table}

\begin{table}
        \centering
        \caption{Computed adsorption  energies (in eV) of the FePc molecule to graphene and different types of defected graphene.}
\begin{tabular}{l|lll}
  & Periodic & Cluster & Pyrene \\
\hline 
 FePc/Gr & -2.1 & -2.03 & -1.00 \\
 FePc/Gr-SW & -2.25 & -2.02 & -0.98 \\
 FePc/Gr-B & -2.46 & -2.05 & -1.14 \\
 FePc/Gr-SW-B & -2.09 & -2.13 & -1.06 \\
 FePc/Gr-N & -2.75 & -2.01 & -1.80 \\
 FePc/Gr-SW-N & -1.56 & -2.00 & -1.09 \\
 FePc/Gr-S & -3.52 & -4.70 & -4.07 \\
 FePc/Gr-SW-S & -4.02 & -4.40 & -2.71 \\
 \hline
\end{tabular}
        \label{table:FePcGrEnergies}
        \end{table}

\subsubsection{FePc on Doped Graphene and Doped Graphene-SW}

Doping of graphene with the boron atom also does not distort the flat geometry of the structure (Fig. \ref{fig:FePcGrBOpt}, \ref{fig:FePcGrBPer}). The system's geometry generally has the same form; the molecule remains bound to the graphene surface only by van der Waals forces. The same applies to the case of doping a boron atom into the Stone-Wales defect (Fig. \ref{fig:FePcGr57BOpt}, \ref{fig:FePcGr57BPer}). The position of the FePc molecule over the defects is also retained in both cases. Adsorption energies are comparable to the cases without defects for all studied graphene representations. 

FePc/Graphene hybrid structures with nitrogen doping (Fig. \ref{fig:FePcGrNOpt}, \ref{fig:FePcGr57NOpt}, \ref{fig:FePcGrNPer}, \ref{fig:FePcGr57NPer}) also remain flat but the geometric and energetic parameters change more significantly. For the periodic FePc/Gr-N structure the adsorption energy is about 30\% higher, while for FePc/Gr-SW-N it is 26\% lower. The increase of adsorption energy for FePc/Pyrene-N is about 80\%, while there is no energy decrease for FePc/Pyrene-SW-N case. In comparison with periodic and pyrene models, adsorption energies of  B- and N- defected FePc/Graphene cluster models do not really differ from non-defected FePc/Graphene.

Doping the graphene layer with the sulphur atom, in turn, leads to serious changes in the geometry of the hybrid system. The sulphur atom moves out of the graphene plane and forms a bond with the iron atom. Figures \ref{fig:FePcGrSOpt}, \ref{fig:FePcGr57SOpt}, \ref{fig:FePcGrSPer}, \ref{fig:FePcGr57SPer} show side projections, which show the geometric transformations of these systems. In the case of the system with simple sulphur doping, the graphene layer retains a flat view, while the graphene with the sulphur atom doped into the Stone-Wales defect has a wavy shape. Due to the formation of the Fe-S bond the adsorption energy of the molecule onto the substrate increases substantially. It is observed for all substrate representations and the value of adsorption energies is similar are comparable. 

\subsubsection{The Density of States Analysis}
The analysis of the density of states (DOS) and the projected density of states (PDOS) was carried out for all studied defects. 
In the section of Supplementary Materials DOS's of graphene and FePc/Graphene, and PDOS's of FePc/Graphene are placed. 
Impurities and defects influence the conductive properties of graphene and make it possible to create devices with a controlled small band gap. Thus, the replacement of one of the carbon atoms with a boron or nitrogen atom leads to p- and n-conductivity, respectively.\cite{panchakarla2009synthesis,wei2009synthesis} According to theoretical studies, the formation of the SW defect leads to the appearance of a small band gap (0.08 eV), while the material is still characterized by high conductivity, and the graphene-SW-B and graphene-SW-S systems show conductive properties.\cite{zhou2020effect}  According to the calculations performed, the band gap for graphene with the SW defect is 0.16 eV. Doping the SW defect with boron and sulfur atoms (i.e., creating SW-B and SW-S defects) reduces the band gap to 0.1 eV in both cases. The band gaps slightly differ from  the literature ones but these results depend on the concentration of defects on the surface. For periodic structures studied graphene impurities do not provoke the appearance of magnetisation.


The density of states of the periodic FePc/Gr systems and the projected density of states of the iron atom d-orbital in the FePc molecule indicate that the presence of the molecule does not change the band gap of graphene. The van der Waals interaction between FePc and graphene does not fundamentally disturb the electronic structure of pristine and defected graphene layers. In turn, various defects in graphene lead to redistributions of the density of states of the iron atom d-orbitals in the FePc, while not changing the total magnetic moment. It should be noted that in the case of the systems with boron doping into pure graphene and into graphene with the Stone-Wales defect, the state of the d-orbital of the iron atom is observed near the Fermi energy. In such cases, a charge transfer occurs from the graphene layer to the FePc molecule. The d-orbital acts as an impurity state that is inside the band gap.

\subsubsection{Spin Distribution in Clusters}

In the case of the cluster approach, it is possible to track changes occurring in the distribution of atomic orbitals, since the wave functions are represented in the LCAO form. As a result, the d-shell of the iron atom in FePc retains its configuration ${d_{xy}}^2{d_{xz}}^2{d_{yz}}^1{d_{z^2}}^1{d_{x^2-y^2}}^0$ in the cases of the presence of the molecule on pure graphene, graphene with the Stone-Wales defect, and also in both cases of doping with boron and nitrogen atoms.

The case with a doped sulphur atom deserves separate consideration. These systems were considered in singlet, triplet, and quintet states. Moreover, in each of these systems, there is no spin moment on the sulphur and iron atoms. As the multiplicity of the system increases, the spin moment is distributed on the graphene surface, and the total energy of the system increases. These results are inconsistent with the results we obtained with the periodic approach, but on the other hand, similar results were obtained in another study.\cite{sarmah2019computational}

Also, existing studies of the solid-state structures of iron sulfide\cite{Jain2013,Lai2015,Bertaut1965,Kwon2011,Uda1968,Berner1962} show that such compounds do not exhibit magnetic properties. In turn, the magnetic order of iron boride\cite{Jain2013,Kolmogorov2010} is ferromagnetic. These results suggest that the nature of the interaction of the FePc molecule with S-doped graphene is rather nonmagnetic, but this requires experimental studies and calculations based on the multiconfigurational approach.

\subsection{Multiconfigurational analysis}

\begin{figure}
    \centering
    \includegraphics[width=0.5\textwidth]{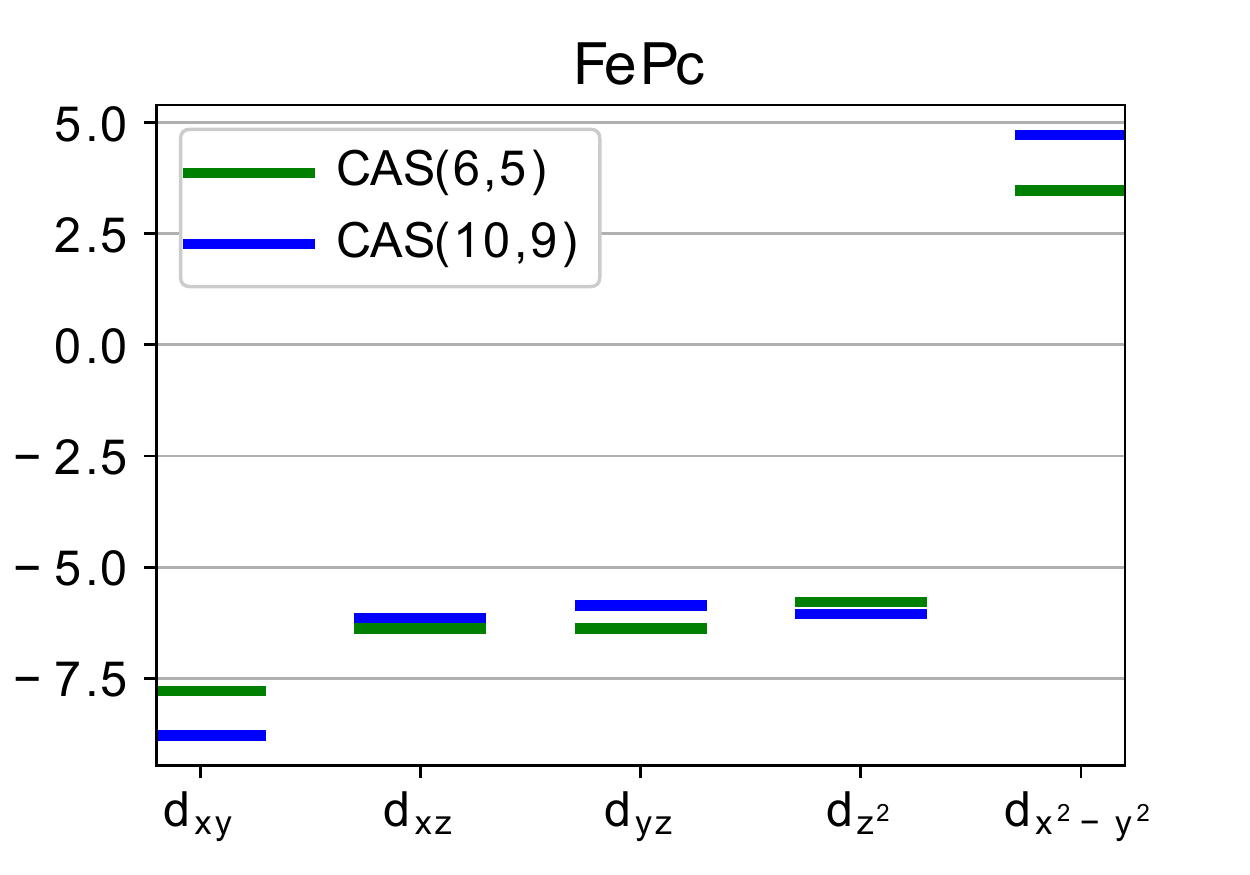}
    \caption{Energies (in eV) of the iron $d$-shell in the FePc molecule; green lines present the results using CAS(6,5), where all 5 orbitals in the active space are iron $d$-orbitals, blue lines correspond to the results using CAS(10,9) where 4 ligand orbitals were added.}
    \label{fig:FePc_Fedshell}
\end{figure}

The free standing FePc molecule has been studied using different complete active spaces: CAS(6,5) and CAS(10,9). In the first case, only iron $d$-orbitals are in the active space, in the second case two orbitals from the core space and two from the set of valence orbitals are added. The active molecular orbitals for CAS(10,9) are shown in Supplementary Materials. The graphical comparison between iron $d$-orbitals energies in CAS(6,5) and CAS(10,9) is presented in Fig. \ref{fig:FePc_Fedshell}. Unfortunately, the addition of energetically nearest ligand orbitals to the active space destroys the symmetry and $d_{xz}$ and $d_{yz}$ orbitals are not degenerated anymore. Also, the expansion of CAS moves $d_{xy}$ and $d_{x^{2} -y^{2}}$ apart. Frequencies of molecular electronic transitions (Table \ref{table:FePc_transitions}) decrease when the active space is expanded. During first transitions, the electron configuration changes only on the iron $d$-shell and the expansion of CAS does not change their order.

\begin{table*}
        \centering
       \caption{FePc NEVPT2 transition energies $\nu$ (in $cm^{-1}$), descriptions of ground and excited states calculated using CAS(6,5) and CAS(10,9), and their multiplicities 2S+1.  } 
       \label{table:FePc_transitions}
\begin{tabular}{c ccc||ccc}
\hline
  & \multicolumn{3}{c||}{\makecell{CAS(6,5)}} & \multicolumn{3}{c}{\makecell{CAS(10,9)}} \\
\hline 
 System & $\nu$ & State & 2S+1 & $\nu$ & State & 2S+1 \\

\hline 
 \multirow{6}{*}{\makecell{FePc}} & --- & $\displaystyle d_{xy}^{2} d_{xz}^{1} d_{yz}^{1} d_{z^{2}}^{2} d_{x^{2} -y^{2}}^{0}$ & 3 &  & $\displaystyle d_{xy}^{2} d_{xz}^{1} d_{yz}^{1} d_{z^{2}}^{2} d_{x^{2} -y^{2}}^{0}$ & 3 \\

   & 738 & $\displaystyle d_{xy}^{2} d_{xz}^{1} d_{yz}^{2} d_{z^{2}}^{1} d_{x^{2} -y^{2}}^{0}$ & 3 & 615 & $\displaystyle d_{xy}^{2} d_{xz}^{1} d_{yz}^{2} d_{z^{2}}^{1} d_{x^{2} -y^{2}}^{0}$ & 3 \\

   & 738 & $\displaystyle d_{xy}^{2} d_{xz}^{2} d_{yz}^{1} d_{z^{2}}^{1} d_{x^{2} -y^{2}}^{0}$ & 3 & 677 & $\displaystyle d_{xy}^{2} d_{xz}^{2} d_{yz}^{1} d_{z^{2}}^{1} d_{x^{2} -y^{2}}^{0}$ & 3 \\

   & 3307 & $\displaystyle d_{xy}^{1} d_{xz}^{1} d_{yz}^{1} d_{z^{2}}^{2} d_{x^{2} -y^{2}}^{1}$ & 5 & 3226 & $\displaystyle d_{xy}^{1} d_{xz}^{1} d_{yz}^{1} d_{z^{2}}^{2} d_{x^{2} -y^{2}}^{1}$ & 5 \\

   & 4978 & $\displaystyle d_{xy}^{1} d_{xz}^{1} d_{yz}^{2} d_{z^{2}}^{1} d_{x^{2} -y^{2}}^{1}$ & 5 & 4442 & $\displaystyle d_{xy}^{1} d_{xz}^{1} d_{yz}^{2} d_{z^{2}}^{1} d_{x^{2} -y^{2}}^{1}$ & 5 \\

   & 4978 & $\displaystyle d_{xy}^{1} d_{xz}^{2} d_{yz}^{1} d_{z^{2}}^{1} d_{x^{2} -y^{2}}^{1}$ & 5 & 4531 & $\displaystyle d_{xy}^{1} d_{xz}^{2} d_{yz}^{1} d_{z^{2}}^{1} d_{x^{2} -y^{2}}^{1}$ & 5 \\
   \hline
\end{tabular}

\end{table*}

\begin{figure*}
    \centering
    \includegraphics[width=\linewidth]{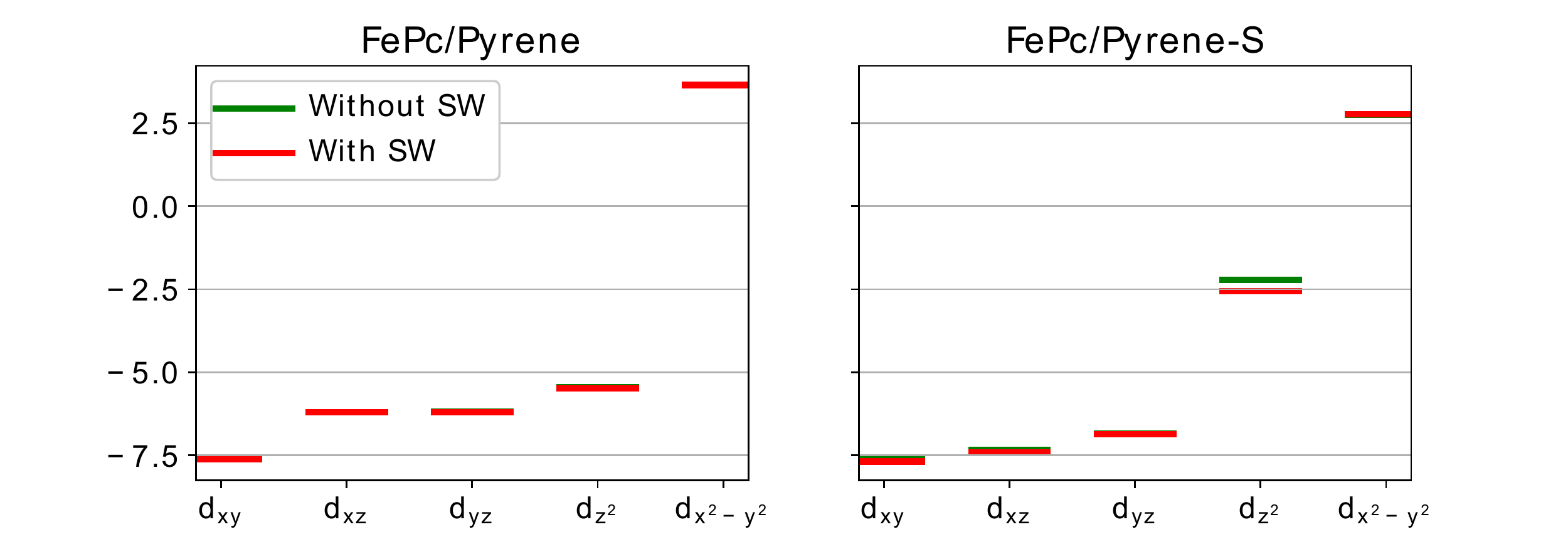}
    \caption{Energies (in eV) of the iron $d$-shell levels in FePc/Pyrene (left panel), and FePc/Pyrene-S, i.e., S-doped pyrene (right panel) hybrid structures.
    Calculations were performed employing CAS(6,5), where 5 orbitals in the active space are iron $d$-orbitals. Green and red lines correspond to pyrene without and with the SW defect, respectively.}
    \label{fig:Fedshell65}
\end{figure*}

\begin{figure*}
    \centering
    \includegraphics[width=\linewidth]{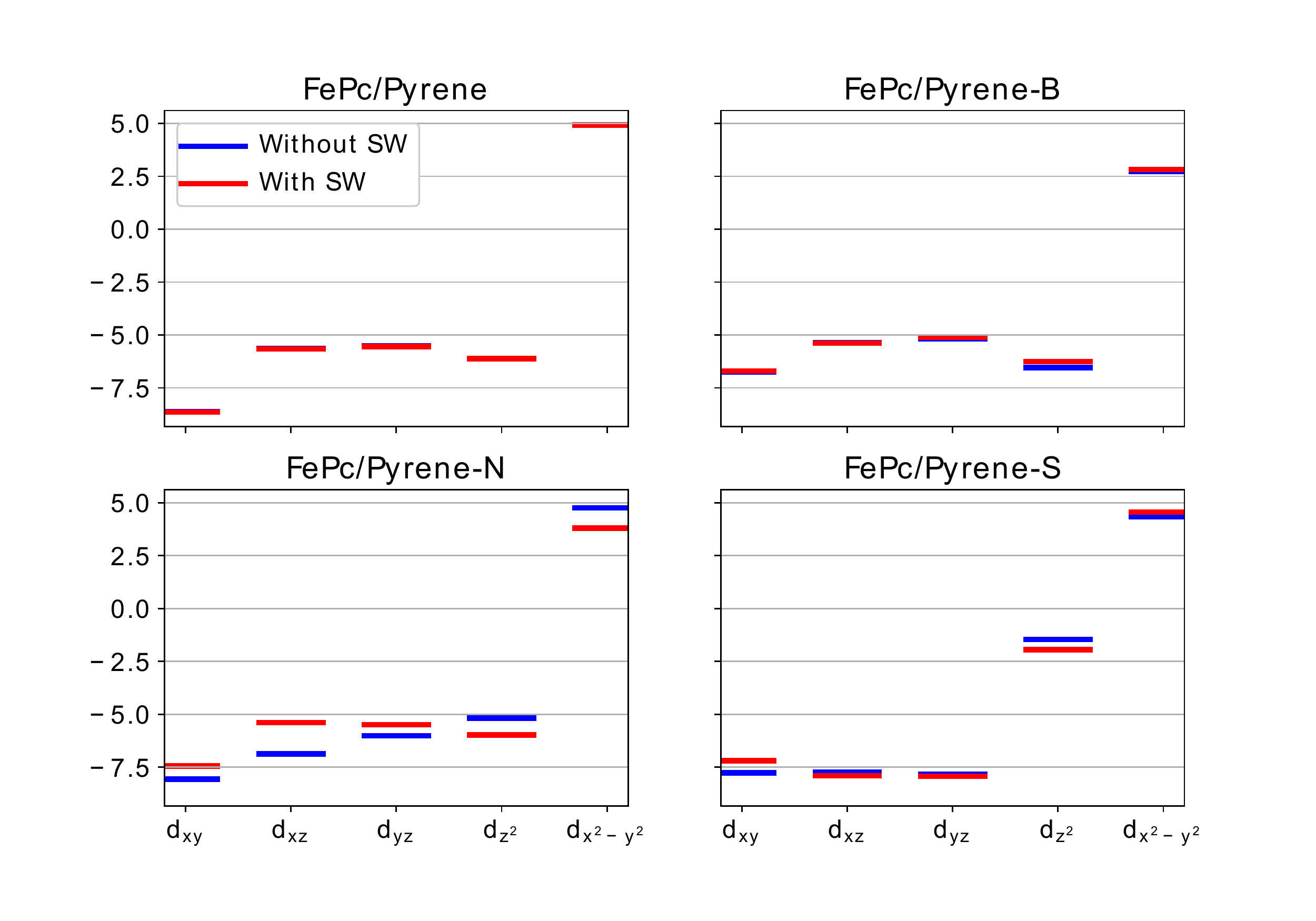}
    \caption{Energies (in eV) of the iron $d$-shell levels in FePc/Pyrene hybrid systems calculated using CAS(10,9) for FePc/Pyrene and FePc/Pyrene-S; CAS(11,9) for FePc/Pyrene-B and FePc/Pyrene-N. Five orbitals in the active space are iron $d$-orbitals and four orbitals are ligand ones. Blue and red lines correspond to pyrene without and with the SW defect, respectively.}
    \label{fig:Fedshell109}
\end{figure*}

Multi-reference calculations of the FePc/Pyrene system using CAS(6,5) were done only for the cases of pure pyrene, pyrene with the SW defect, S-doping, and the combined SW-S defect. The active molecular orbitals energies (energies of iron $d$-orbitals) are graphically shown in Fig. \ref{fig:Fedshell65} and the description of molecular electronic transitions are in Table \ref{table:FePcPyrene65_transitions}. The results with extended CAS(10,9) for the system with these defects and for FePc/Pyrene(-SW)-B and FePc/Pyrene(-SW)-N using CAS(11,9) are in Fig. \ref{fig:Fedshell109} and Table \ref{table:FePcPyrene109_transitions}. Views of 9 active molecular orbitals for FePc/Pyrene and all studied defects are presented in Supplementary Materials.

In general, the presence of the usual pyrene molecule barely changes the electronic properties of FePc. The energies of iron $d$-orbitals in FePc/Pyrene do not change a lot by introducing the SW defect. Electron occupations remain the same for ground and the first five excited states, exactly as in the case of the pure FePc molecule. Molecular orbitals corresponding to pyrene are not inside the expanded active space CAS(10,9) and their energy levels are quite far from the expanded active space energy levels. It is only noticed that in the case of the SW defect, the transition energies to excited states slightly increase.

The FePc/Pyrene hybrid system with B-doped pyrene has a denser distribution of iron $d$-orbitals energies in comparison with the FePc/Pyrene with undoped pyrene. The $d_{z^2}$ orbital is on the same energy level as $d_{xy}$. The boron atom has an unpaired electron in the ground state. First excited electronic states are characterised by the changing of the iron $d$-orbitals occupation, while electron occupation of the boron atom remains the same. The first excited state and the ground state have small transition energy between each other and different multiplicities (the ground state is doublet $2S+1=2$ and the excited state is quartet $2S+1=4$). For the system with the SW defect, the states have the same iron $d$-orbital occupation $d_{xy}^{2} d_{xz}^{1} d_{yz}^{1} d_{z^{2}}^{2} d_{x^{2} -y^{2}}^{0}$ and $\nu = 26 cm^{-1}$. The transition energy between these states is quite small and it is hard to recognise the real ground state spin moment of the system (previous researchers predicted\cite{sarmah2019computational} the total magnetic moment equals 3 $\mu_B$ for FePc/Graphene-B periodic structure which corresponds to a quartet while we obtained doublet).In contrast to this, for pure FePc the transition frequency between the states $d_{xy}^{2} d_{xz}^{1} d_{yz}^{1} d_{z^{2}}^{2} d_{x^{2} -y^{2}}^{0}$ with different multiplicities (triplet and singlet) is 11628 $cm^{-1}$.

The complexes with N-doped atoms exhibit similar behaviour, the complex without the SW defect has the same iron electron occupations $d_{xy}^{2} d_{xz}^{1} d_{yz}^{1} d_{z^{2}}^{2} d_{x^{2} -y^{2}}^{0}$ in the ground and first excited states, and the transition energy between them is just $25 cm^{-1}$. The excitation energy of the complex with the SW defect is much higher. 
The nitrogen atom has an unpaired electron in the ground state and in first excited states by analogy with doped boron complexes. 
The SW defect changes the order of molecular orbitals energies corresponding to $d_{xz} d_{yz} d_{z^{2}}$ iron orbitals from ascending to descending. This can be associated with changes in geometry.  

The doping of a sulfur atom into the pyrene molecule also strongly changes the electronic properties of the FePc/Pyrene complex. The iron $d_{z^2}$ orbital is not on the energy level of $d_{xz} d_{yz}$ anymore, and it is much higher. In turn, $d_{xz}$ and $d_{yz}$ orbitals become closer to $d_{xy}$. As a result, in the ground state, the iron $d$-shell has the occupation $d_{xy}^{2} d_{xz}^{2} d_{yz}^{2} d_{z^{2}}^{0} d_{x^{2} -y^{2}}^{0}$, and the ground state is not spin-polarised. The energy of the first excitation is quite high here in comparison with other studied cases, and the first excited state corresponds to a triplet. The first excitation energy differs a lot due to the presence of the SW defect (6514 $cm^{-1}$ without the defect and 2520 $cm^{-1}$ with it) and it should be visible well in an experiment.

\begin{table*}
        \centering
       
        \caption{NEVPT2 transition energies $\nu$ (in $cm^{-1}$), $d$-shell configurations, and state multiplicities 2S+1 for the ground and excited states of the studied FePc/Pyrene hybrid systems without the SW defect (left panel) and with the SW defect (right panel). All presented results were obtained employing CAS(6,5), where 5 orbitals in the active space are iron $d$-orbitals.} 
       \label{table:FePcPyrene65_transitions}
\begin{tabular}{c ccc||ccc}
\hline
  & \multicolumn{3}{c||}{\makecell{Without the SW defect}} & \multicolumn{3}{c}{\makecell{With the SW defect}} \\
\hline 
 System & $\nu$ & State & 2S+1 & $\nu$ & State & 2S+1 \\
\hline 

 \multirow{6}{*}{\makecell{FePc-Pyrene}} & --- & $\displaystyle Fe\left( d_{xy}^{2} d_{xz}^{1} d_{yz}^{1} d_{z^{2}}^{2} d_{x^{2} -y^{2}}^{0}\right)$ & 3 & --- & $\displaystyle Fe\left( d_{xy}^{2} d_{xz}^{1} d_{yz}^{1} d_{z^{2}}^{2} d_{x^{2} -y^{2}}^{0}\right)$ & 3 \\

   & 176 & $\displaystyle Fe\left( d_{xy}^{2} d_{xz}^{1} d_{yz}^{2} d_{z^{2}}^{1} d_{x^{2} -y^{2}}^{0}\right)$ & 3 & 185 & $\displaystyle Fe\left( d_{xy}^{2} d_{xz}^{1} d_{yz}^{2} d_{z^{2}}^{1} d_{x^{2} -y^{2}}^{0}\right)$ & 3 \\

   & 197 & $\displaystyle Fe\left( d_{xy}^{2} d_{xz}^{2} d_{yz}^{1} d_{z^{2}}^{1} d_{x^{2} -y^{2}}^{0}\right)$ & 3 & 232 & $\displaystyle Fe\left( d_{xy}^{2} d_{xz}^{2} d_{yz}^{1} d_{z^{2}}^{1} d_{x^{2} -y^{2}}^{0}\right)$ & 3 \\

   & 3284 & $\displaystyle Fe\left( d_{xy}^{1} d_{xz}^{1} d_{yz}^{1} d_{z^{2}}^{2} d_{x^{2} -y^{2}}^{1}\right)$ & 5 & 3310 & $\displaystyle Fe\left( d_{xy}^{1} d_{xz}^{1} d_{yz}^{1} d_{z^{2}}^{2} d_{x^{2} -y^{2}}^{1}\right)$ & 5 \\

   & 4379 & $\displaystyle Fe\left( d_{xy}^{1} d_{xz}^{1} d_{yz}^{2} d_{z^{2}}^{1} d_{x^{2} -y^{2}}^{1}\right)$ & 5 & 4458 & $\displaystyle Fe\left( d_{xy}^{1} d_{xz}^{1} d_{yz}^{2} d_{z^{2}}^{1} d_{x^{2} -y^{2}}^{1}\right)$ & 5 \\

   & 4384 & $\displaystyle Fe\left( d_{xy}^{1} d_{xz}^{2} d_{yz}^{1} d_{z^{2}}^{1} d_{x^{2} -y^{2}}^{1}\right)$ & 5 & 4463 & $\displaystyle Fe\left( d_{xy}^{1} d_{xz}^{2} d_{yz}^{1} d_{z^{2}}^{1} d_{x^{2} -y^{2}}^{1}\right)$ & 5 \\
 \hline
  \multirow{6}{*}{\makecell{FePc-Pyrene-S}} & --- & $\displaystyle Fe\left( d_{xy}^{2} d_{xz}^{2} d_{yz}^{2} d_{z^{2}}^{0} d_{x^{2} -y^{2}}^{0}\right)$ & 1 & --- & $\displaystyle Fe\left( d_{xy}^{2} d_{xz}^{2} d_{yz}^{2} d_{z^{2}}^{0} d_{x^{2} -y^{2}}^{0}\right)$ & 1 \\

   & 4217 & $\displaystyle Fe\left( d_{xy}^{2} d_{xz}^{1} d_{yz}^{2} d_{z^{2}}^{1} d_{x^{2} -y^{2}}^{0}\right)$ & 3 & 2610 & $\displaystyle Fe\left( d_{xy}^{2} d_{xz}^{1} d_{yz}^{2} d_{z^{2}}^{1} d_{x^{2} -y^{2}}^{0}\right)$ & 3 \\

   & 4313 & $\displaystyle Fe\left( d_{xy}^{2} d_{xz}^{2} d_{yz}^{1} d_{z^{2}}^{1} d_{x^{2} -y^{2}}^{0}\right)$ & 3 & 2782 & $\displaystyle Fe\left( d_{xy}^{2} d_{xz}^{2} d_{yz}^{1} d_{z^{2}}^{1} d_{x^{2} -y^{2}}^{0}\right)$ & 3 \\

   & 6683 & $\displaystyle Fe\left( d_{xy}^{1} d_{xz}^{2} d_{yz}^{2} d_{z^{2}}^{1} d_{x^{2} -y^{2}}^{0}\right)$ & 3 & 5254 & $\displaystyle Fe\left( d_{xy}^{1} d_{xz}^{2} d_{yz}^{2} d_{z^{2}}^{1} d_{x^{2} -y^{2}}^{0}\right)$ & 3 \\

   & 8003 & $\displaystyle Fe\left( d_{xy}^{1} d_{xz}^{1} d_{yz}^{2} d_{z^{2}}^{1} d_{x^{2} -y^{2}}^{1}\right)$ & 5 & 6453 & $\displaystyle Fe\left( d_{xy}^{1} d_{xz}^{1} d_{yz}^{2} d_{z^{2}}^{1} d_{x^{2} -y^{2}}^{1}\right)$ & 5 \\

   & 8104 & $\displaystyle Fe\left( d_{xy}^{1} d_{xz}^{2} d_{yz}^{1} d_{z^{2}}^{1} d_{x^{2} -y^{2}}^{1}\right)$ & 5 & 6591 & $\displaystyle Fe\left( d_{xy}^{1} d_{xz}^{2} d_{yz}^{1} d_{z^{2}}^{1} d_{x^{2} -y^{2}}^{1}\right)$ & 5 \\
 \hline
\end{tabular}

\end{table*}

\begin{table*}[h!t]
        \centering
       \caption{NEVPT2 transition energies $\nu$ (in $cm^{-1}$), $d$-shell configurations, and state multiplicities 2S+1 for the ground and excited states of the studied FePc/Pyrene hybrid systems without the SW defect (left panel) and with the SW defect (right panel). All presented results were obtained employing CAS(10,9) for FePc/Pyrene(-S) and CAS(11,9) for FePc/Pyrene-B(-N), 
       where iron $d$-orbitals and closet HOMO and LUMO ligand state are in the active space.} 
       \label{table:FePcPyrene109_transitions}
\begin{tabular}{c ccc|| ccc}
\hline
  & \multicolumn{3}{c||}{\makecell{Without the SW defect}} & \multicolumn{3}{c}{\makecell{With the Stone-Wales defect}} \\
\hline 
 System & $\nu$ & State & 2S+1 & $\nu$ & State & 2S+1 \\
\hline 
  \multirow{6}{*}{\makecell{FePc-Pyrene}} & --- & $\displaystyle Fe\left( d_{xy}^{2} d_{xz}^{1} d_{yz}^{1} d_{z^{2}}^{2} d_{x^{2} -y^{2}}^{0}\right)$ & 3 & --- & $\displaystyle Fe\left( d_{xy}^{2} d_{xz}^{1} d_{yz}^{1} d_{z^{2}}^{2} d_{x^{2} -y^{2}}^{0}\right)$ & 3 \\

   & 188 & $\displaystyle Fe\left( d_{xy}^{2} d_{xz}^{1} d_{yz}^{2} d_{z^{2}}^{1} d_{x^{2} -y^{2}}^{0}\right)$ & 3 & 255 & $\displaystyle Fe\left( d_{xy}^{2} d_{xz}^{1} d_{yz}^{2} d_{z^{2}}^{1} d_{x^{2} -y^{2}}^{0}\right)$ & 3 \\

   & 253 & $\displaystyle Fe\left( d_{xy}^{2} d_{xz}^{2} d_{yz}^{1} d_{z^{2}}^{1} d_{x^{2} -y^{2}}^{0}\right)$ & 3 & 306 & $\displaystyle Fe\left( d_{xy}^{2} d_{xz}^{2} d_{yz}^{1} d_{z^{2}}^{1} d_{x^{2} -y^{2}}^{0}\right)$ & 3 \\

   & 3216 & $\displaystyle Fe\left( d_{xy}^{1} d_{xz}^{1} d_{yz}^{1} d_{z^{2}}^{2} d_{x^{2} -y^{2}}^{1}\right)$ & 5 & 3242 & $\displaystyle Fe\left( d_{xy}^{1} d_{xz}^{1} d_{yz}^{1} d_{z^{2}}^{2} d_{x^{2} -y^{2}}^{1}\right)$ & 5 \\

   & 3955 & $\displaystyle Fe\left( d_{xy}^{1} d_{xz}^{1} d_{yz}^{2} d_{z^{2}}^{1} d_{x^{2} -y^{2}}^{1}\right)$ & 5 & 4043 & $\displaystyle Fe\left( d_{xy}^{1} d_{xz}^{1} d_{yz}^{2} d_{z^{2}}^{1} d_{x^{2} -y^{2}}^{1}\right)$ & 5 \\

   & 4035 & $\displaystyle Fe\left( d_{xy}^{1} d_{xz}^{2} d_{yz}^{1} d_{z^{2}}^{1} d_{x^{2} -y^{2}}^{1}\right)$ & 5 & 4081 & $\displaystyle Fe\left( d_{xy}^{1} d_{xz}^{2} d_{yz}^{1} d_{z^{2}}^{1} d_{x^{2} -y^{2}}^{1}\right)$ & 5 \\
 \hline
 
  \multirow{6}{*}{\makecell{FePc-Pyrene-B}} & --- & $\displaystyle Fe(d_{xy}^{2} d_{xz}^{1} d_{yz}^{2} d_{z^{2}}^{1} d_{x^{2} -y^{2}}^{0}) \ B\left( p_{z}^{1}\right)$ & 2 & --- & $\displaystyle Fe(d_{xy}^{2} d_{xz}^{1} d_{yz}^{1} d_{z^{2}}^{2} d_{x^{2} -y^{2}}^{0} \ B\left( p_{z}^{1}\right))$ & 2 \\

   & 126 & $\displaystyle Fe(d_{xy}^{2} d_{xz}^{1} d_{yz}^{1} d_{z^{2}}^{2} d_{x^{2} -y^{2}}^{0}) \ B\left( p_{z}^{1}\right)$ & 4 & 26 & $\displaystyle Fe(d_{xy}^{2} d_{xz}^{1} d_{yz}^{1} d_{z^{2}}^{2} d_{x^{2} -y^{2}}^{0}) \ B\left( p_{z}^{1}\right)$ & 4 \\

   & 724 & $\displaystyle Fe(d_{xy}^{2} d_{xz}^{2} d_{yz}^{1} d_{z^{2}}^{1} d_{x^{2} -y^{2}}^{0}) \ B\left( p_{z}^{1}\right)$ & 2 & 148 & $\displaystyle Fe(d_{xy}^{2} d_{xz}^{1} d_{yz}^{2} d_{z^{2}}^{1} d_{x^{2} -y^{2}}^{0}) \ B\left( p_{z}^{1}\right)$ & 2 \\

   & 908 & $\displaystyle Fe(d_{xy}^{2} d_{xz}^{1} d_{yz}^{1} d_{z^{2}}^{2} d_{x^{2} -y^{2}}^{0}) \ B\left( p_{z}^{1}\right)$ & 2 & 164 & $\displaystyle Fe(d_{xy}^{2} d_{xz}^{1} d_{yz}^{2} d_{z^{2}}^{1} d_{x^{2} -y^{2}}^{0}) \ B\left( p_{z}^{1}\right)$ & 4 \\

   & 1245 & $\displaystyle Fe(d_{xy}^{2} d_{xz}^{1} d_{yz}^{2} d_{z^{2}}^{1} d_{x^{2} -y^{2}}^{0}) \ B\left( p_{z}^{1}\right)$ & 4 & 252 & $\displaystyle Fe(d_{xy}^{2} d_{xz}^{2} d_{yz}^{1} d_{z^{2}}^{1} d_{x^{2} -y^{2}}^{0}) \ B\left( p_{z}^{1}\right)$ & 4 \\

   & 1776 & $\displaystyle Fe(d_{xy}^{2} d_{xz}^{2} d_{yz}^{1} d_{z^{2}}^{1} d_{x^{2} -y^{2}}^{0}) \ B\left( p_{z}^{1}\right)$ & 4 & 280 & $\displaystyle Fe(d_{xy}^{2} d_{xz}^{2} d_{yz}^{1} d_{z^{2}}^{1} d_{x^{2} -y^{2}}^{0}) \ B\left( p_{z}^{1}\right)$ & 2 \\
    \hline
 \multirow{6}{*}{\makecell{FePc-Pyrene-N}} & --- & $\displaystyle Fe\left( d_{xy}^{2} d_{xz}^{1} d_{yz}^{1} d_{z^{2}}^{2} d_{x^{2} -y^{2}}^{0}\right) \ N\left( p_{z}^{1}\right)$ & 2 & --- & $\displaystyle Fe\left( d_{xy}^{2} d_{xz}^{1} d_{yz}^{2} d_{z^{2}}^{1} d_{x^{2} -y^{2}}^{0}\right) \ N\left( p_{z}^{1}\right)$ & 2 \\

   & 25 & $\displaystyle Fe\left( d_{xy}^{2} d_{xz}^{1} d_{yz}^{1} d_{z^{2}}^{2} d_{x^{2} -y^{2}}^{0}\right) \ N\left( p_{z}^{1}\right)$ & 4 & 329 & $\displaystyle Fe\left( d_{xy}^{2} d_{xz}^{1} d_{yz}^{1} d_{z^{2}}^{2} d_{x^{2} -y^{2}}^{0} \ \right) N\left( p_{z}^{1}\right)$ & 2 \\

   & 154 & $\displaystyle Fe\left( d_{xy}^{2} d_{xz}^{2} d_{yz}^{1} d_{z^{2}}^{1} d_{x^{2} -y^{2}}^{0}\right) \ N\left( p_{z}^{1}\right)$ & 2 & 429 & $\displaystyle Fe\left( d_{xy}^{2} d_{xz}^{1} d_{yz}^{2} d_{z^{2}}^{1} d_{x^{2} -y^{2}}^{0}\right) \ N\left( p_{z}^{1}\right)$ & 4 \\

   & 421 & $\displaystyle Fe\left( d_{xy}^{2} d_{xz}^{2} d_{yz}^{1} d_{z^{2}}^{1} d_{x^{2} -y^{2}}^{0}\right) \ N\left( p_{z}^{1}\right)$ & 4 & 472 & $\displaystyle Fe\left( d_{xy}^{2} d_{xz}^{1} d_{yz}^{1} d_{z^{2}}^{2} d_{x^{2} -y^{2}}^{0}\right) \ N\left( p_{z}^{1}\right)$ & 4 \\

   & 465 & $\displaystyle Fe\left( d_{xy}^{2} d_{xz}^{1} d_{yz}^{2} d_{z^{2}}^{1} d_{x^{2} -y^{2}}^{0}\right) \ N\left( p_{z}^{1}\right)$ & 2 & 1238 & $\displaystyle Fe\left( d_{xy}^{2} d_{xz}^{2} d_{yz}^{1} d_{z^{2}}^{1} d_{x^{2} -y^{2}}^{0}\right) \ N\left( p_{z}^{1}\right)$ & 2 \\

   & 508 & $\displaystyle Fe\left( d_{xy}^{2} d_{xz}^{1} d_{yz}^{2} d_{z^{2}}^{1} d_{x^{2} -y^{2}}^{0}\right) \ N\left( p_{z}^{1}\right)$ & 4 & 1335 & $\displaystyle Fe\left( d_{xy}^{2} d_{xz}^{2} d_{yz}^{1} d_{z^{2}}^{1} d_{x^{2} -y^{2}}^{0}\right) \ N\left( p_{z}^{1}\right)$ & 4 \\
 \hline
 \multirow{6}{*}{\makecell{FePc-Pyrene-S}} & --- & $\displaystyle Fe\left( d_{xy}^{2} d_{xz}^{2} d_{yz}^{2} d_{z^{2}}^{0} d_{x^{2} -y^{2}}^{0}\right)$ & 1 & --- & $\displaystyle Fe\left( d_{xy}^{2} d_{xz}^{2} d_{yz}^{2} d_{z^{2}}^{0} d_{x^{2} -y^{2}}^{0}\right)$ & 1 \\

   & 6514 & $\displaystyle Fe\left( d_{xy}^{2} d_{xz}^{1} d_{yz}^{2} d_{z^{2}}^{1} d_{x^{2} -y^{2}}^{0}\right)$ & 3 & 2520 & $\displaystyle Fe\left( d_{xy}^{2} d_{xz}^{1} d_{yz}^{2} d_{z^{2}}^{1} d_{x^{2} -y^{2}}^{0}\right)$ & 3 \\

   & 6542 & $\displaystyle Fe\left( d_{xy}^{2} d_{xz}^{2} d_{yz}^{1} d_{z^{2}}^{1} d_{x^{2} -y^{2}}^{0}\right)$ & 3 & 2615 & $\displaystyle Fe\left( d_{xy}^{2} d_{xz}^{2} d_{yz}^{1} d_{z^{2}}^{1} d_{x^{2} -y^{2}}^{0}\right)$ & 3 \\

   & 8885 & $\displaystyle Fe\left( d_{xy}^{1} d_{xz}^{2} d_{yz}^{2} d_{z^{2}}^{1} d_{x^{2} -y^{2}}^{0}\right)$ & 3 & 4833 & $\displaystyle Fe\left( d_{xy}^{1} d_{xz}^{2} d_{yz}^{2} d_{z^{2}}^{1} d_{x^{2} -y^{2}}^{0}\right)$ & 3 \\

   & 10055 & $\displaystyle Fe\left( d_{xy}^{1} d_{xz}^{1} d_{yz}^{2} d_{z^{2}}^{1} d_{x^{2} -y^{2}}^{1}\right)$ & 5 & 5972 & $\displaystyle Fe\left( d_{xy}^{1} d_{xz}^{1} d_{yz}^{2} d_{z^{2}}^{1} d_{x^{2} -y^{2}}^{1}\right)$ & 5 \\

   & 10109 & $\displaystyle Fe\left( d_{xy}^{1} d_{xz}^{2} d_{yz}^{1} d_{z^{2}}^{1} d_{x^{2} -y^{2}}^{1}\right)$ & 5 & 6035 & $\displaystyle Fe\left( d_{xy}^{1} d_{xz}^{2} d_{yz}^{1} d_{z^{2}}^{1} d_{x^{2} -y^{2}}^{1}\right)$ & 5 \\

 \hline
\end{tabular}
        \end{table*}


The ground state of the FePc molecule is the triplet, and it does not change when the molecule is adsorbed to pyrene. The presence of the SW defect in the pyrene also does not change the spin state of the system. Performed NEVPT2 calculations show that studied systems with B- and N-doping have a doublet ground state, whereas systems with sulphur are singlet (Table \ref{table:FePcPyrene109_transitions}).
Since the zero field splitting effect appears only in systems with a triplet multiplicity and higher, zero field splitting parameters only for systems with triplet states (FePc, FePc/Pyrene, FePc/Pyrene-SW) are presented in Table \ref{table:FePcPyreneZFS}. The FePc molecule has the zero field splitting axial component $D=90 cm^{-1}$ and the transversal component $E=0 cm^{-1}$, while the active space consists of only iron $d$-shell orbitals. The expansion of the active space leads to the increase of the transversal component because new HOMO-LUMO ligand active orbitals break the $D_{4h}$ symmetry of the active space. In FePc/Pyrene complexes pyrene presence lowers the zero field splitting parameter $D$  by several $cm^{-1}$ in comparison to the $D$ value for the free standing FePc molecule. Also, the value of the transversal component $E$ increases owing to the reduced symmetry of the hybrid system.

\begin{table}[h!t]
        \centering
       \caption{FePc, FePc/Pyrene, and FePc/Pyrene-SW zero-field splitting parameters.}
        \label{table:FePcPyreneZFS}
\begin{tabular}{c cc cc}

\hline
  & \multicolumn{2}{c}{CAS(6,5)} & \multicolumn{2}{c}{CAS(10,9)} \\
\hline
  System & D, $cm^{-1}$  & E/D  & D, $cm^{-1}$  & E/D  \\
\hline
 FePc & 90.0 & 0 & 91.6 & 0.06 \\

 FePc/Pyrene & 85.1 & 0.1 & 87.8 & 0.13 \\

 FePc/Pyrene-SW & 86.2 & 0.14 & 89.9 & 0.09 \\
 \hline
\end{tabular}

\end{table}

\section{Summary and Conclusions}

    The interaction between the FePc molecule and the graphene substrate has been studied using several approaches: based on periodic plane wave DFT, LCAO DFT with a cluster representation of graphene, and multiconfigurational methods with the pyrene molecule presented as a miniaturised graphene flake.
    Cluster representation of the graphene surface in the FePc/Graphene complex reproduces the results of the periodic model quite well. When using similar DFT parameters in the two models, the geometric and energy parameters differ insignificantly. 
    The model with the pyrene molecule instead of the graphene sheet shows really similar geometry in the vicinity of the FePc's iron atom and it could be used in demanding multireference calculations.
    The presence of FePc on the pure graphene layer does not significantly change the properties of both FePc and graphene despite their good van der Waals connection. The molecule on the top of the graphene layer does not open the band gap and the FePc iron $d$-shell does not significantly change its distribution in the FePc/Graphene hybrid system.
    The presence of the Stone-Wales defect in graphene and pyrene barely changes the FePc adsorption parameters and the iron d-shell distribution.
    FePc/Pyrene complexes with B, N-doping and SW-B, SW-N combined defects have a really small energy difference between states with different multiplicities around the ground state. It was found, that the ground state is unstable and the spin state of the system is subject to thermal fluctuations. Taking that into account, our results match the previous investigation.\cite{sarmah2019computational} While there were found only ground states of systems because of the one Slater determinant restriction, the behaviour of systems near the ground state was not taken into account. This restriction has been overcome in our studies by using multireference methods of calculation. 
    On the contrary, the multireference analysis shows that FePc/Pyrene complexes with S-doping and combined SW-S defects have a big first excitation frequency. The FePc ground state and the ground state of the whole systems are not triplets but singlets, these defects make the complexes non-magnetic. The presence of the sulphur atom in graphene also sufficiently increases the FePc-Graphene adsorption energy and the sulphur atom moves out of the graphene plane. The defects crucially change the morphology of the FePc/Graphene hybrid system; they modify the magnetic moment of the system, its geometry, and its energetic parameters.

\section{Acknowledgement}
This work has been supported by National Science Centre Poland (UMO 2016/23/B/ST3/03567).

\bibliography{Article}

\end{document}


\preprint{AIP/123-QED}

\title{Supplemental material}%

\maketitle

\begin{figure}[ht!]
\centering
\begin{subfigure}{0.4\linewidth}
  \centering
  \includegraphics[width=\textwidth]{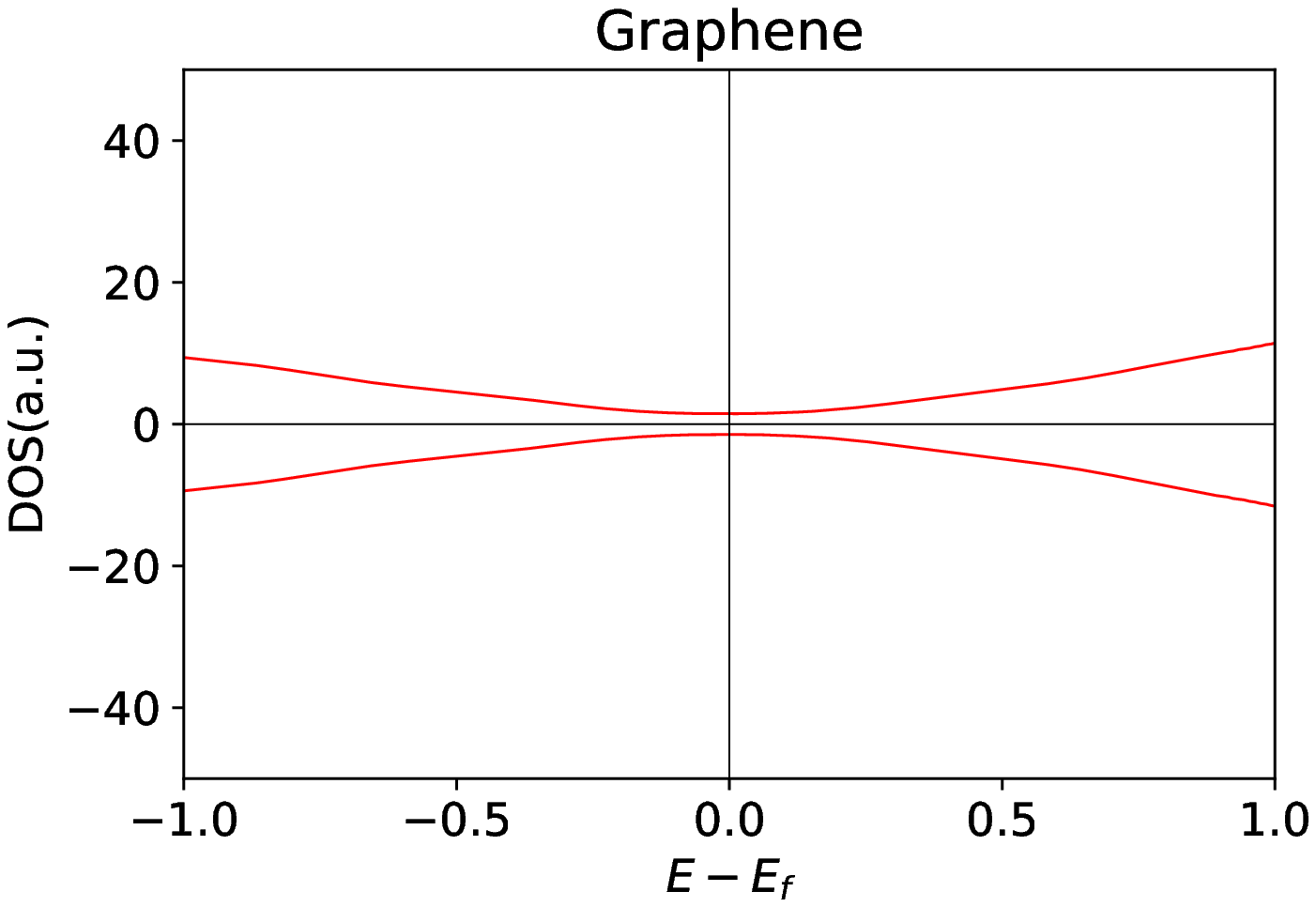}
  \caption{}
  \label{fig:GrDOS}
\end{subfigure}
\hfill
\begin{subfigure}{0.4\linewidth}
  \centering
  \includegraphics[width=\textwidth]{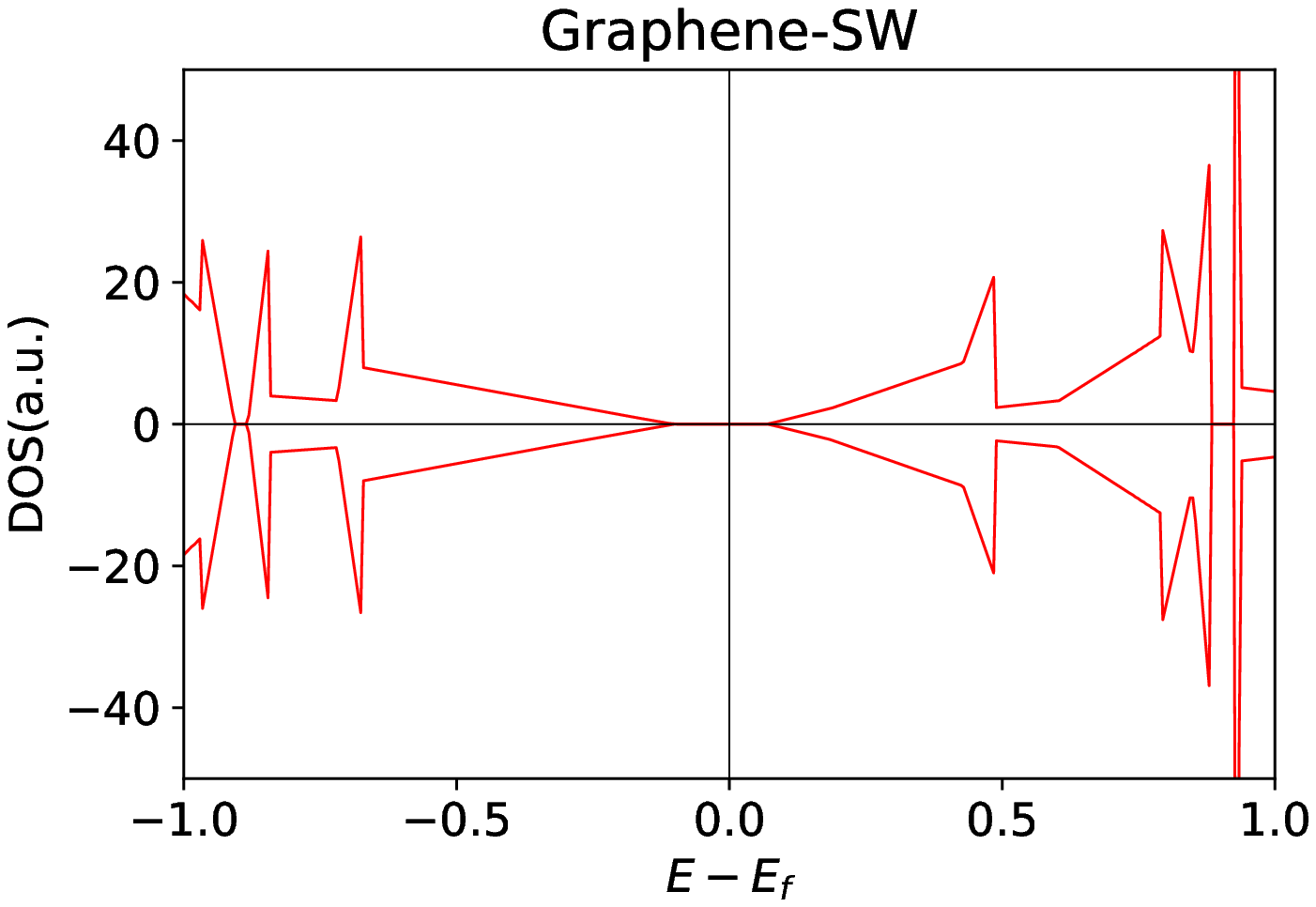}
  \caption{}
  \label{fig:Gr57DOS}
\end{subfigure}

\begin{subfigure}{0.4\linewidth}
  \centering
  \includegraphics[width=\linewidth]{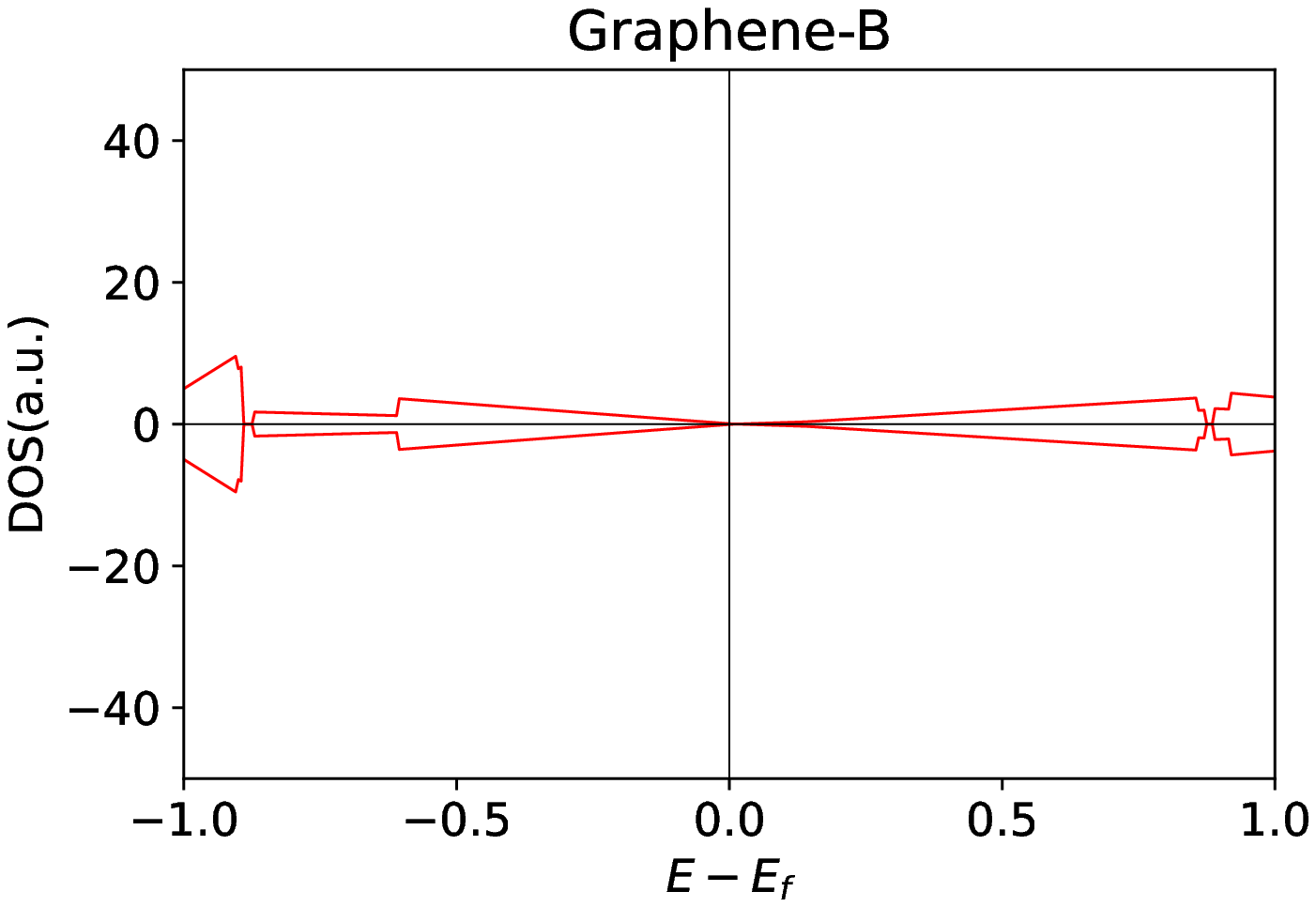}
  \caption{}
  \label{fig:GrBDOS}
\end{subfigure}
\hfill
\begin{subfigure}{0.4\linewidth}
  \centering
  \includegraphics[width=\linewidth]{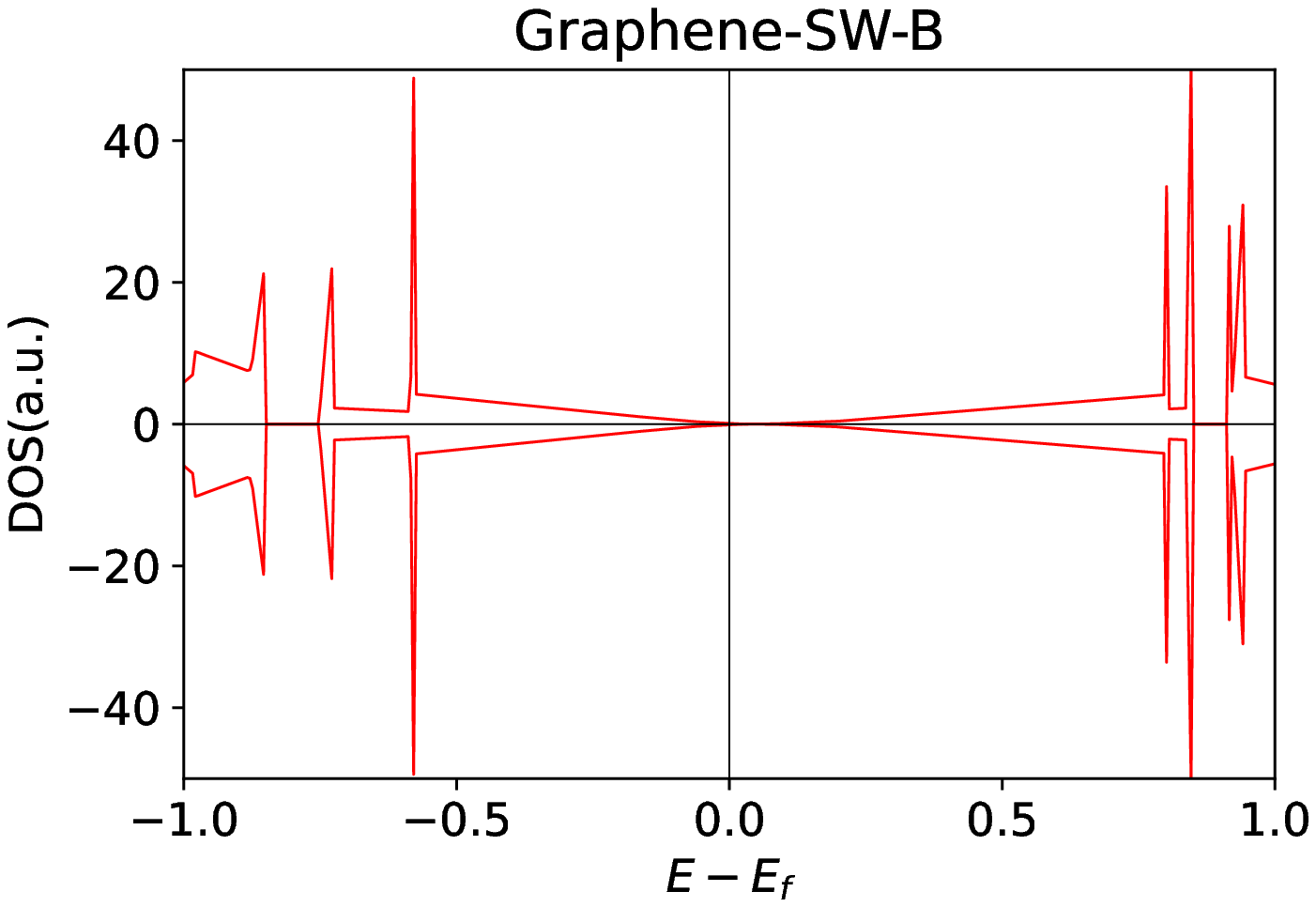}
  \caption{}
  \label{fig:Gr57BDOS}
\end{subfigure}

\begin{subfigure}{0.4\linewidth}
  \centering
  \includegraphics[width=\linewidth]{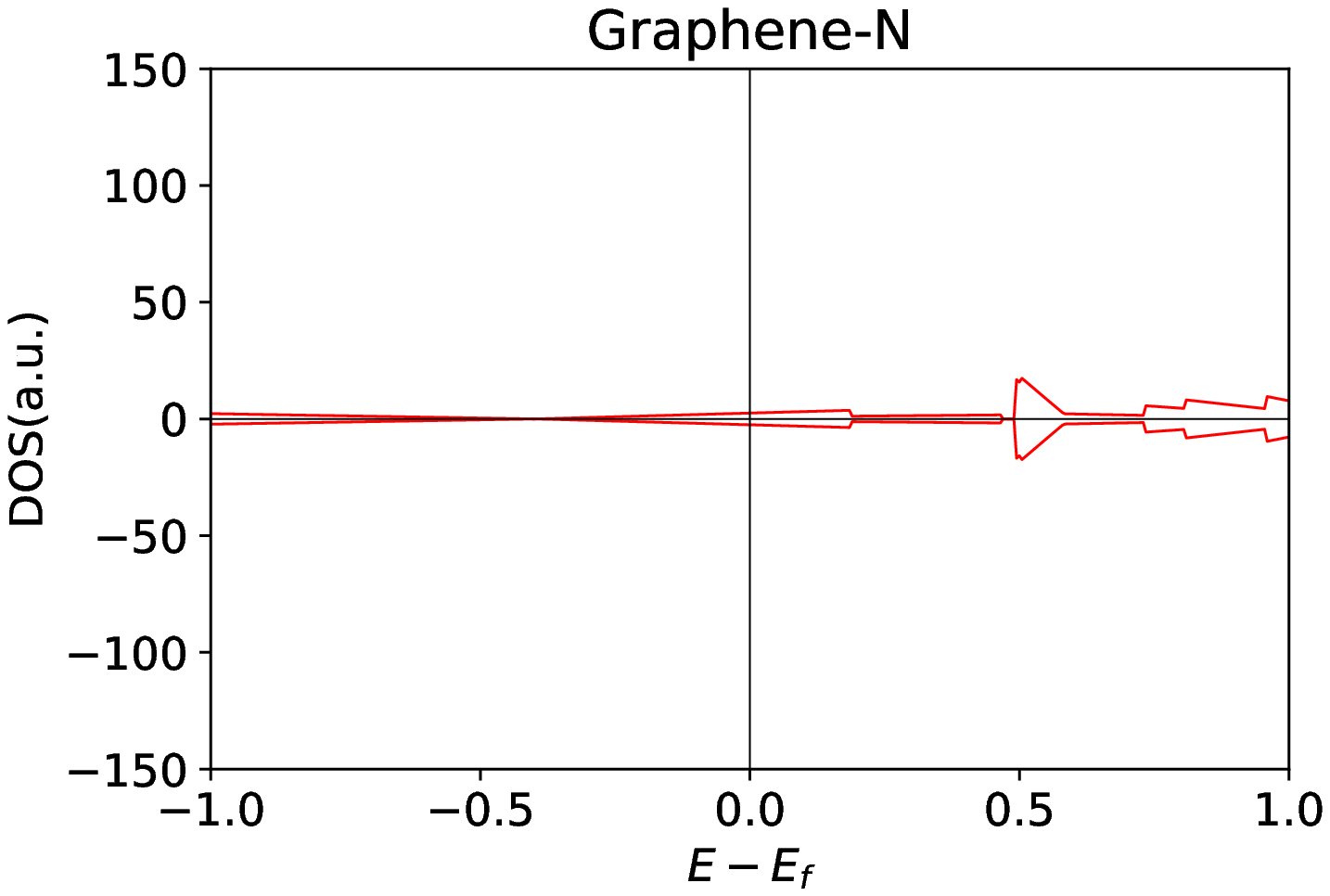}
  \caption{}
  \label{fig:GrNDOS}
\end{subfigure}
\hfill
\begin{subfigure}{0.4\linewidth}
  \centering
  \includegraphics[width=\linewidth]{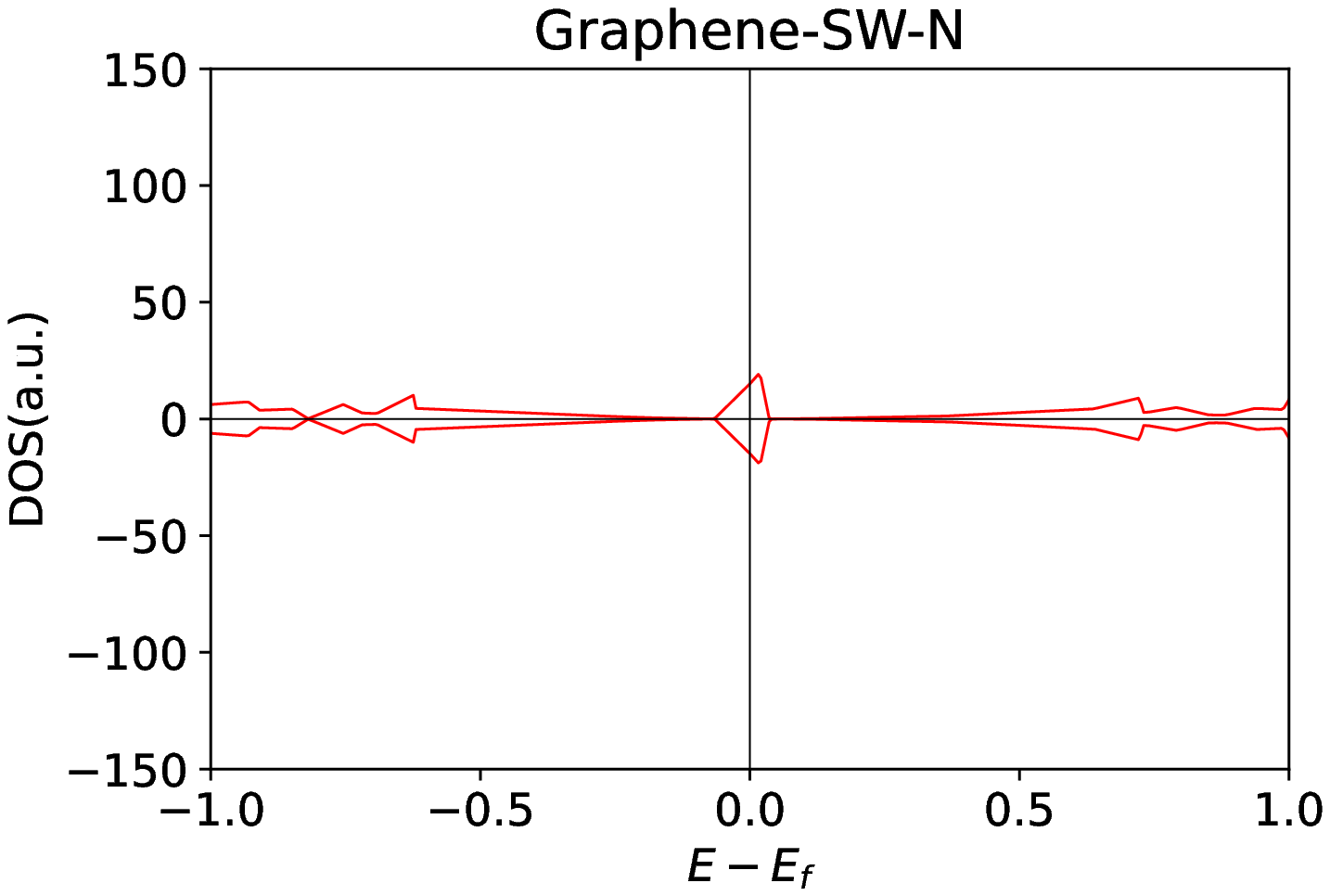}
  \caption{}
  \label{fig:Gr57NDOS}
\end{subfigure}

\begin{subfigure}{0.4\linewidth}
  \centering
  \includegraphics[width=\linewidth]{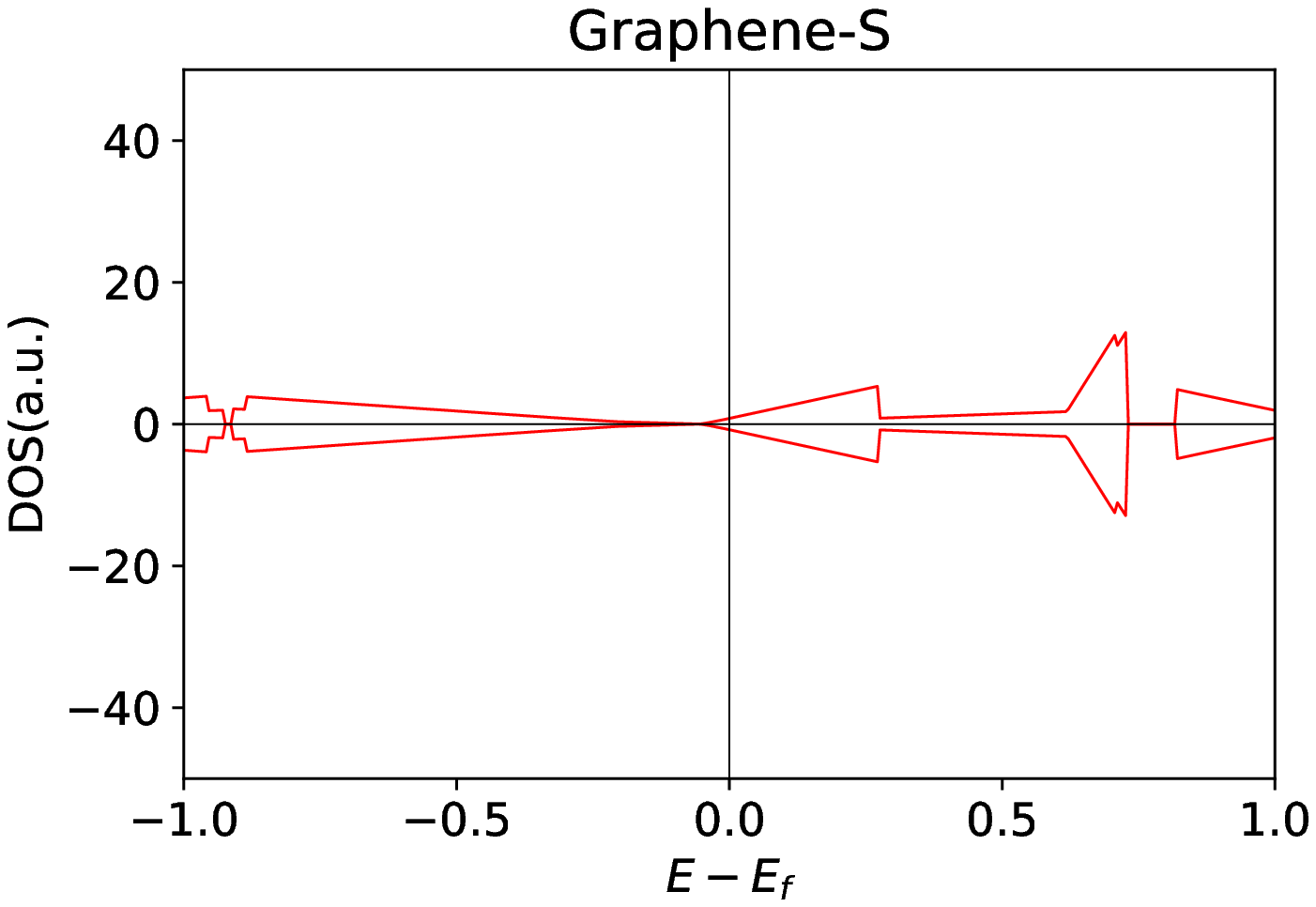}
  \caption{}
  \label{fig:GrSDOS}
\end{subfigure}
\hfill
\begin{subfigure}{0.4\linewidth}
  \centering
  \includegraphics[width=\linewidth]{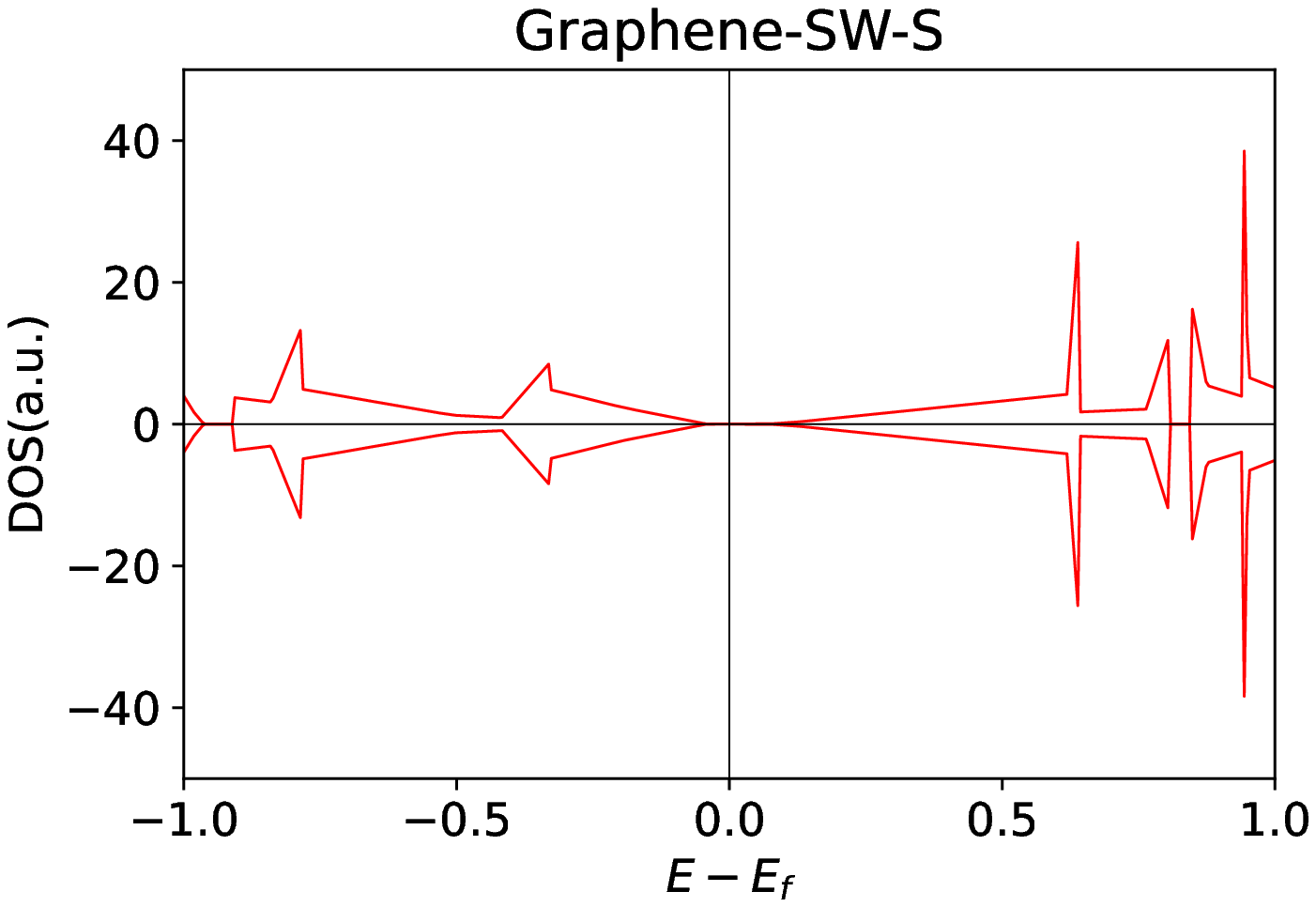}
  \caption{}
  \label{fig:Gr57SDOS}
\end{subfigure}

\caption{Densities of states for studied pristine and defected graphene layers}
\label{fig:GrDOSAll}
\end{figure}

\begin{figure}[ht!]
\centering

\begin{subfigure}{0.4\linewidth}
  \centering
  \includegraphics[width=\linewidth]{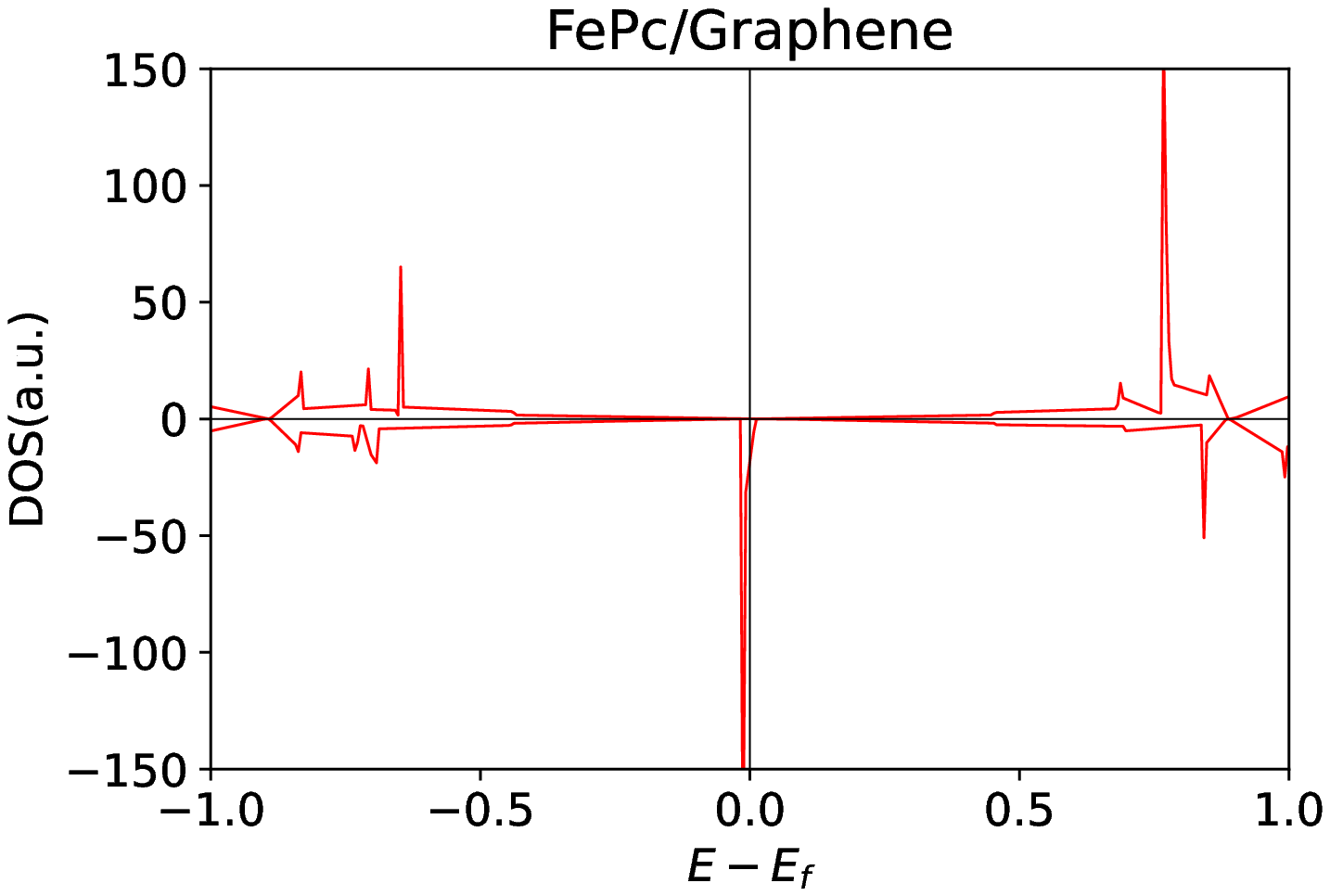}
  \caption{}
  \label{fig:FePcGrDOS}
\end{subfigure}
\hfill
\begin{subfigure}{0.4\linewidth}
  \centering
  \includegraphics[width=\linewidth]{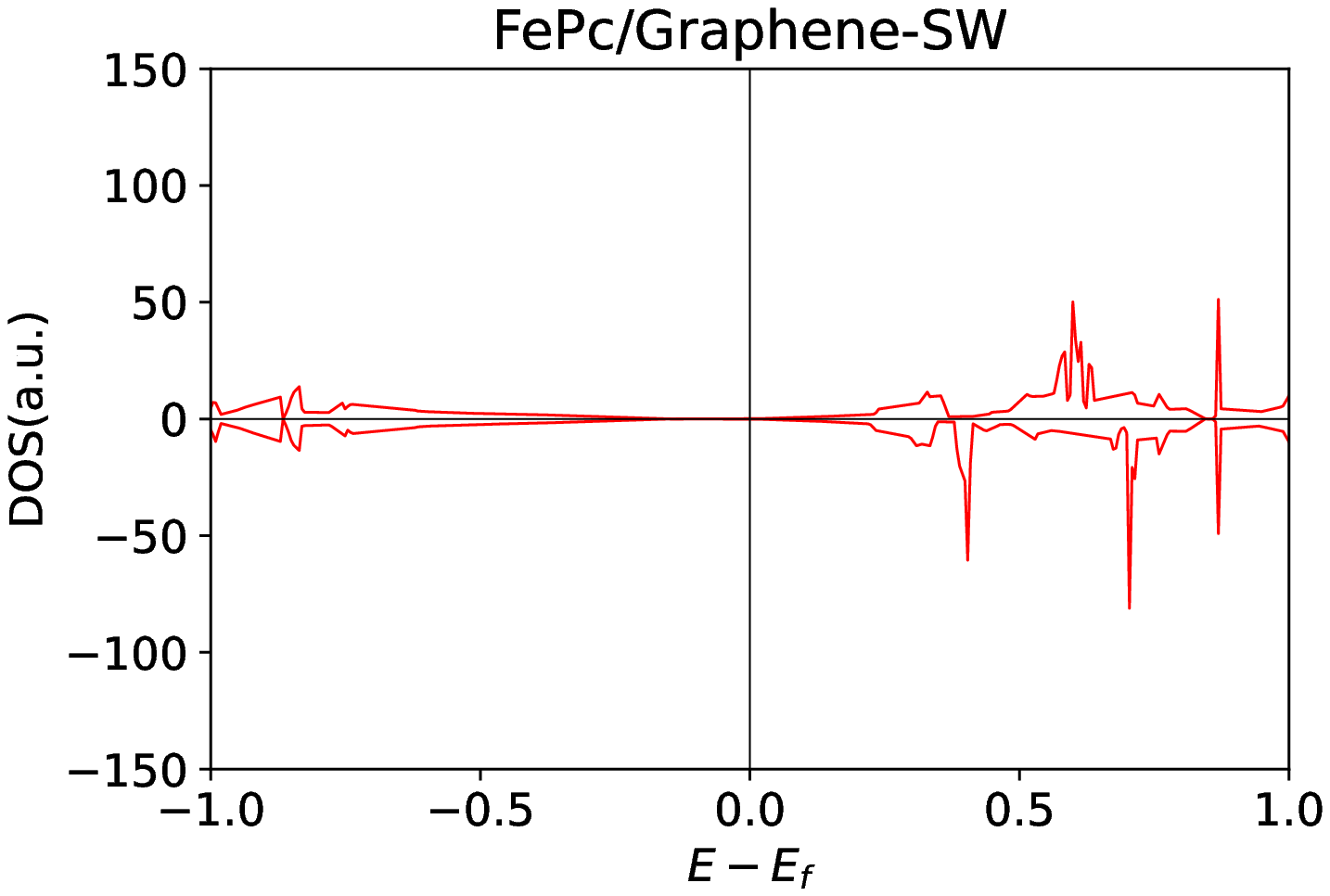}
  \caption{}
  \label{fig:FePcGr57DOS}
\end{subfigure}

\begin{subfigure}{0.4\linewidth}
  \centering
  \includegraphics[width=\linewidth]{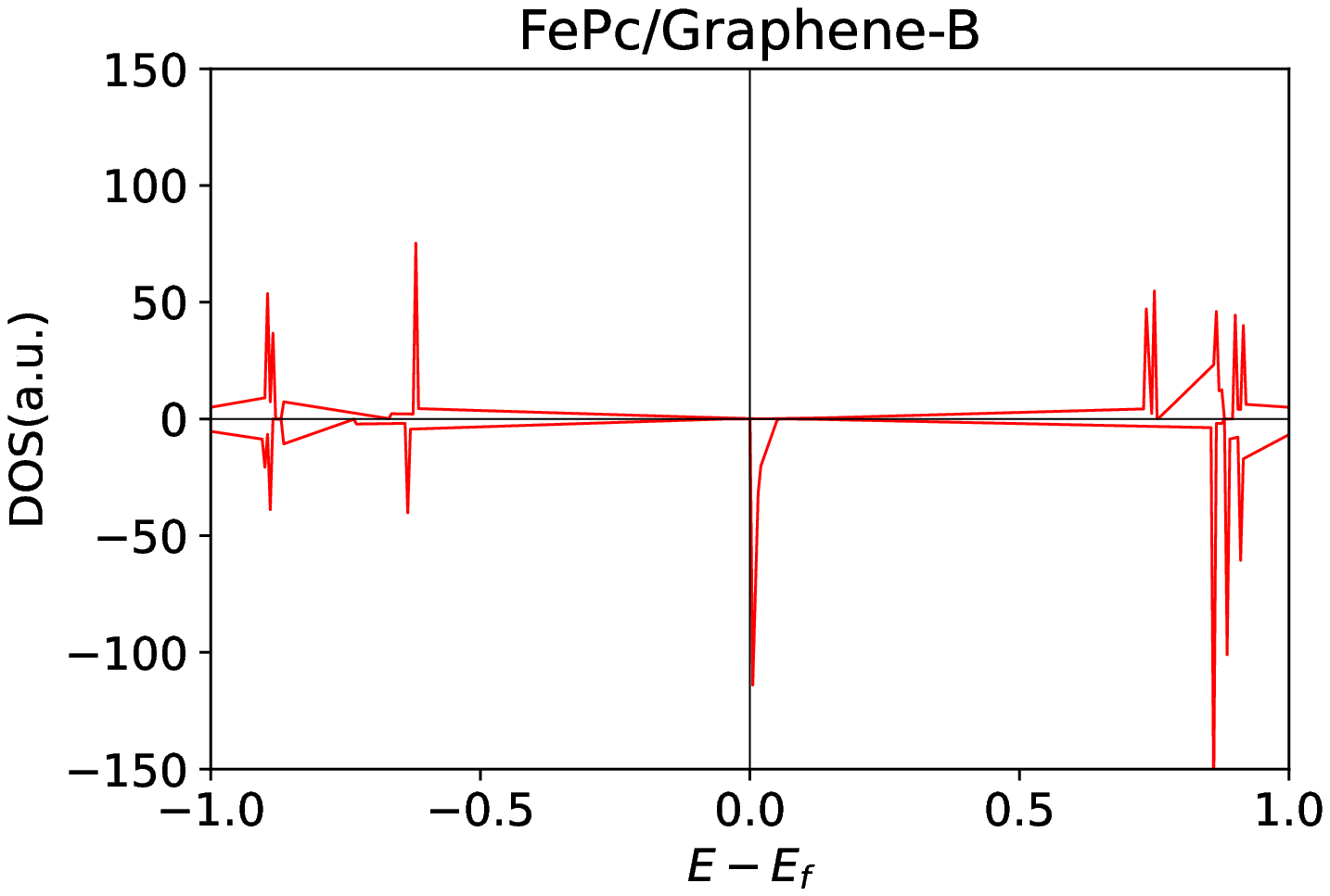}
  \caption{}
  \label{fig:FePcGrBDOS}
\end{subfigure}
\hfill
\begin{subfigure}{0.4\linewidth}
  \centering
  \includegraphics[width=\linewidth]{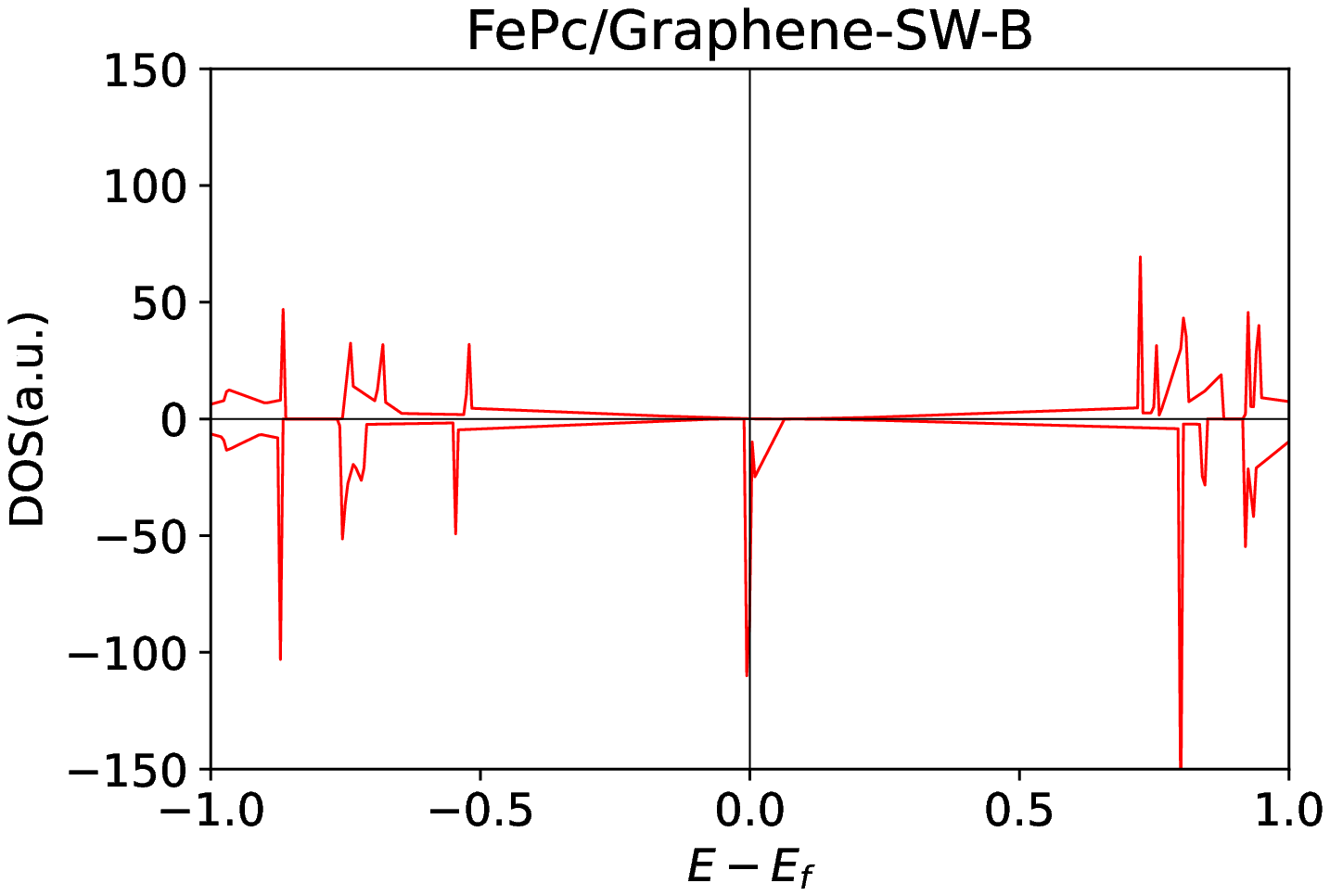}
  \caption{}
  \label{fig:FePcGr57BDOS}
\end{subfigure}

\begin{subfigure}{0.4\linewidth}
  \centering
  \includegraphics[width=\linewidth]{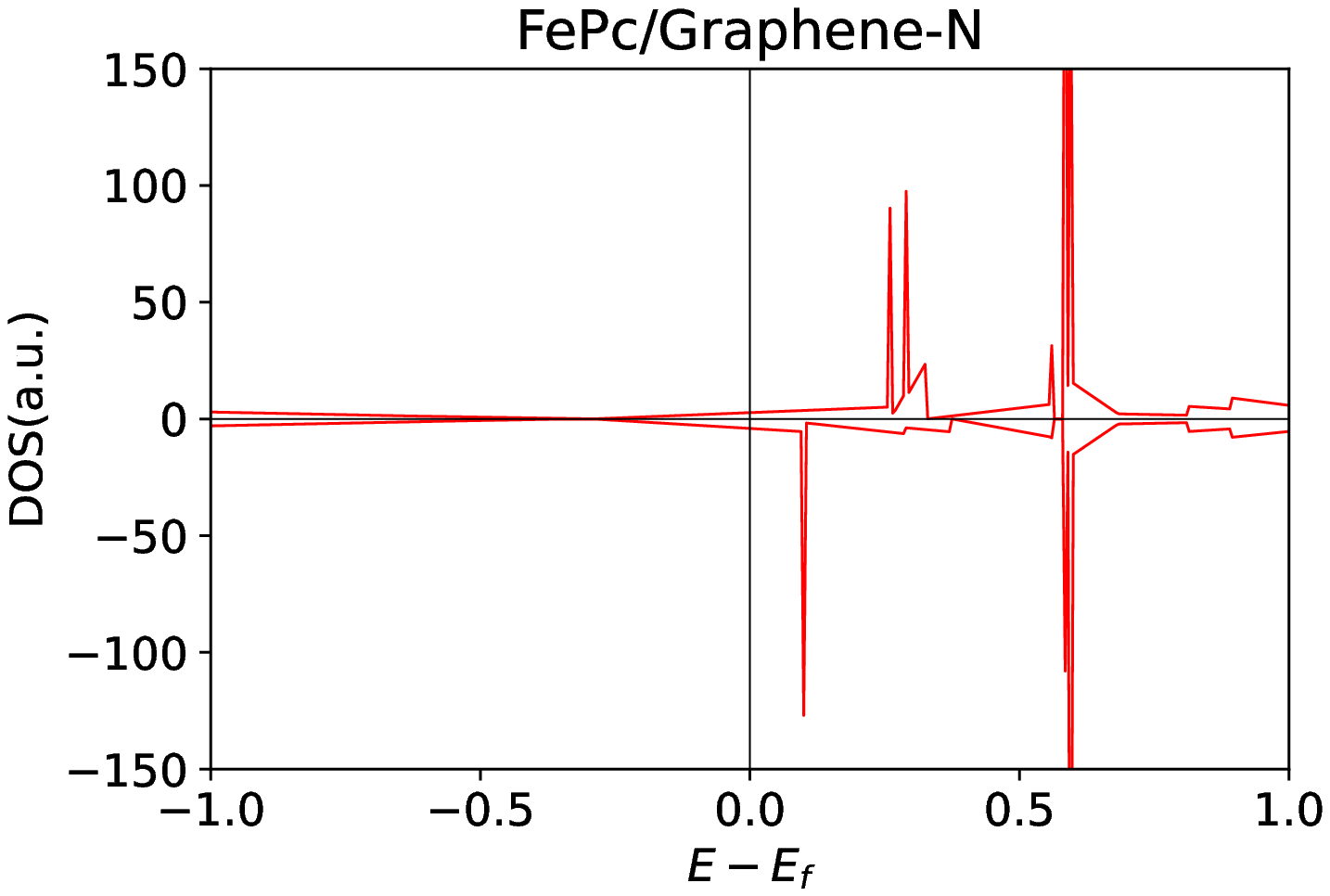}
  \caption{}
  \label{fig:FePcGrNDOS}
\end{subfigure}
\hfill
\begin{subfigure}{0.4\linewidth}
  \centering
  \includegraphics[width=\linewidth]{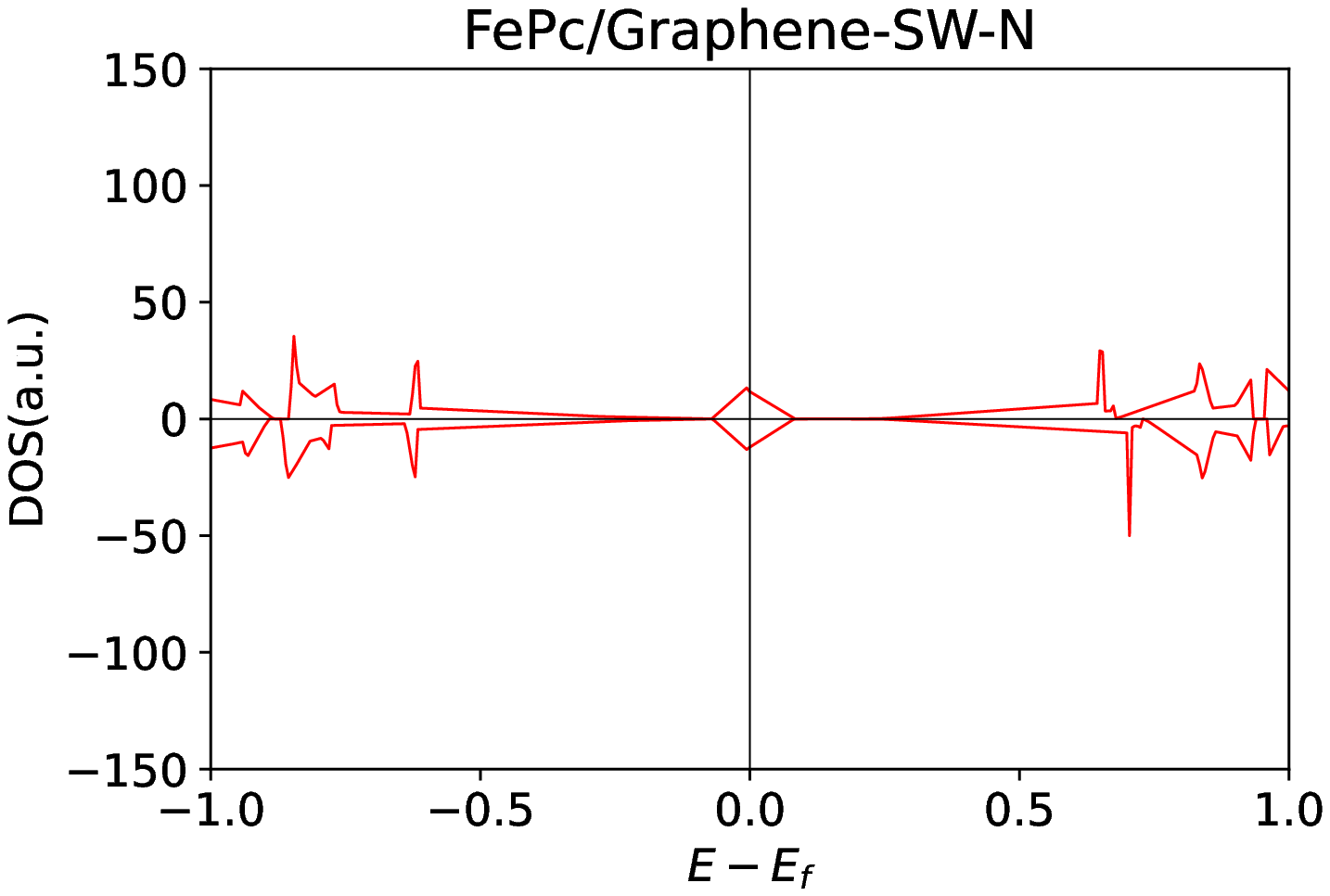}
  \caption{}
  \label{fig:FePcGr57NDOS}
\end{subfigure}

\begin{subfigure}{0.4\linewidth}
  \centering
  \includegraphics[width=\linewidth]{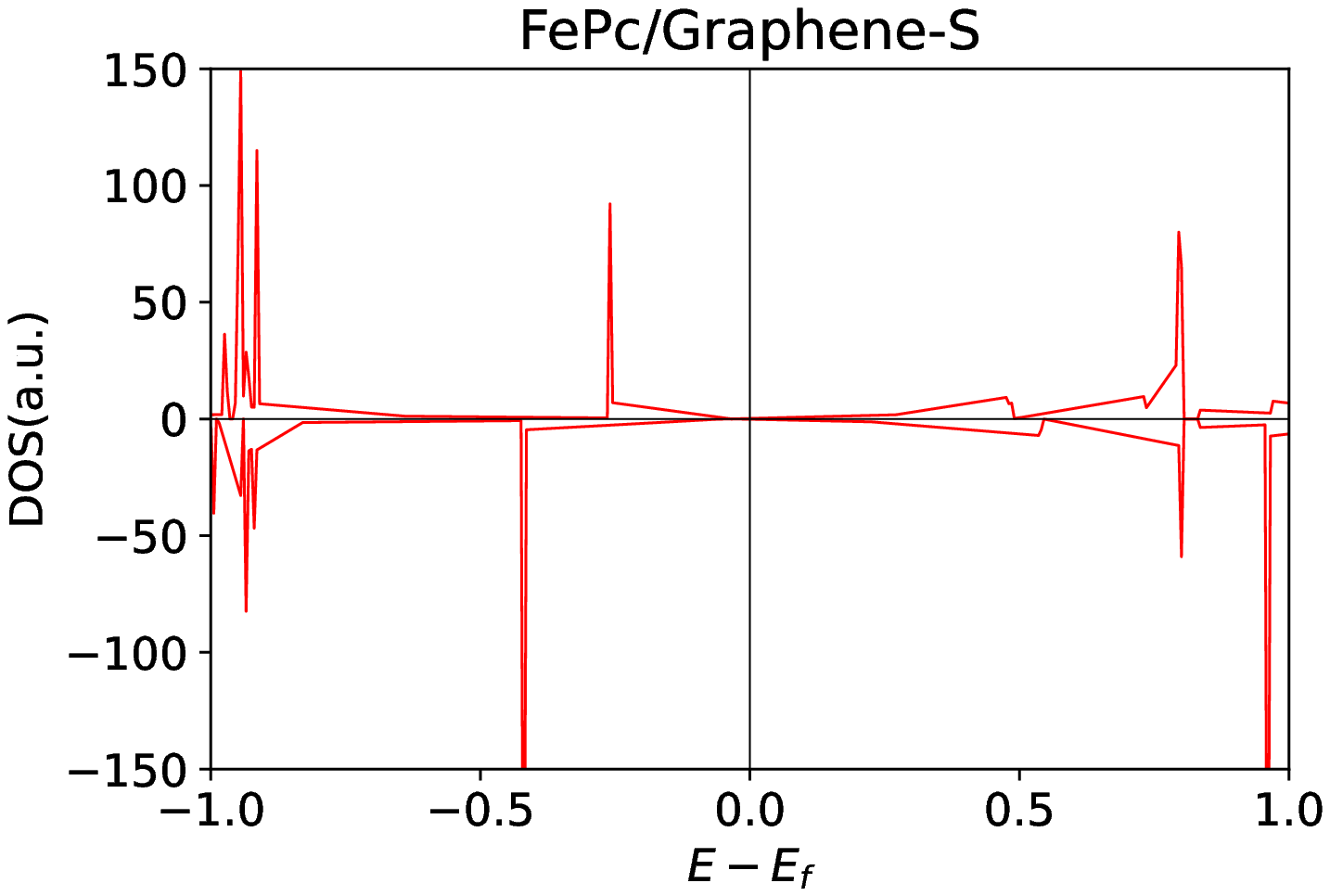}
  \caption{}
  \label{fig:FePcGrSDOS}
\end{subfigure}
\hfill
\begin{subfigure}{0.4\linewidth}
  \centering
  \includegraphics[width=\linewidth]{FePcGrPictures/DOS/DOSGrSWS.eps}
  \caption{}
  \label{fig:FePcGr57SDOS}
\end{subfigure}

\caption{Densities of states for studied hybrid systems consisting of FePc molecule adsorbed to pristine and defected graphene layers.}
\label{fig:AllFePcGrDOS}
\end{figure}

\begin{figure}[ht!]
\centering

\begin{subfigure}{0.4\linewidth}
  \centering
  \includegraphics[width=\linewidth]{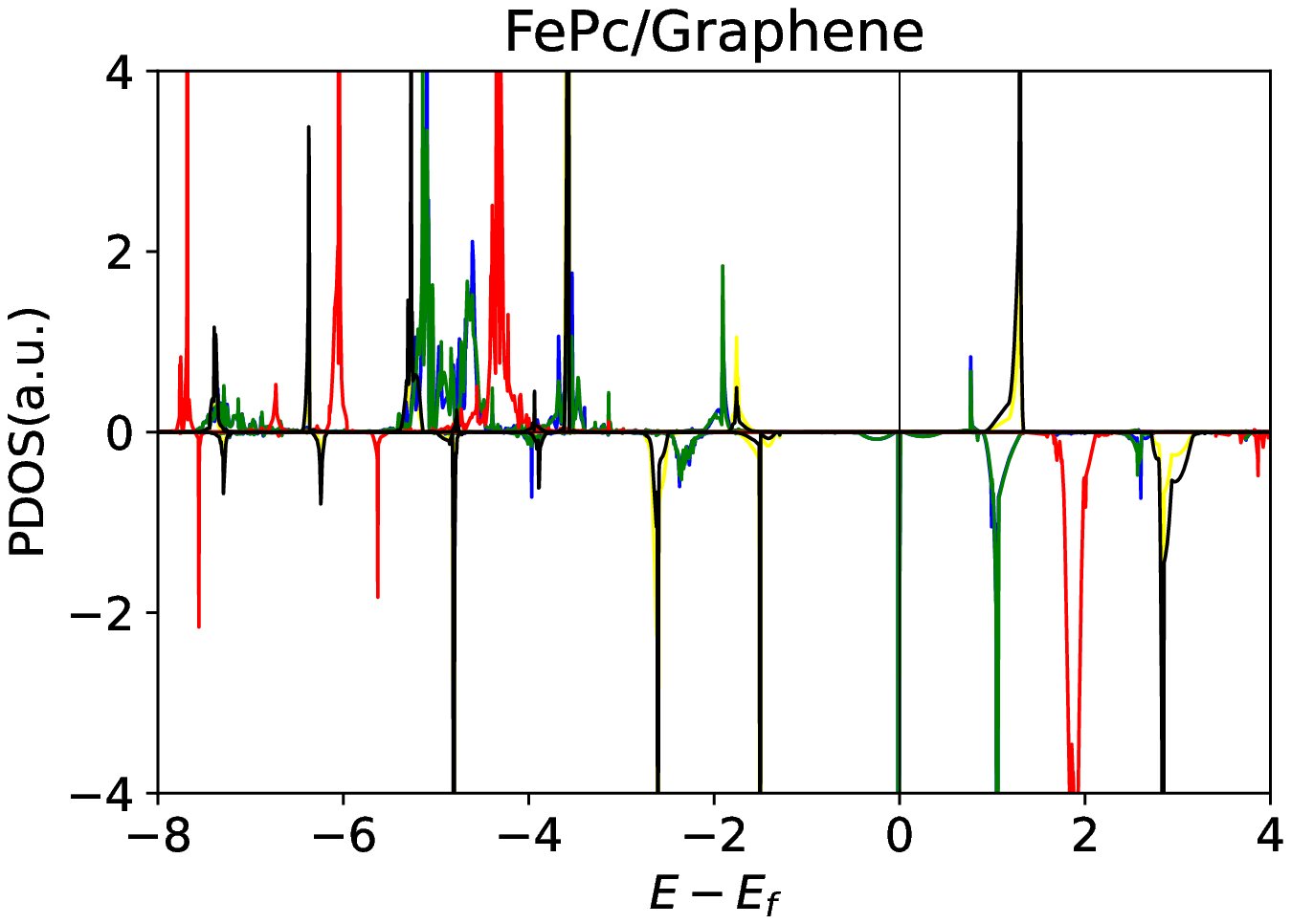}
  \caption{}
  \label{fig:FePcGrFePDOS}
\end{subfigure}
\hfill
\begin{subfigure}{0.4\linewidth}
  \centering
  \includegraphics[width=\linewidth]{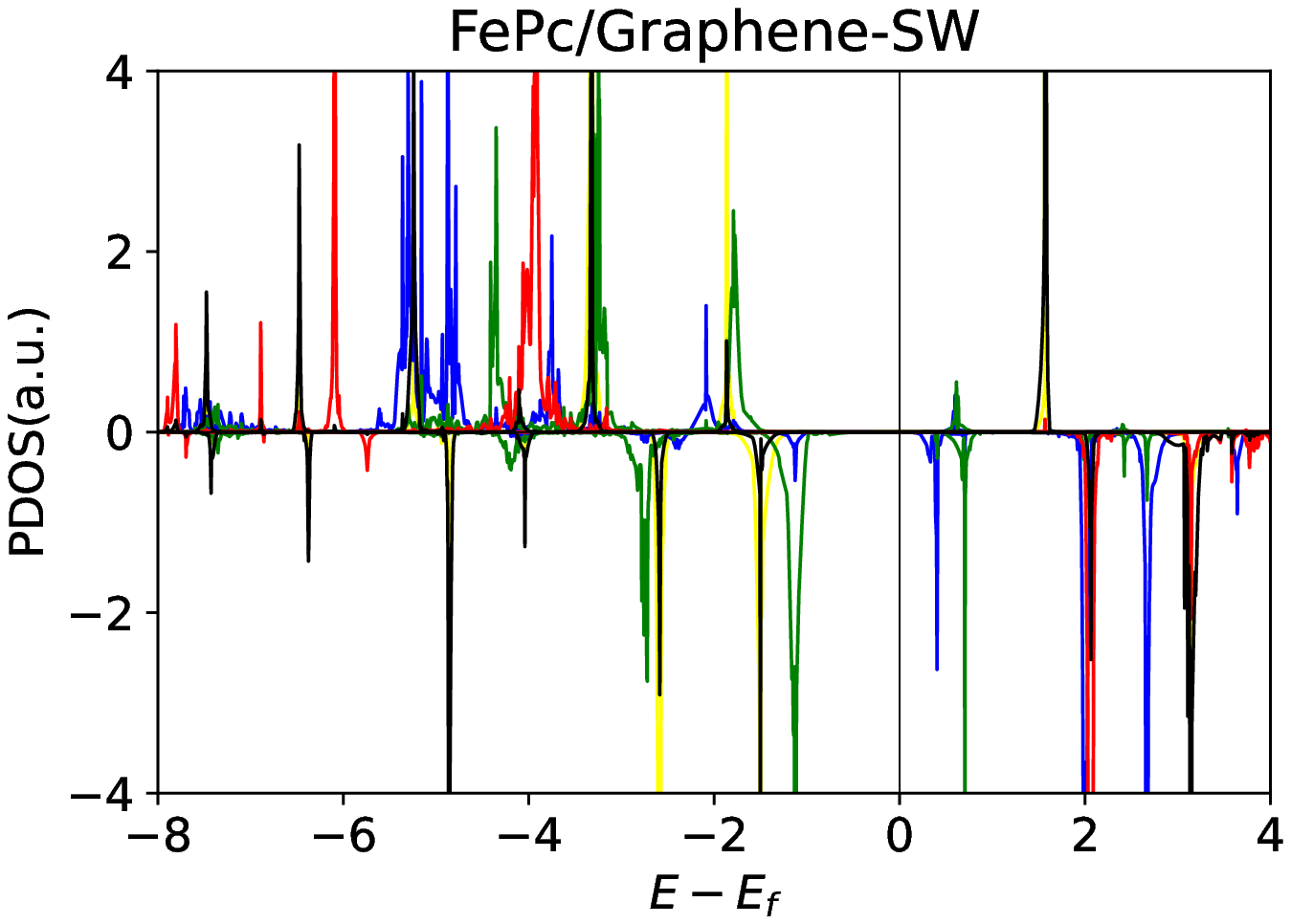}
  \caption{}
  \label{fig:FePcGr57FePDOS}
\end{subfigure}

\begin{subfigure}{0.4\linewidth}
  \centering
  \includegraphics[width=\linewidth]{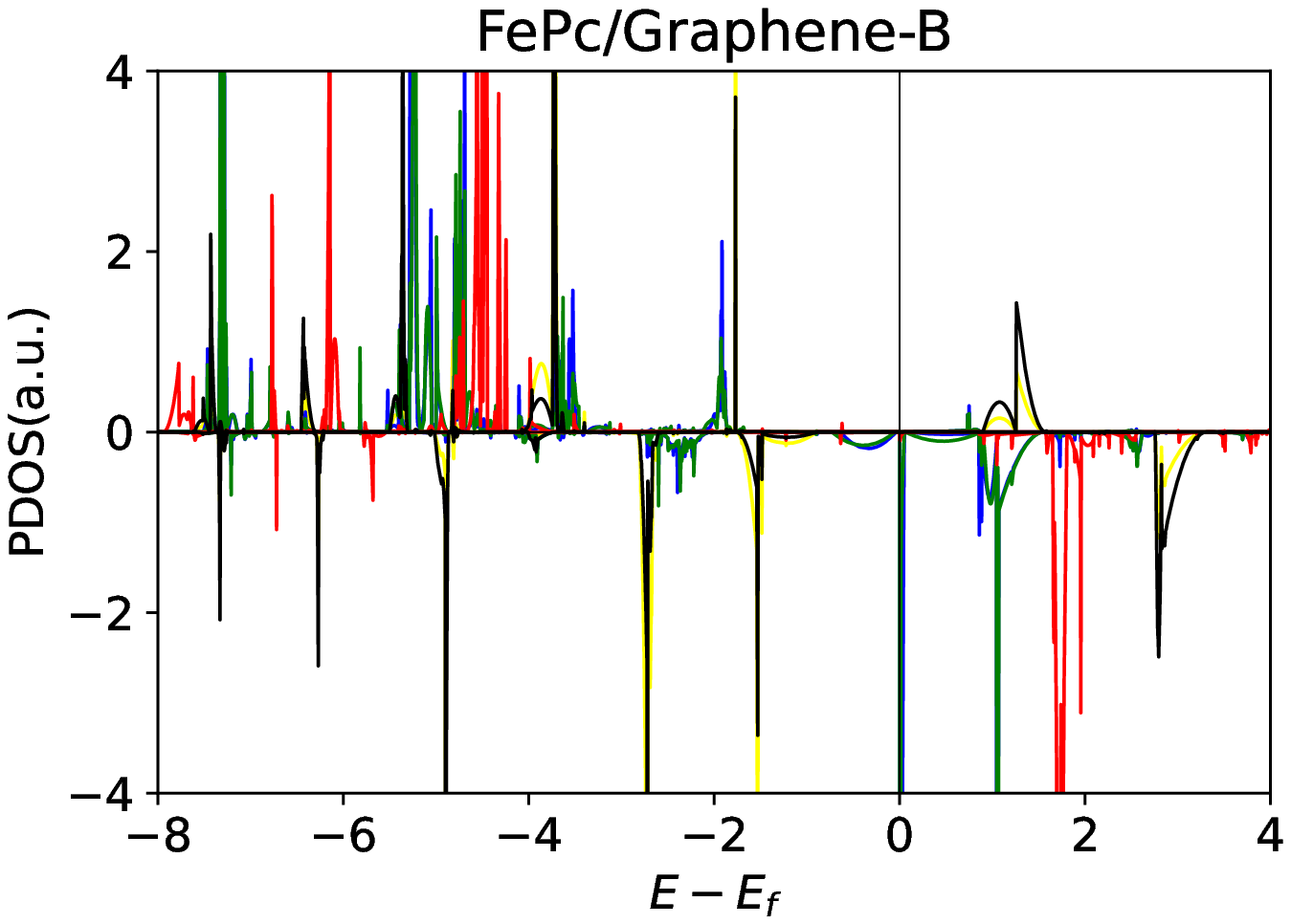}
  \caption{}
  \label{fig:FePcGrBFePDOS}
\end{subfigure}
\hfill
\begin{subfigure}{0.4\linewidth}
  \centering
  \includegraphics[width=\linewidth]{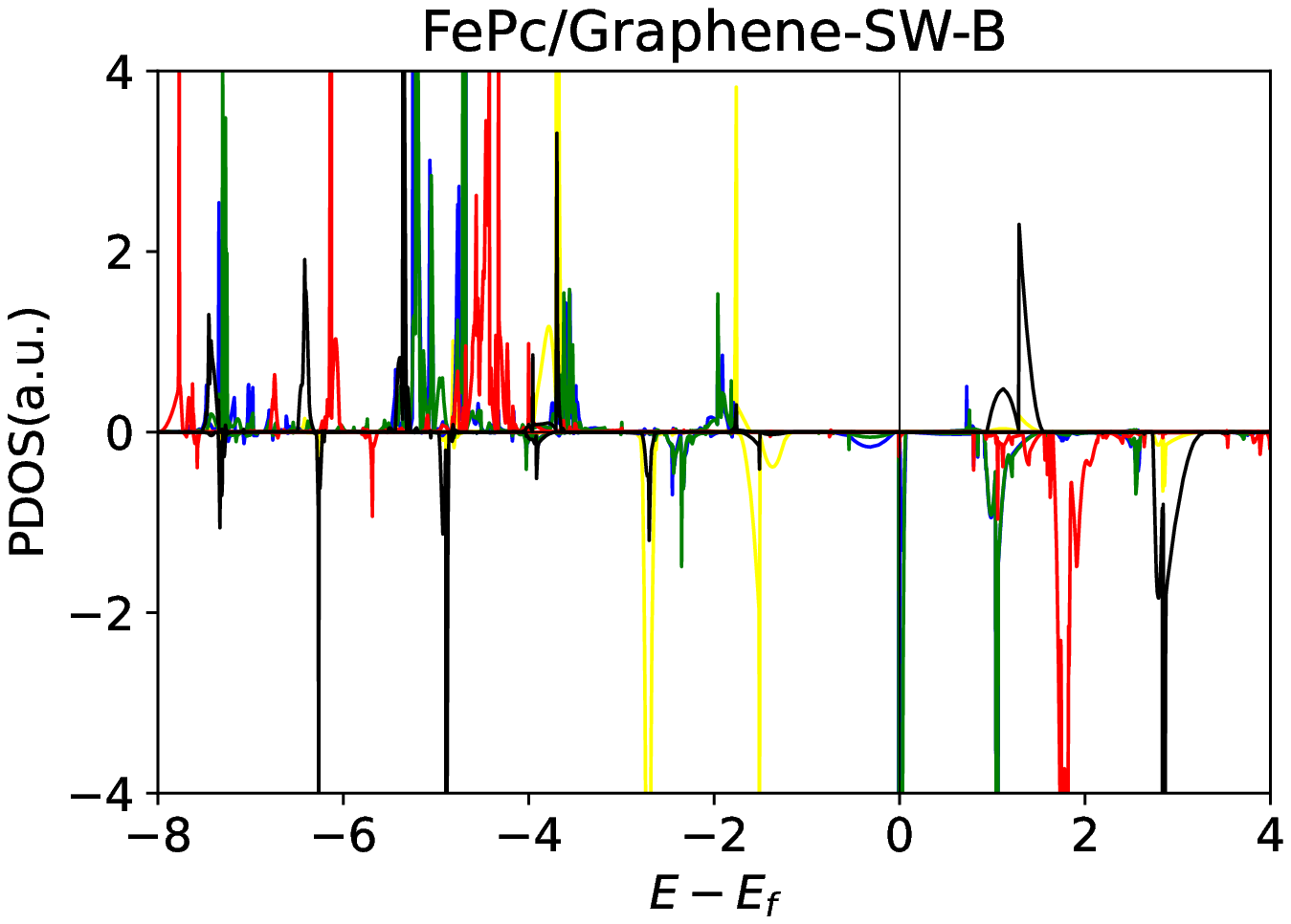}
  \caption{}
  \label{fig:FePcGr57BFePDOS}
\end{subfigure}

\begin{subfigure}{0.4\linewidth}
  \centering
  \includegraphics[width=\linewidth]{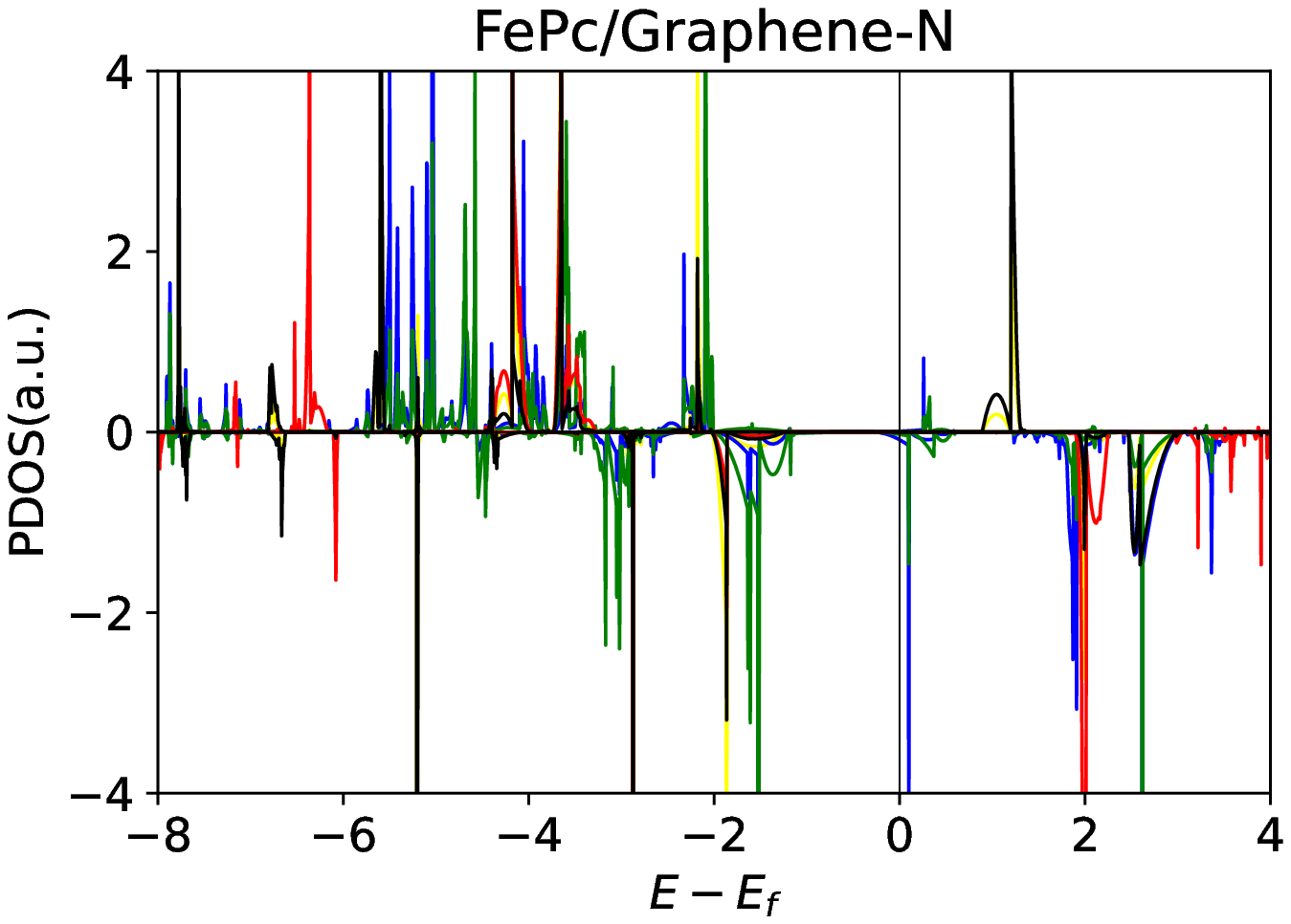}
  \caption{}
  \label{fig:FePcGrNFePDOS}
\end{subfigure}
\hfill
\begin{subfigure}{0.4\linewidth}
  \centering
  \includegraphics[width=\linewidth]{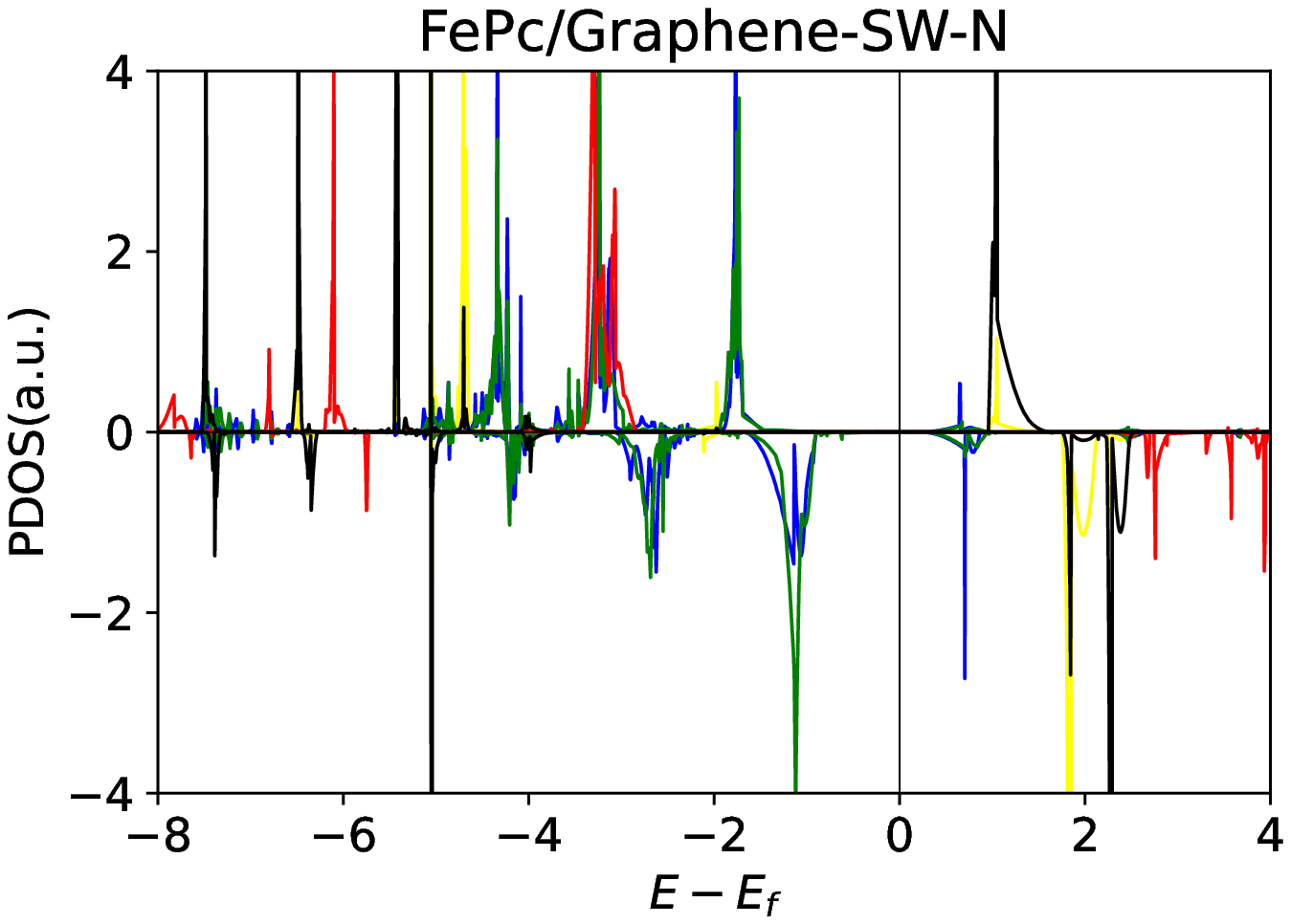}
  \caption{}
  \label{fig:FePcGr57NFePDOS}
\end{subfigure}

\begin{subfigure}{0.4\linewidth}
  \centering
  \includegraphics[width=\linewidth]{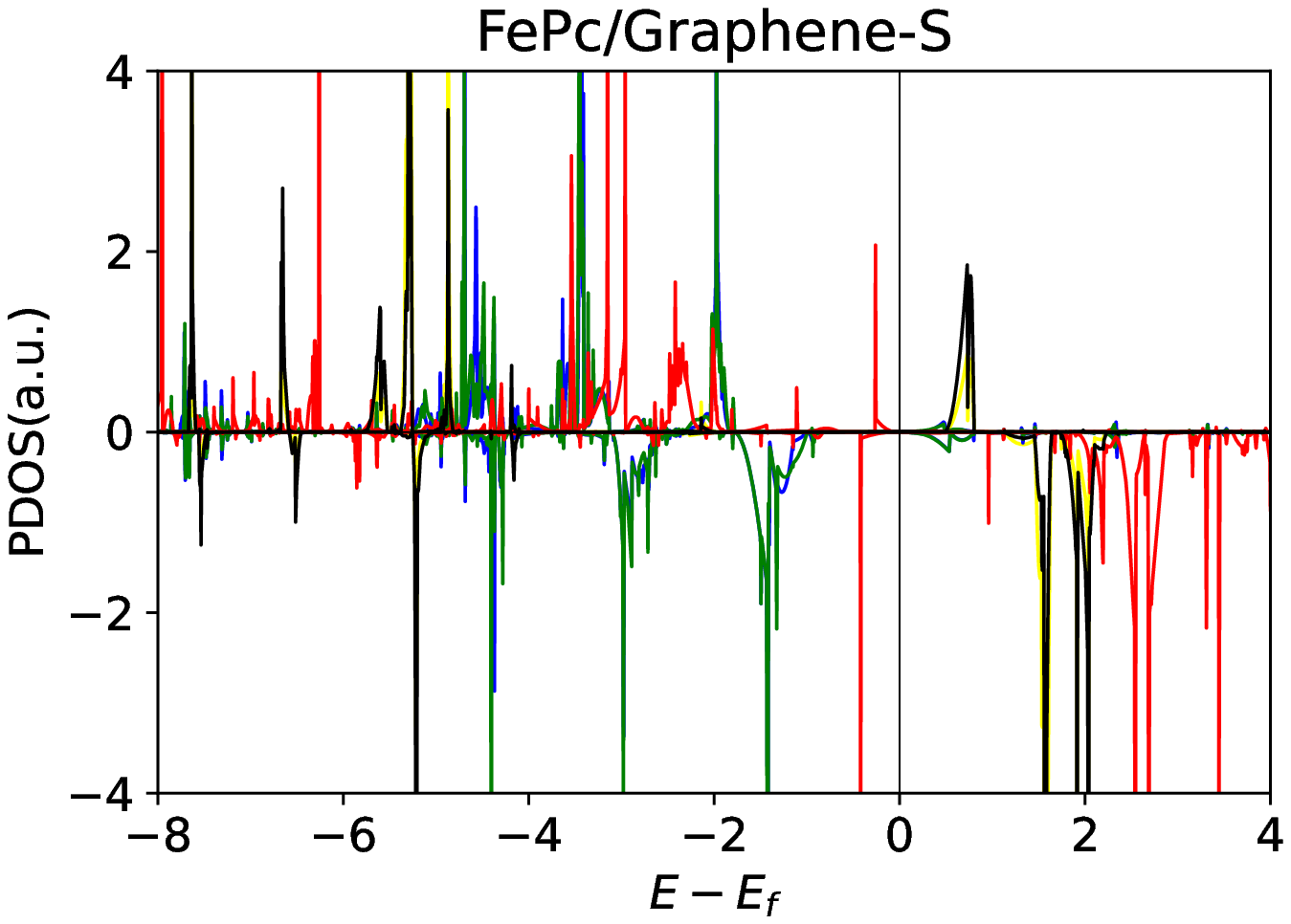}
  \caption{}
  \label{fig:FePcGrSFePDOS}
\end{subfigure}
\hfill
\begin{subfigure}{0.4\linewidth}
  \centering
  \includegraphics[width=\linewidth]{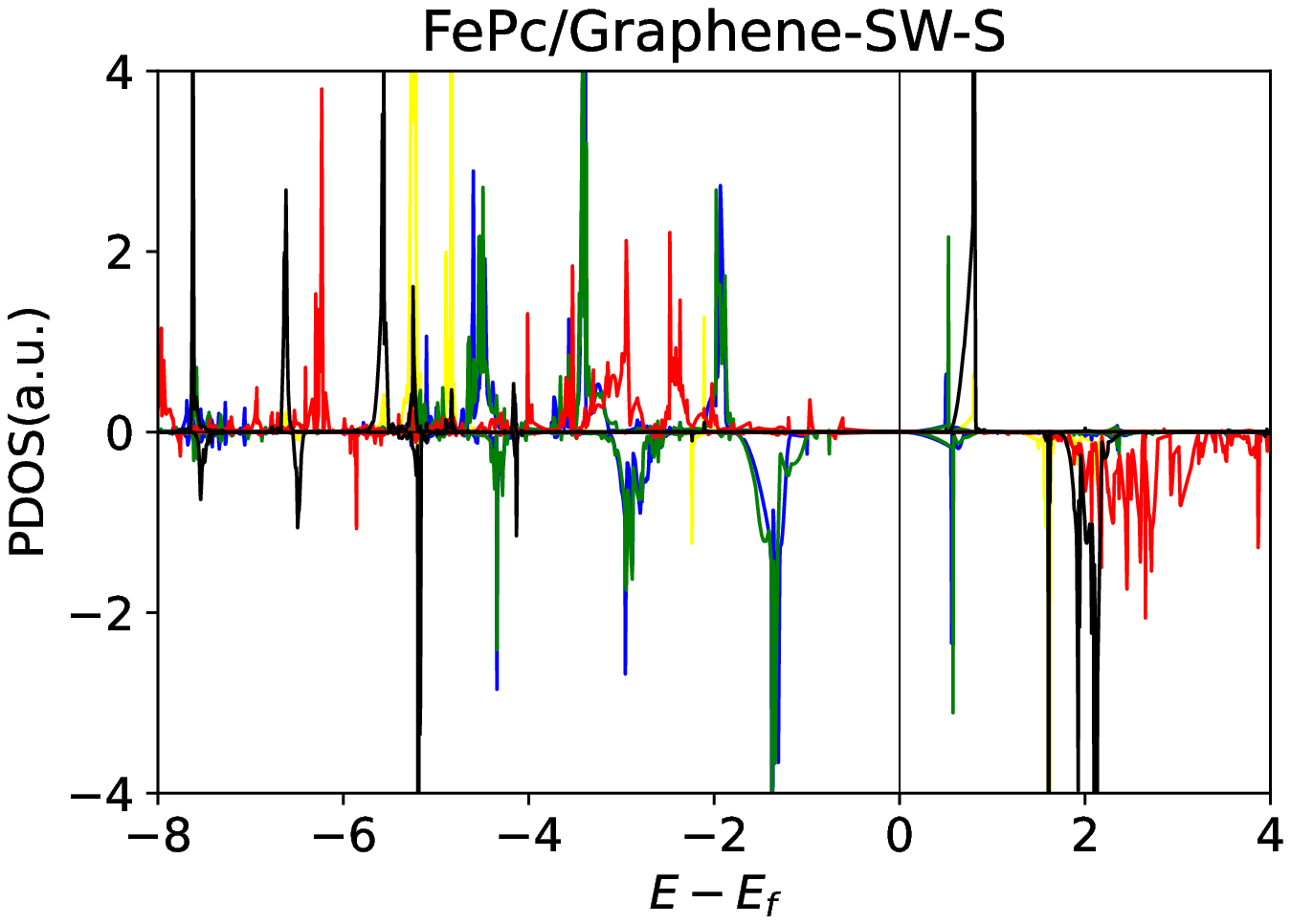}
  \caption{}
  \label{fig:FePcGr57SFePDOS}
\end{subfigure}

\caption{Projected densities of states for the iron atom d-orbitals in studied FePc/Graphene layers (yellow - ${d_{xy}}$, blue - ${d_{xz}}$, green - $d_{yz}$, red - $d_{z^2}$, black - $d_{x^2-y^2}$). For clarity of the pictures, the gaussian broadening parameter of the order of the energy grid step (0.005 eV) has been chosen.}
\label{fig:AllFePcGrFePDOS}
\end{figure}

\begin{figure}[ht!]
    \centering
    \includegraphics[width=\linewidth]{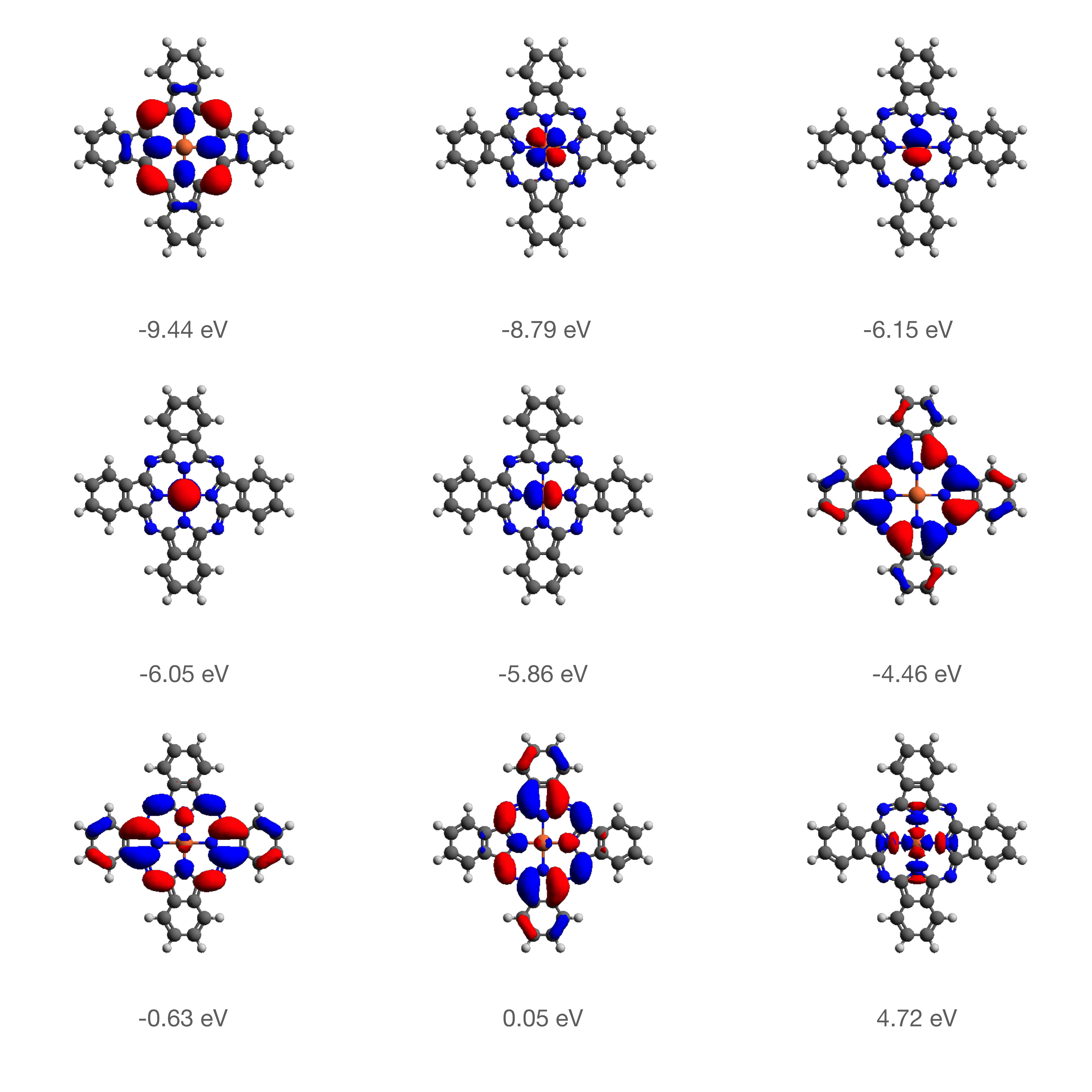}
    \caption{FePc active molecular orbitals in CAS(10,9) and their energies. Red and blue colours mean positive and negative charged densities, respectively.}
    \label{fig:FePcOrbitals}
\end{figure}

\newpage

\begin{figure}[ht!]
    \centering
    \includegraphics[width=\linewidth]{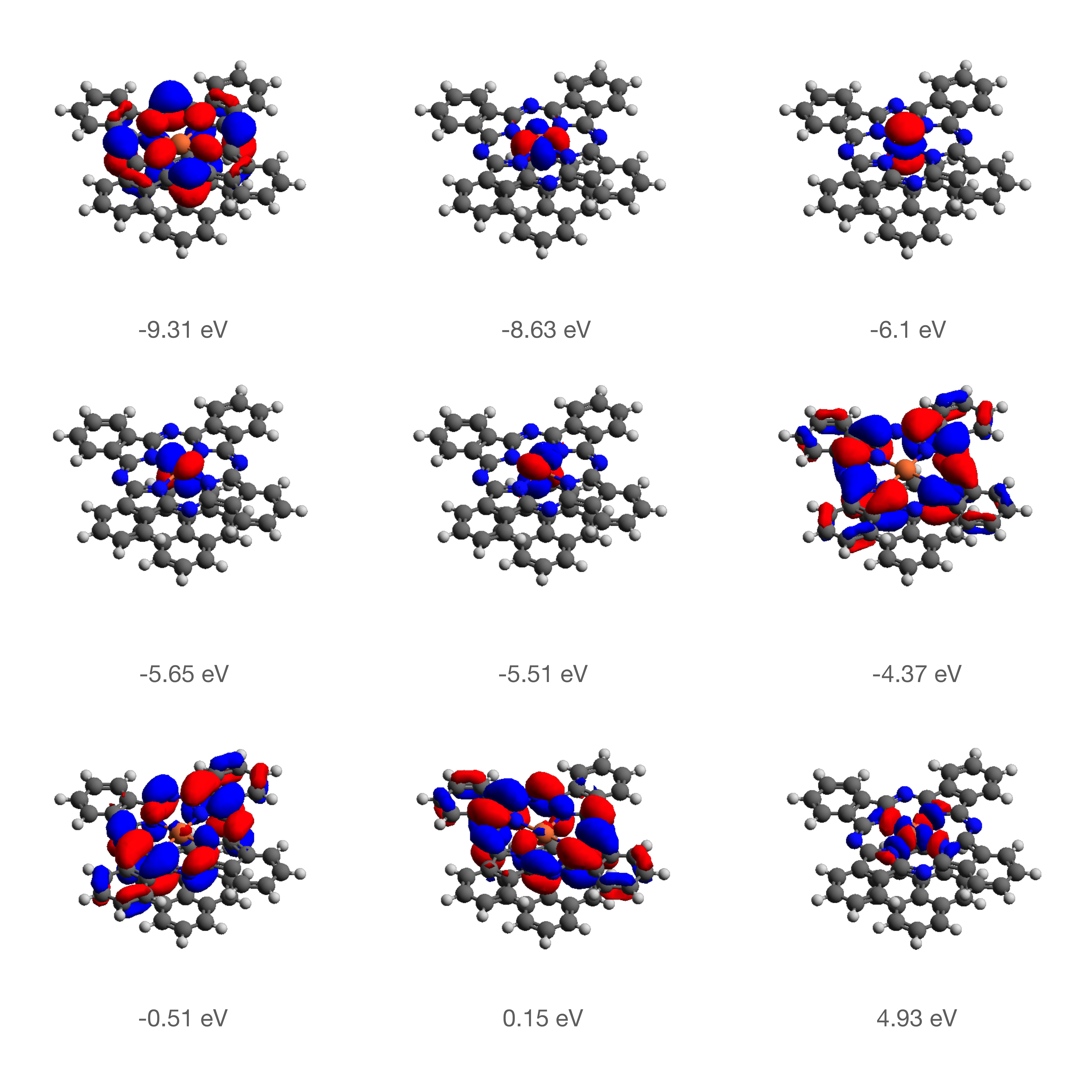}
    \caption{FePc/Pyrene active molecular orbitals in CAS(10,9) and their energies. Red and blue colours mean positive and negative charged densities, respectively.}
    \label{fig:FePcPyreneOrbitals}
\end{figure}

\begin{figure}[ht!]
    \centering
    \includegraphics[width=\linewidth]{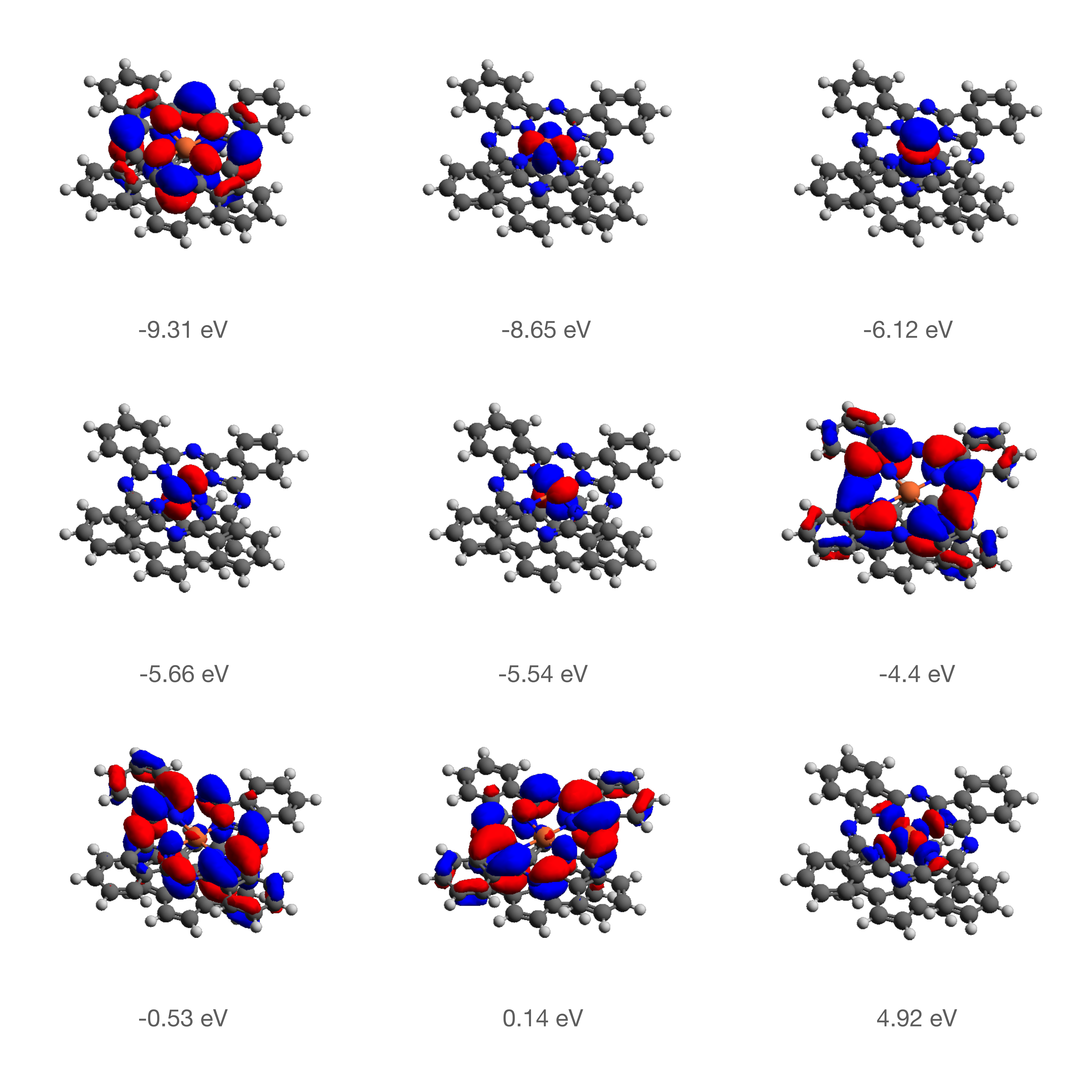}
    \caption{FePc/Pyrene-SW active molecular orbitals in CAS(10,9) and their energies. Red and blue colours mean positive and negative charged densities, respectively.}
    \label{fig:FePcPyrene-SWOrbitals}
\end{figure}

\begin{figure}[ht!]
    \centering
    \includegraphics[width=\linewidth]{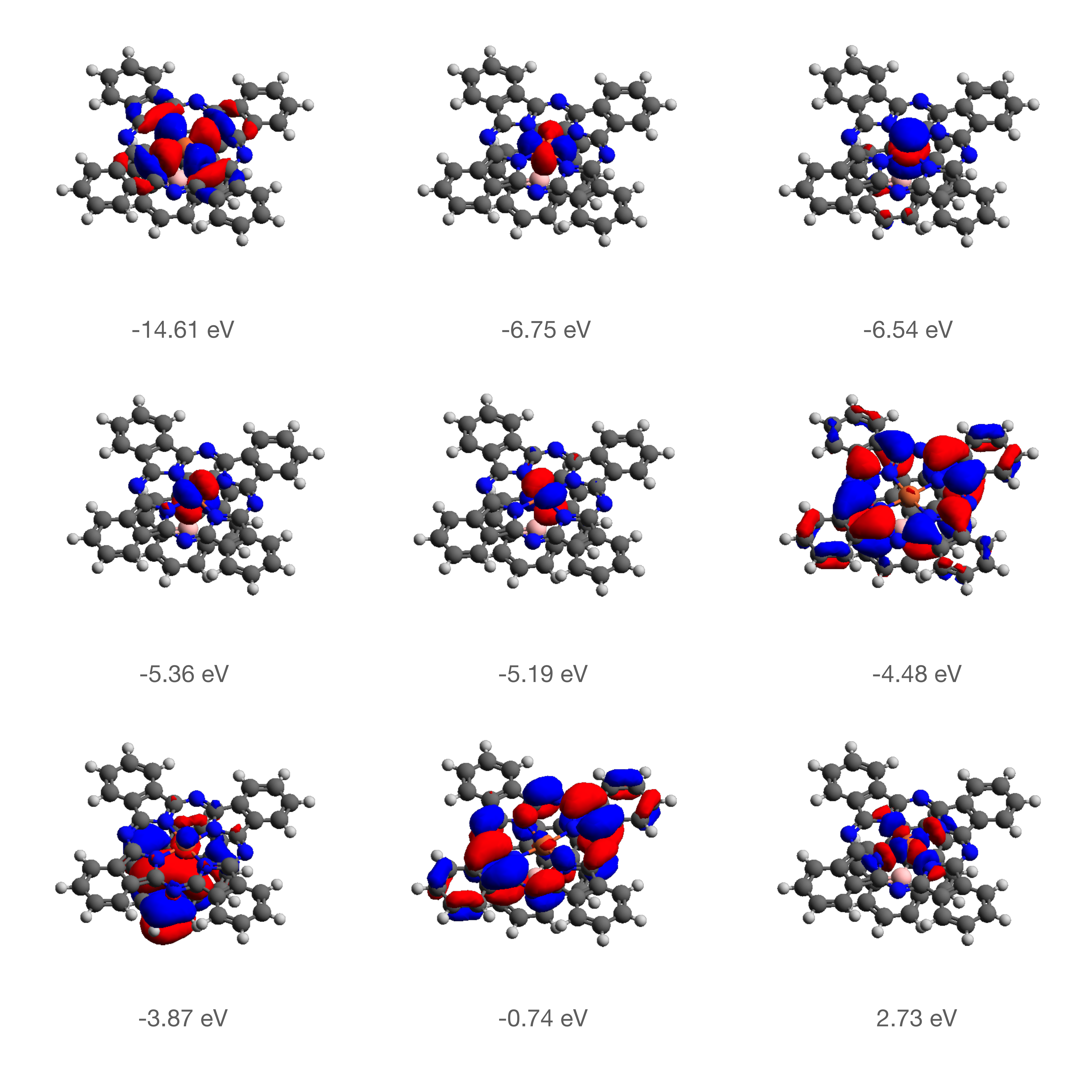}
    \caption{FePc/Pyrene-B active molecular orbitals in CAS(11,9) and their energies. Red and blue colours mean positive and negative charged densities, respectively.}
    \label{fig:FePcPyrene-BOrbitals}
\end{figure}

\begin{figure}[ht!]
    \centering
    \includegraphics[width=\linewidth]{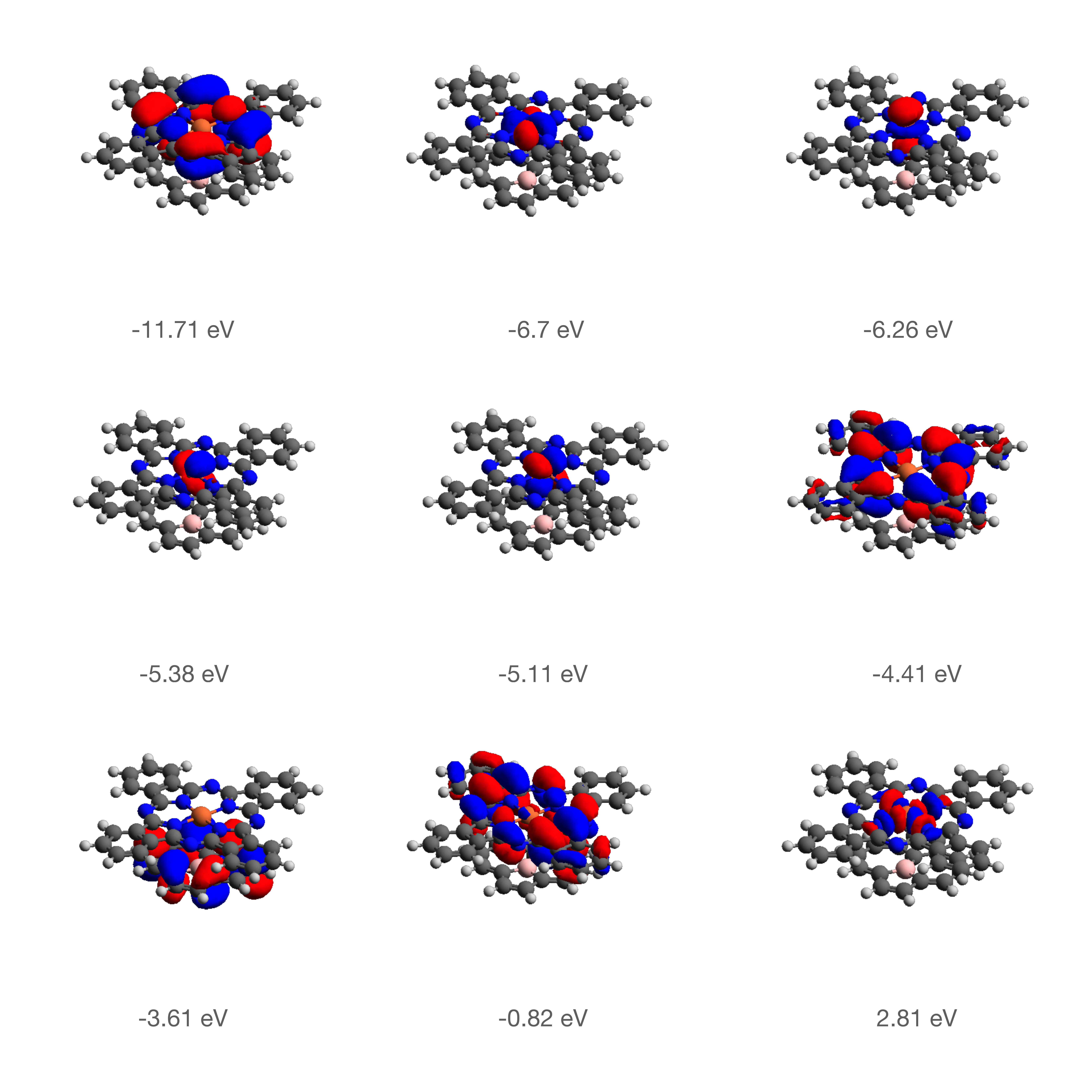}
    \caption{FePc/Pyrene-SW-B active molecular orbitals in CAS(11,9) and their energies. Red and blue colours mean positive and negative charged densities, respectively.}
    \label{fig:FePcPyrene-SW-BOrbitals}
\end{figure}

\begin{figure}[ht!]
    \centering
    \includegraphics[width=\linewidth]{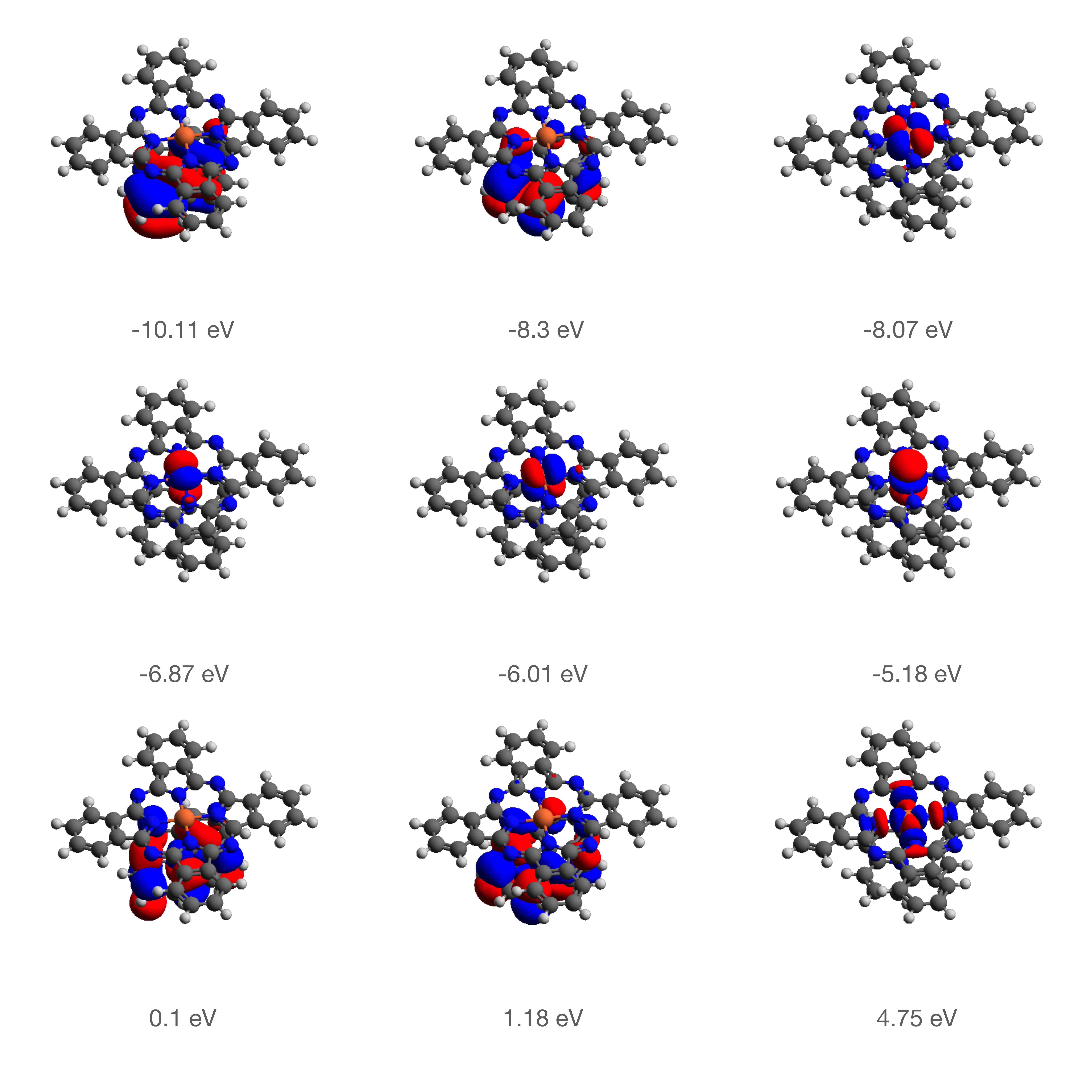}
    \caption{FePc/Pyrene-N active molecular orbitals in CAS(11,9) and their energies. Red and blue colours mean positive and negative charged densities, respectively.}
    \label{fig:FePcPyrene-NOrbitals}
\end{figure}

\begin{figure}[ht!]
    \centering
    \includegraphics[width=\linewidth]{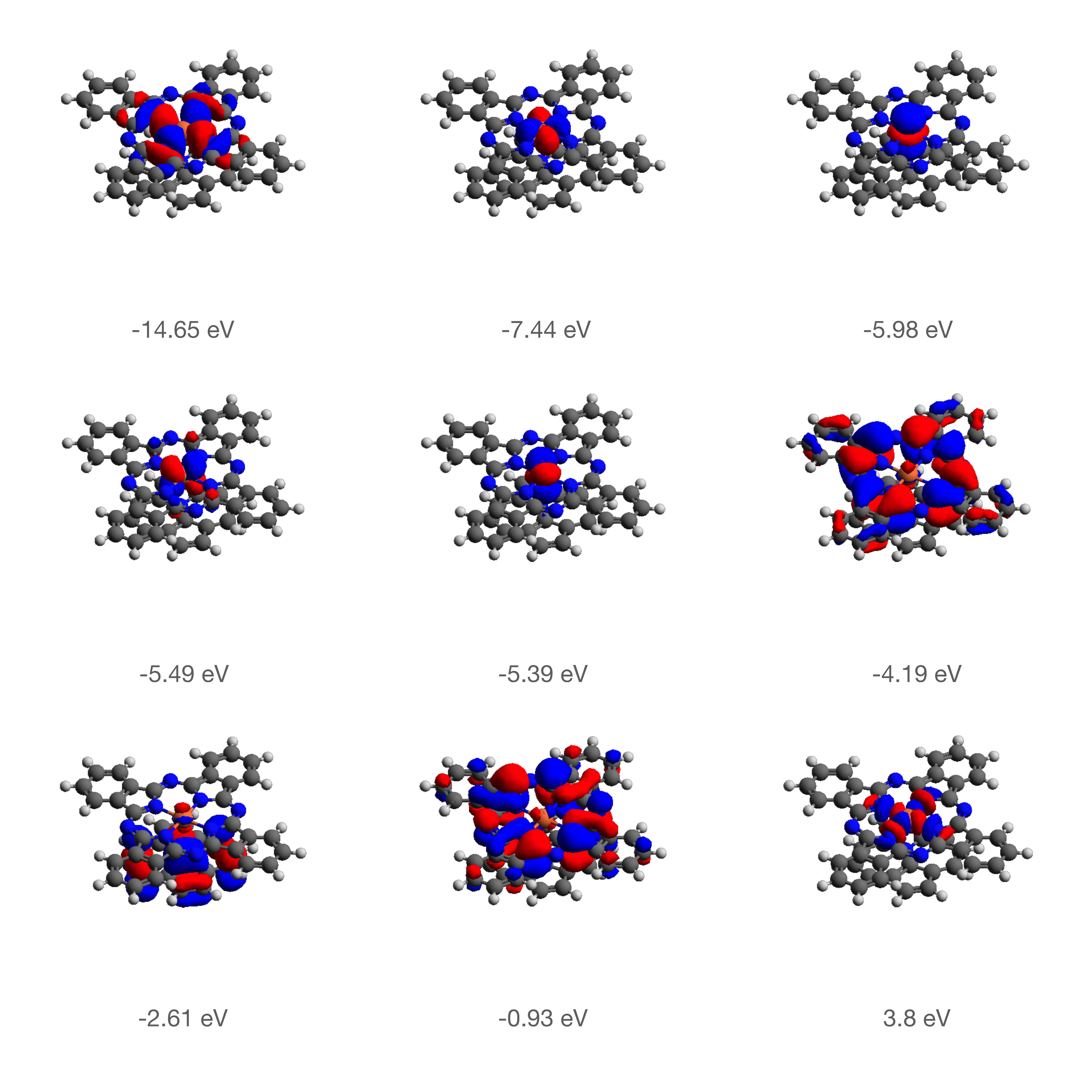}
    \caption{FePc/Pyrene-SW-N active molecular orbitals in CAS(11,9) and their energies. Red and blue colours mean positive and negative charged densities, respectively.}
    \label{fig:FePcPyrene-SW-NOrbitals}
\end{figure}

\begin{figure}[ht!]
    \centering
    \includegraphics[width=\linewidth]{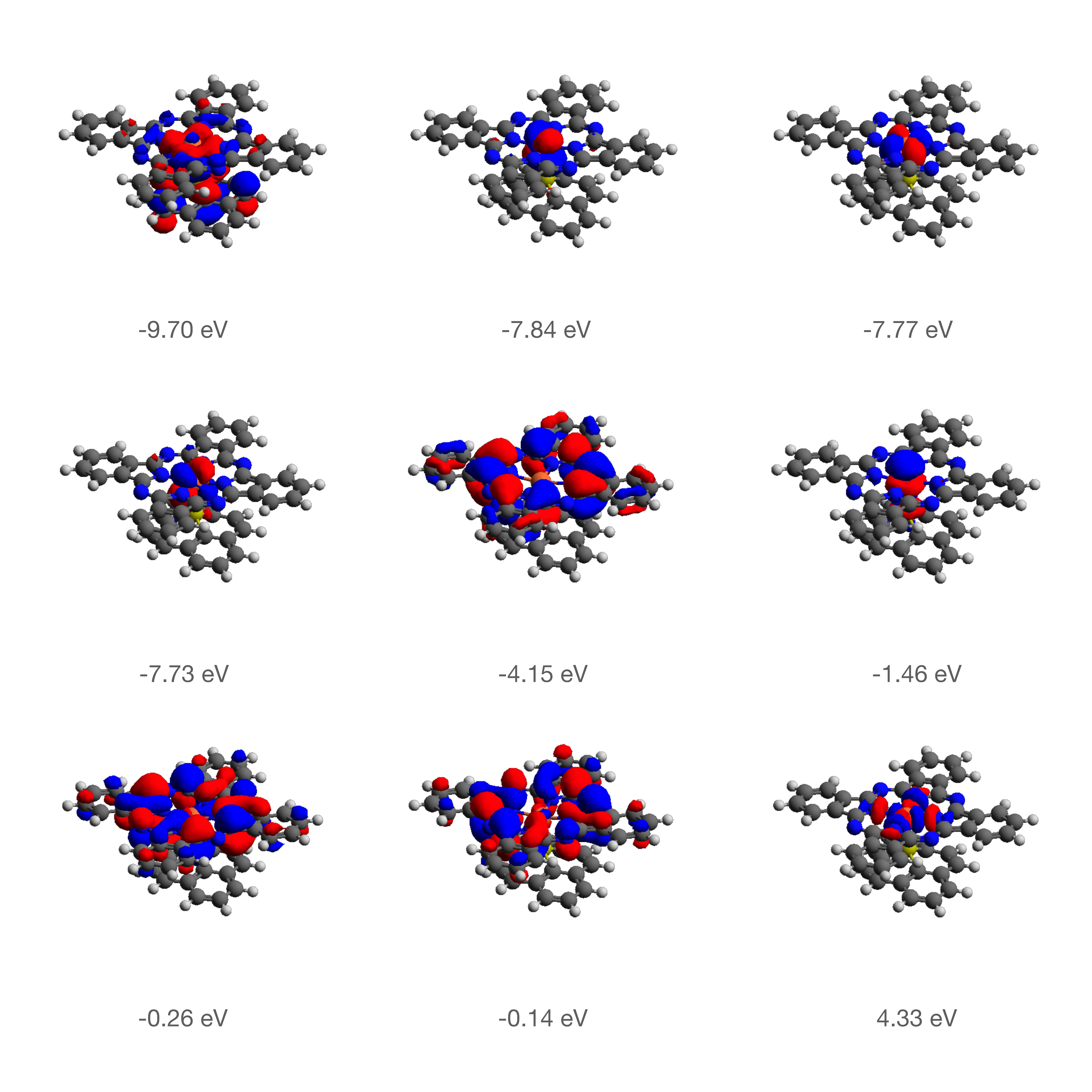}
    \caption{FePc/Pyrene-S active molecular orbitals in CAS(10,9) and their energies. Red and blue colours mean positive and negative charged densities, respectively.}
    \label{fig:FePcPyrene-SOrbitals}
\end{figure}

\begin{figure}[ht!]
    \centering
    \includegraphics[width=\linewidth]{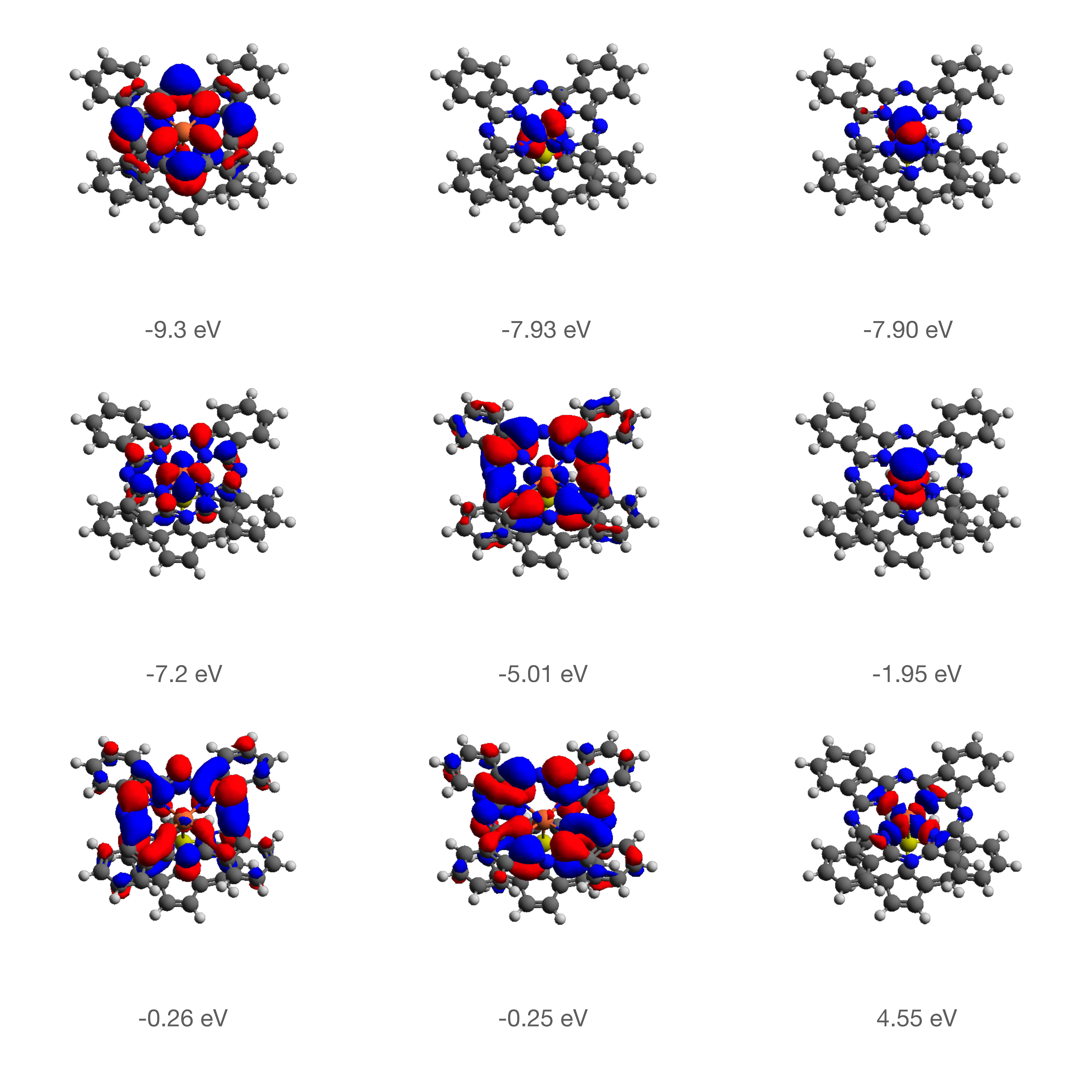}
    \caption{FePc/Pyrene-SW-S active molecular orbitals in CAS(10,9) and their energies. Red and blue colours mean positive and negative charged densities, respectively.}
    \label{fig:FePcPyrene-SW-SOrbitals}
\end{figure}